\documentclass[journal]{IEEEtran}
\usepackage{amsfonts}
\usepackage{amssymb}
\usepackage{amsthm}
\usepackage{amsmath,amsfonts,amssymb}
\usepackage{graphicx}
\usepackage{epstopdf}
\usepackage{verbatim}
\usepackage{bm}
\usepackage[ruled,vlined]{algorithm2e}
\usepackage{cite}
\usepackage{color}
\usepackage{multirow}
\usepackage{setspace}

\usepackage{gensymb}
\usepackage{booktabs}
\usepackage{threeparttable}
\usepackage{tabularx}
\usepackage{multirow}
\usepackage{subfigure}
\usepackage{array}
\newcommand{\PreserveBackslash}[1]{\let\temp=\\#1\let\\=\temp}
\newcolumntype{C}[1]{>{\PreserveBackslash\centering}p{#1}}
\newcolumntype{R}[1]{>{\PreserveBackslash\raggedleft}p{#1}}
\newcolumntype{L}[1]{>{\PreserveBackslash\raggedright}p{#1}}

\usepackage{url}

\newcommand{\figwidth}{8.8}
\IEEEoverridecommandlockouts

\begin{document}
	\title{A Survey on Millimeter-Wave Beamforming Enabled UAV Communications and Networking}
	
	\author{Zhenyu Xiao,~\IEEEmembership{Senior Member,~IEEE,}
		Lipeng Zhu, ~\IEEEmembership{Graduate Student Member,~IEEE,}
		Yanming Liu, ~\IEEEmembership{Graduate Student Member,~IEEE,}
		Pengfei Yi, ~\IEEEmembership{Graduate Student Member,~IEEE,}
		Rui Zhang,~\IEEEmembership{Fellow,~IEEE,}
		Xiang-Gen Xia,~\IEEEmembership{Fellow,~IEEE,}
		and Robert Schober,~\IEEEmembership{Fellow,~IEEE}
		\thanks{This work was supported in part by the National Key Research and Development Program (Grant Nos. 2016YFB1200100), and the National Natural Science Foundation of China (NSFC) (Grant Nos. 61827901 and 91738301).}
		\thanks{Z. Xiao, L. Zhu, Y. Liu, and P. Yi are with the School of Electronic and Information Engineering, Beihang University, Beijing 100191, China. (\{xiaozy,zhulipeng,liuyanming,yipengfei\}@buaa.edu.cn)}
		\thanks{R. Zhang is with the Department of Electrical and Computer Engineering, National University of Singapore, 117583, Singapore. (elezhang@nus.edu.sg)}
		\thanks{X.-G. Xia is with the Department of Electrical and Computer Engineering, University of Delaware, Newark, DE 19716, USA. (xxia@ee.udel.edu)}
		\thanks{R. Schober is with the Institute for Digital Communications, Friedrich-Alexander University of Erlangen-Nuremberg, Erlangen 91054, Germany. (robert.schober@fau.de)}
	}
	\vspace{-10mm}
	\maketitle
	\vspace{-10mm}
	\begin{abstract}
		Unmanned aerial vehicles (UAVs) have found widespread commercial, civilian, and military applications. Wireless communication has always been one of the core technologies for UAV. However, the communication capacity is becoming a bottleneck for UAV to support more challenging application scenarios. The heavily-occupied sub-6 GHz frequency band is not sufficient to meet the ultra high-data-traffic requirements. The utilization of the millimeter-wave (mmWave) frequency bands is a promising direction for UAV communications, where large antenna arrays can be packed in a small area on the UAV to perform three-dimensional (3D) beamforming. On the other hand, UAVs serving as aerial access points or relays can significantly enhance the coverage and quality of service of the terrestrial mmWave cellular networks. In this paper, we provide a comprehensive survey on mmWave beamforming enabled UAV communications and networking. The technical potential of and challenges for mmWave-UAV communications are presented first. Then, we provide an overview on relevant mmWave antenna structures and channel modeling. Subsequently, the technologies and solutions for UAV-connected mmWave cellular networks and mmWave-UAV ad hoc networks are reviewed, respectively. Finally, we present open issues and promising directions for future research in mmWave beamforming enabled UAV communications and networking.
	\end{abstract}


	\section{Introduction}
	\IEEEPARstart{U}{nmanned} aerial vehicles (UAVs) are aircrafts that are controlled by remote radio or an autonomous program without a human onboard. The embryonic stage of UAV can be traced back to more than 100 years ago. In the 1910s, the first automatic gyroscopic stabilizer was invented, which enabled an aircraft to keep its balance autonomously when flying forward. Thereafter, UAV technology rapidly evolved and UAVs were originally applied for military purposes, such as target drones, reconnaissance planes, and fighter aircrafts \cite{giones2017fromto}. In the past few decades, the industrial chains of chips, batteries, sensors, controllers, and communications are becoming more mature. UAV platforms are gradually developing towards miniaturization and low-power consumption, which greatly reduce the manufacturing cost. With the opening of low-altitude airspace, small-scale and medium-scale drones have been increasingly used in civilian fields, including geological prospecting, disaster rescue, forest fire prevention, power grid inspection, remote sensing, aerial photography, express delivery, and agricultural irrigation. The investment scale of the UAV industries around the world has increased thirtyfold during the past 20 years \cite{giones2017fromto,whitepaperUAV}. In the foreseeable future, the application of UAVs will become more widespread and promote the development of different kinds of linkage industries.
	
	Wireless communication is one of the most important technologies for UAV. The communication for UAVs can be categorized into a command \& control link and a data link \cite{zeng2016wirele}. The command \& control link supplies necessary environmental information, self-status reports, and control instructions to ensure the safe operation of a UAV, and thus requires high reliability and low latency. The data link usually carries the mission-related data, and should support higher data rates compared to the command \& control link. As the resolution of the on-board sensors becomes higher and the tasks become more strenuous and arduous, the traffic of mission-related UAV data is rapidly growing, especially in the backbone network \cite{yu2018cluste}. Some promising applications of UAVs, such as virtual reality (VR), augmented reality (AR), hologram, device-to-device (D2D) communications, intelligent transportation, and smart city, impose enormous demands on the communication system. However, the extremely congested sub-6 GHz frequency band is not sufficient to meet the increasing data rate requirements. In contrast, millimeter-wave (mmWave) communication with its abundant spectrum resource has the potential to support the high-throughput and low-latency requirements of various UAV application scenarios \cite{xiao2017millim,wang2018millim}. In addition, with the deployment acceleration of the fifth-generation (5G) mobile network, the standardization and commercialization of 5G mmWave cellular network are imminent \cite{wang2018millim,uwaechia2020acompre}.
Thus, UAV-assisted wireless communication is becoming an important research direction for enhancing service ability of beyond 5G (B5G) and sixth-generation (6G) networks \cite{zhang2019asurve,zeng2019access}.

    \subsection{Potentials, Applications, and Challenges}
	For the evolution from 5G to 6G, both UAV communication and mmWave techniques play important roles. UAVs are an indispensable component of the three-dimensional (3D) heterogeneous network architecture envisioned for 6G \cite{geraci2021whatwi,dao2021survey}, which aims to facilitate 3D seamless communications. In particular, UAVs provide a higher probability for line-of-sight (LoS) links, and can be flexibly deployed in a cost-effective manner to support high-rate communication for remote areas or emergency situations. However, the use of UAVs also introduces challenges for power consumption, mobility, and long-distance transmission.
	
	The wideband transmission in the mmWave frequency bands is widely considered to be a key feature of 5G new radio (NR), and will play even a more far-reaching role in 6G. MmWave systems commendably cater to the requirement of extremely high bit rates, and the emergence of smart surfaces and environments in 6G \cite{saad2020avisio}. However, mmWave signals suffer from high propagation loss, comprising the free-space path loss, atmospheric and molecular absorption, and rain attenuation. Thus, directional antennas or antenna arrays should be utilized in mmWave communication systems to acquire high beam gains and improve the transmission range, where beamforming plays an important role \cite{niu2015asurve,qi2020hierar}. Besides, beamforming potentially improves the spectrum efficiency through coherent combining and an increased antenna aperture~\cite{ahmed2018asurve}.
	
	The combination of UAV and mmWave technologies may provide more opportunities for future communication networks, where mmWave beamforming enabled UAV communication systems and networks have a high technical potential and wide applications. Meanwhile, there are still important challenges to be addressed. In the following, the main potentials, applications, and challenges of mmWave-UAV communications are summarized.
	
	\subsubsection{Potentials}
	Enabling mmWave communication for UAV platforms and utilizing UAVs to assist mmWave cellular networks provide the following unique advantages.
	
	\begin{figure*}[t]
		\begin{center}
			\includegraphics[width=16 cm]{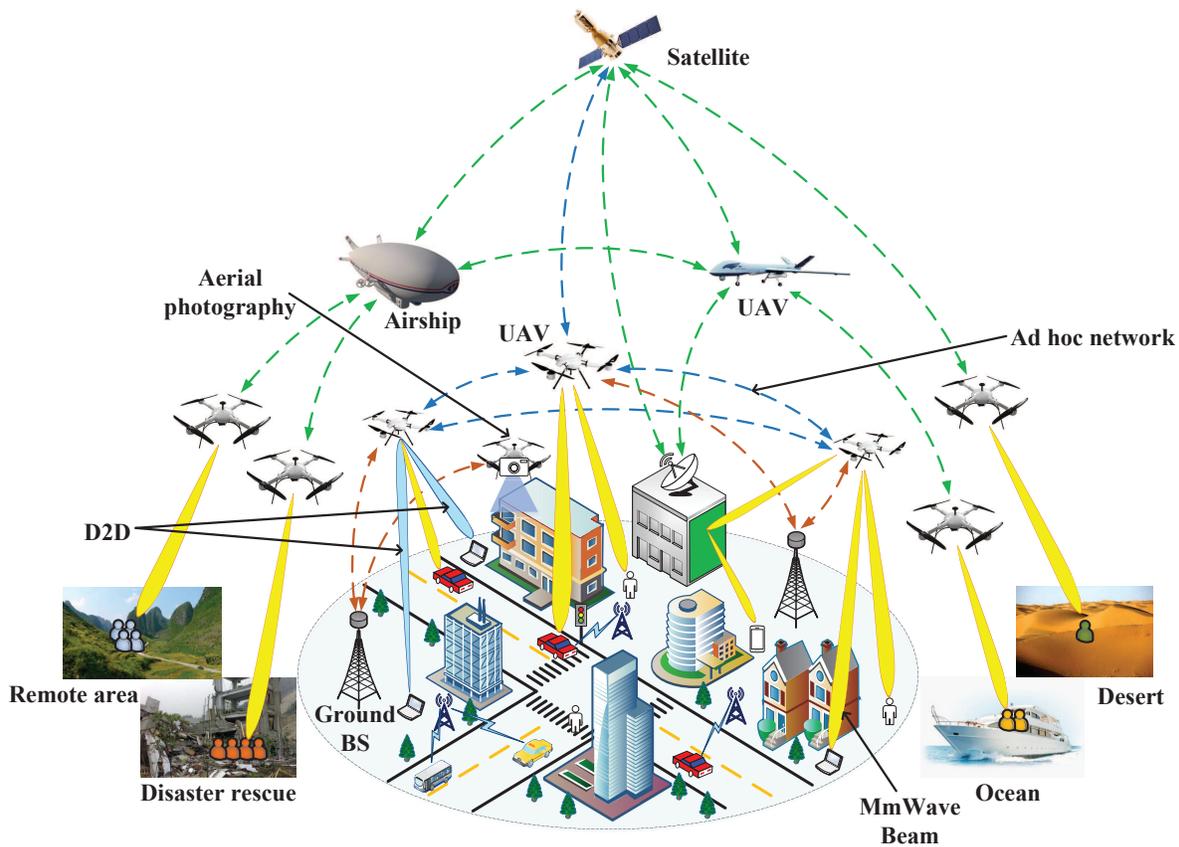}
			\caption{Illustration of the application scenarios for mmWave beamforming enabled UAV communications and networking.}
			\label{Fig:network_architecture}
		\end{center}
	\end{figure*}
	
    \begin{itemize}
    \item[$\bullet$] \emph{Broad Bandwidth:} MmWave communication is a promising technology for overcoming the spectrum scarcity in UAV communications and networking. Excluding unfavorable bands, such as the oxygen absorption band and the water vapor absorption band, the available bandwidth in the mmWave bands is more than 150\,GHz \cite{xiao2017millim}. There are several continuous frequency bands that can be potentially utilized for UAV communications, including the 24 GHz band (24.25-27.5 GHz), 28 GHz band (27.5-29.5 GHz), 38 GHz band (37-40 GHz), 45\,GHz (42.3-47.3\,GHz or 47.2-48.4\,GHz), and E band (71-76\,GHz, 81-86\,GHz, and 92-95\,GHz) \cite{feng2019spectr}. Recently, Facebook Inc. demonstrated an air-to-ground (A2G) E-band communication link, which was shown to achieve a peak data rate of 40\,Gbps bi-directionally \cite{tang2019demons}. The broad bandwidth in the mmWave frequency band facilitates the support of ultra high data traffic and diverse UAV applications.

    \item[$\bullet$] \emph{Short Wavelength:} The wavelength of mmWave signals typically ranges from 1 mm to 10 mm, which makes it possible to equip a large number of antennas in a small area \cite{xiao2017millim,wang2018millim}. For example, for a half-wavelength spaced antenna array at\,38 GHz, more than 600 antennas can be packed in an area of 1 square decimeter. The small-sized components in the mmWave frequency band are tailor-made for UAV platforms which are size and weight constrained. Large-scale arrays can provide considerable beam gains to improve the channel quality in UAV communication systems.

    \item[$\bullet$] \emph{Spatial Sparsity:} MmWave channels exhibit sparsity in the spatial domain. It was shown that, for mmWave transceivers on the ground, an average of 3-4 clusters of multi-path components (MPCs) are relevant \cite{rappaport2015wideba}. The sparsity of mmWave channels is more dominant for UAV communications because there are fewer potential reflectors at the UAV side of the link \cite{xiao2016enabli,zhang2019resear}. Besides, the pencil-like mmWave beams enable directional communication and enhance the spatial sparsity of mmWave-UAV communications. These characteristics facilitate the reuse of the spectrum resource in the spatial domain \cite{singh2020spectr}.

    \item[$\bullet$] \emph{Flexible Beamforming:} The combination of mmWave communication and massive MIMO increases the degrees of freedom (DoFs) in UAV communication systems enabling spatial multiplexing and diversity gains. Due to the high altitudes of UAVs, the interference from the ground BSs, ground user equipments (UEs), and other aerial platforms is severe \cite{zhang2019asurve,zeng2019access}. The flexible mmWave beamforming provides DoFs for handling the dominant interference in the spatial domain. For example, the well-known zero-forcing (ZF) method can be utilized for digital beamforming to eliminate multi-user interference. It was shown that analog beamforming can also achieve a considerable performance gain by suppressing interference, even though only the phases of the antenna weights can be adjusted \cite{zhu2020millim}. Moreover, mmWave beams can be quickly adjusted to adapt to the mobility of UAVs and realize flexible coverage of the target region \cite{zhu20193dbeam}.

    \item[$\bullet$] \emph{High Operation Altitude:} MmWave channels are highly vulnerable to obstacles because the penetration and reflection losses of mmWave signals are very high. It was shown that LoS paths are typically more than 20 dB stronger than non-LoS (NLoS) paths in the mmWave frequency bands \cite{rappaport2015wideba,xiao2017millim}. Due to the high altitude of the UAV, it is likely that LoS links can be established for both A2G and air-to-air (A2A) communication scenarios. Even in situation where the link is blocked by obstacles, UAVs can quickly adjust their positions and attitudes to improve the channel quality.
    
    \item[$\bullet$] \emph{Controllable Mobility:} Due to the high propagation loss of mmWave signals, their transmission distance is limited. Although the dense deployment of micro and pico BSs can overcome this issue, this entails high hardware and labor costs, which is undesirable for providing wireless coverage in remote regions. Thanks to their controllable mobility, UAVs can be rapidly deployed as aerial access points (APs) and flexibly relocated according to the requirements of the ground UEs \cite{zhang2019asurve,zeng2019access}. The deployment of UAV APs can efficiently enlarge the coverage area of mmWave cellular networks, especially in emergency communication scenarios.
    \end{itemize}

    \subsubsection{Applications}
By leveraging the abundant bandwidth and directional beamforming, mmWave-UAV communications can be applied to a host of different scenarios as shown in Fig. \ref{Fig:network_architecture}. For a UAV serving as an aerial BS, the backhaul link is a performance bottleneck. For ground BSs, in general, wire backhaul links over fiber are employed, which are not possible for UAV BSs. High-capacity wireless backhaul links for UAV BSs may be enabled by exploiting mmWave communications \cite{gao2015mmwave}. Besides, UAVs can be used as aerial relays for increasing the data rate between two or multiple ground nodes in the mmWave frequency bands. Since mmWave signals have a poor diffraction capability and their penetration loss is very high, UAV relays flexibly adjusting their positions can be deployed to establish high-quality communication links \cite{zhu2020millim}. When UAVs are used for aerial photography, surveillance, and remote sensing, the uplink data traffic is usually high because they need to transmit visual and sensor information to the control center \cite{zeng2019cellul}. In this case, mmWave access links can be established for the UAVs to connect with the core networks via ground BSs or satellites. Furthermore, multiple aircrafts can carry out complex tasks cooperatively, where high-capacity communication and networking are crucial for supporting the information exchange between different nodes \cite{gupta2016survey}. For a UAV swarm, the directional transmission in the mmWave frequency bands can efficiently improve communication security under hostile attacks such as jamming and eavesdropping \cite{feng2019spectr}.

	\subsubsection{Challenges}
The combination of mmWave techniques and UAV communications still faces many inherent and new challenges as detailed in the following.

    \begin{itemize}
    \item[$\bullet$] \emph{Complex Antenna Design:} Compared to terrestrial infrastructures, UAVs have size, weight, and power (SWAP) constraints and come in various shapes. The design of mmWave antenna arrays for UAV platforms entails strict requirements in terms of accuracy, size, and deployment. On the one hand, on-board antennas should be appropriately designed to meet the constraints of shape, size, and weight for UAV platforms. On the other hand, when deploying antennas and transceivers, electromagnetic compatibility should be particularly considered for UAVs which employ various avionics devices in different frequency bands.
    
    \item[$\bullet$] \emph{Complex Communication Characteristic:} Due to the high altitude and mobility, UAVs are more likely to be exposed to complex electromagnetic environments. The communication characteristics of mmWave-UAV communications are more complex than those of conventional ground communication systems. Compared to the low-frequency bands, the signals in the mmWave frequency bands are more sensitive to small-scale changes of the environment. For UAV systems, airflow disturbances and engine vibrations may cause fuselage swaying and jittering, which significantly impact the channel characteristics. Thus, the complex communication characteristics pose a non-negligible challenge for channel modeling in mmWave UAV communications.
    
    \item[$\bullet$] \emph{High Mobility:} To compensate for the high path loss in mmWave-UAV communications, narrow beams are usually shaped and the beams at transmitters and receivers must be aligned to achieve high array gains. However, due to the fast movement and random swaying of UAVs, it is challenging for a UAV to adequately and timely adjust its beam to track the varying channel. 
    	Moreover, because of the high frequencies, the Doppler shift, which results from the fast movement of UAVs, will be more significant compared to sub-6~GHz frequencies. This may result in severe inter-carrier interference and fast fading of the channel.
    
    \item[$\bullet$] \emph{Limited Coverage:} In mmWave-UAV communications, although forming a narrow beam at the UAVs helps to increase the array gain, it meanwhile reduces coverage on the ground. In general, the narrower the beam is, the smaller the coverage area is. Hence, in order to serve multiple dispersed users, a wider beam needs to be formed to cover all the users, or alternatively, multiple narrow beams need to be shaped to cover all the  users. Although a wider beam or multiple narrow beams are not difficult to shape with digital beamforming, it is challenging with analog beamforming, which is typically required at UAVs due to the SWAP constraints.
    
    \item[$\bullet$] \emph{Self Organization:} To accomplish more complex tasks, it may be necessary to form a mmWave UAV ad hoc network for the benefits of high autonomy, flexiblility, and self-healing. However, compared with the low-frequency ad hoc networks, the organization of mmWave UAV ad hoc networks is more difficult. One key reason is that the narrow beams make the neighbor discovery and routing more challenging. Meanwhile, resource management involves not only time-frequency blocks, but also spatial beams, and is usually also coupled with the routing strategy, which further intensifies the challenge.
    \end{itemize}

	\subsection{Existing Works and Our Contributions}
	Recently, a number of excellent survey and tutorial papers on UAV communications have been published. The authors of \cite{zeng2016wirele} provided an overview of UAV-aided wireless communications, discussing ubiquitous coverage, relaying, and information dissemination. The basic networking architecture, channel characteristics, and key design considerations for UAV communications were elaborated. It was demonstrated that with the aid of controlled mobility, UAVs could act as aerial relays to improve the throughput, reliability, and coverage of terrestrial communication systems. The authors of \cite{gupta2016survey} conducted a comprehensive survey on UAV networks, where the characteristics of UAV ad hoc mesh networks, routing under constrained circumstances, automating control via software defined networking (SDN), seamless handovers, and greening of UAV networks were discussed in depth. Focusing on the design mechanisms and protocols for airborne communication networks, the authors of \cite{cao2018airbor} reviewed important aspects of low-altitude platform (LAP) based communication networks, high-altitude platform (HAP) based communication networks, and integrated satellite-HAP-LAP networks. The authors of \cite{khuwaja2018asurve} provided a survey on channel modeling for UAV communications, where measurement methods and various channel characterizations were discussed. The authors also examined real-world challenges in UAV communications, such as airframe shadowing and channel non-stationarity. In \cite{khawaja2019asurve}, the same authors focused on A2G propagation channel modeling for UAVs, including large-scale fading, small-scale fading, MIMO channel characteristics and models, and channel simulations. A tutorial on UAV communications for B5G systems was provided in \cite{zeng2019access}, where the fundamentals of UAV communications were elaborated, including the channel model, antenna model, energy consumption model, and performance metrics. Two typical scenarios, namely UAV-assisted wireless communications and cellular-connected UAV were considered in combination with key technologies and solutions. Furthermore, the advantages, challenges, and promising technologies for cellular-connected UAV were summarized in \cite{zeng2019cellul}. The authors of \cite{mozaffari2019atutor} provided key guidelines for analyzing, designing, and optimizing UAV-based wireless communication systems. The main challenges for UAV communications, including 3D deployment, performance analysis, channel modeling, and energy efficiency, and corresponding mathematical tools, such as optimization theory, machine learning (ML), stochastic geometry, transport theory, and game theory, were comprehensively introduced. The authors of \cite{li2019uavcom} presented an exhaustive survey on UAV communication in 5G and B5G wireless networks, where the space-air-ground integrated network (SAGIN) architecture, physical layer technologies, network layer technologies, and joint communication, computing, and caching were comprehensively reviewed. The authors of \cite{sun2019physicMag} explored the challenges and opportunities for UAV communication systems from the perspective of physical layer security. The joint UAV trajectory design and resource allocation were exploited to guarantee security. In particular, mmWave communication and 3D beamforming were recommended to enhance the security of UAV communications. The authors of \cite{zhang2019cooper} investigated the cooperation mechanisms for cellular-connected UAV networks, where a cooperative sense-and-send protocol, trajectory design, and radio resource management were proposed to enhance the quality of service (QoS) of the cooperative cellular internet of UAVs.
Utilizing a large number of realistic case studies, the authors of \cite{geraci2021whatwi} surveyed UAV cellular communications for 5G NR, illustrated how to resolve the access, interference, and coverage issues by exploiting massive MIMO and mmWave technologies, and described A2A cellular communications. Besides, visions for B5G were put forward and some promising paradigms (i.e., non-terrestrial networks, cell-free architectures, artificial intelligence (AI), reconfigurable intelligent surface (RIS), and THz communication) for UAV cellular communications towards 6G were evaluated and discussed.
	The authors of \cite{dao2021survey} surveyed aerial radio access networks towards a comprehensive 6G access infrastructure in terms of the network features and design, system performance evaluation models, and enabling technologies regarding energy replenishment, operational management, and data delivery planes.
	The authors of \cite{zhang2021asurve} provided an introduction to the integrated access and backhaul (IAB) architecture in 5G NR, and discussed resource allocation, scheduling, caching, optical communications, and non-terrestrial communications in IAB networks.
	The authors of \cite{khan2021thero} analyzed key challenges for the deployment of UAV relays and discussed the optimal deployment of mmWave-enabled UAV relays. In addition, ML-based UAV-assisted networks and path loss models for the mmWave frequency band were briefly reviewed.
	
\begin{table*}[th] \scriptsize
            \begin{center}
			\caption{Summary and comparison of related papers}\label{tab:surv}
            \begin{spacing}{1.1}			\begin{tabular}
            		{|m{0.35cm}<{\centering}|m{0.35cm}<{\centering}|m{0.35cm}<{\centering}|m{0.35cm}<{\centering}|m{0.35cm}<{\centering}|m{0.35cm}<{\centering}|m{0.35cm}<{\centering}|m{12.2cm}<{\centering}|}
            		\hline
            		\textbf{Ref.}  &\textbf{MW} &\textbf{Sec. II} &\textbf{Sec. III} &\textbf{Sec. IV} &\textbf{Sec. V} &\textbf{Sec. VI} &\textbf{Main content/contributions} \\ \toprule[0.5pt]
            		\hline
            		\cite{zeng2016wirele}  &  &  &$\mu$ & &$\mu$  &  &Basic networking architecture and main channel characteristics of UAV-aided communications; three typical use cases, i.e., ubiquitous coverage, relaying, and information dissemination, to enhance system performance  \\
            		\hline
            		\cite{gupta2016survey}  &  &  &  &  &  &$\mu$  &Comprehensive survey on network characteristics, routing, seamless handover, and energy efficiency in UAV networks \\
            		\hline
            		\cite{cao2018airbor}  &  &  &$\mu$ &  &$\mu$  &$\mu$  &Characteristics, mobility control, resource management, and routing protocols of LAP networks; characteristics, channel models for HAP networks; integrated networks\\
            		\hline
            		\cite{khuwaja2018asurve}  &  &  &$\mu$ &  &  &  &Exhaustive survey on measurement  campaigns, propagation characteristics, and channel models for UAV communications  \\
            		\hline
            		\cite{khawaja2019asurve}  &  &  &$\mu$ &  &  &  &Comprehensive survey on measurement  approaches, characteristics, and implementation aspects of A2G channels   \\
            		\hline
            		\cite{zeng2019access}  &$\partial$  &$\partial$  &$\mu$  &$\mu$  &$\mu$  &   &Overview on fundamental mathematical models for channel, antenna, UAV energy consumption, and trajectory design; state-of-the-art results for UAV-assisted terrestrial communications and cellular-connected UAVs  \\
            		\hline
            		\cite{zeng2019cellul}  &$\partial$  &  &$\mu$ &  &$\mu$  &  &Brief discussion of potential benefits, requirements, new design considerations (i.e., channel characteristics and 3D coverage), and promising technologies (i.e., 3D beamforming and NOMA) of cellular-connected UAV communications   \\
            		\hline
            		\cite{mozaffari2019atutor}  &$\partial$  &  &$\mu$ &  &$\mu$  &$\mu$   &Comprehensive tutorial on potential applications, key research directions (i.e., channel modeling, deployment, trajectory, resource management, greening, and cellular UE), and analytical frameworks for UAV communications\\
            		\hline
            		\cite{li2019uavcom}  &$\partial$  &  &  &  &$\mu$  &$\mu$  &Exhaustive review of physical layer techniques (i.e., mmWave, NOMA, cognitive radio, energy harvesting), network layer techniques (i.e., D2D and SDN), and joint communication, computing, and caching in UAV communications\\
            		\hline
            		\cite{sun2019physicMag}  &$\partial$  &  &  &  &  &$\mu$  &Overview on trajectory design, resource allocation, and cooperative UAVs to mitigate eavesdropping; application of NOMA, MIMO, and mmWave to improve spectral efficiency and to guarantee security   \\
            		\hline
            		\cite{zhang2019cooper}  &$\partial$  &  &$\mu$  &  &$\mu$  &  &Overview on cooperative sense-and-send protocol, channel models, and key techniques (i.e., cooperative trajectory design and cooperative radio resource management) in cellular internet of UAVs    \\
            		\hline
            		\cite{geraci2021whatwi}  &$\partial$  &  &$\mu$  &  &$\partial$  &$\mu$  &Comprehensive overview on 5G NR massive MIMO and mmWave, A2A cellular communications, and new paradigms towards 6G (i.e., space-aided networks, cell-free architectures, AI, RIS, and THz) for UAV communications  \\
            		\hline
            		\cite{dao2021survey}  &$\partial$  &  &$\mu$  &$\mu$  &$\partial$  &  &Comprehensive survey on network architecture design, system models (i.e., propagation, energy, latency and mobility), enabling technologies (e.g., charging and data delivery), and applications of aerial access networks  \\
            		\hline
            		\cite{zhang2021asurve} &$\partial$ &  &  &   &$\partial$  &  &Comprehensive survey on IAB networks in terms of network mode, resource allocation, scheduling, caching, optical communications, and support of non-terrestrial networks \\
            		\hline
            		\cite{khan2021thero}  &$\partial$  &  &$\partial$  &  &$\partial$  &  &Review of key challenges for deployment of UAV relays,  optimal deployment of mmWave-enabled UAV relays, ML-based UAV-assisted networks, and path loss models for the mmWave frequency band   \\
            		\hline
            		\cite{xiao2016enabli}  &$\surd$  &  &$\partial$  &  &$\partial$  &  &Overview on UAV-connected mmWave cellular networks covering channel models, beam training and tracking, SDMA, and user discovery  \\
            		\hline
            		\cite{solomitckii2018techno} &$\surd$  &  &  &   &$\partial$  &  &UAV detection by utilizing 5G mmWave infrastructures with MIMO, beamforming, and beam steering techniques  \\
            		\hline
            		\cite{zhang2019resear} &$\surd$  &  &$\partial$ &  &$\partial$  &  &Overview on channel characteristics and challenges regarding channel estimation, beam training and tracking, detection, positioning and deployment, and interference mitigation in UAV-connected mmWave cellular networks  \\
            		\hline
            		\cite{feng2019spectr} &$\surd$  &  &  &   &$\partial$  &  &Spectrum management and consecutive time period optimization for mmWave-enabled UAV cellular networks  \\
            		\hline
            		\cite{wang2019multip}  &$\surd$  &   &  & &$\partial$   &   &CoDMA-enabled architecture, physical layer constellation coding, and MAC layer flexible access for mmWave UAV BSs  \\
            		\hline
            		\cite{tafintsev2020aerial}  &$\surd$  &  &  & &$\partial$  &  &Overview on UAVs for IAB in 5G NR and deployment performance of 5G mmWave networks supporting UAV APs and relays \\
            		\hline
            		\cite{zhang2019asurve}  &$\surd$  &$\partial$  &$\partial$  &  &$\partial$  &  &Antenna techniques, radio propagation channel, multiple access mechanisms, spatial configurations, resource management, security, and performance assessment for UAV terrestrial mmWave cellular networks where UAVs serve as BSs or relays \\
            		\hline
            		\cite{xiao2020uavcom}  &$\surd$  &$\partial$  &$\partial$  &  &$\partial$  &  &Overview on mmWave beamforming enabled UAV communications in terms of antenna structures, typical channel models, and three typical communication scenarios (i.e., communication terminal, AP, and backbone link) \\
            		\hline
            		$\blacktriangledown$ &$\surd$  &$\surd$  &$\surd$  &$\surd$  &$\surd$  &$\surd$  &Comprehensive survey on basic issues, important technologies, and state-of-the-art progress in mmWave beamforming enabled UAV-connected mmWave cellular networks and mmWave-UAV ad hoc networks \\
            		\hline
            	\end{tabular}
            	\begin{tablenotes}
            		\scriptsize
            		\item Ref. = reference, MW = mmWave, Sec. = section, $\mu$ = some issues discussed but not for mmWave, $\partial$: partially mentioned, $\surd$: covered, $\blacktriangledown$: this paper
            	\end{tablenotes}
            \end{spacing}
			\end{center}
		\end{table*}

	In the above survey and/or tutorial papers, the characteristics of mmWave channels and mmWave communication technologies for UAV platforms were not covered (e.g., \cite{zeng2016wirele,gupta2016survey,cao2018airbor,khuwaja2018asurve,khawaja2019asurve}) or only briefly mentioned (e.g., \cite{zeng2019access,zeng2019cellul,mozaffari2019atutor,li2019uavcom,sun2019physicMag,zhang2019cooper, geraci2021whatwi,dao2021survey,zhang2021asurve,khan2021thero}). Enabling mmWave communication to support the high-data-rate requirements of UAV cellular networks was preliminarily studied in \cite{xiao2016enabli}. The channel propagation characteristics for mmWave-UAV communications were shown, and several key technologies, such as fast beam training and tracking, mmWave spatial division multiple access (SDMA), impact of Doppler effect and blockage, and user discovery were analyzed. The authors of \cite{solomitckii2018techno} proposed to utilize 5G mmWave infrastructures for the detection of trespassing amateur drones. The authors of \cite{zhang2019resear} discussed the current state of the art, potentials, and challenges of mmWave-UAV communications. In particular, the channel characteristics and modeling, channel acquisition, precoding design, and recommendation techniques for UAV-connected mmWave cellular networks were introduced and analyzed. The authors of \cite{feng2019spectr} developed a novel spectrum management architecture for mmWave enabled UAV swarm networks to support broadband wireless transmission. Several potential techniques, such as interference detection, interference mitigation, integrated sub-6 GHz and mmWave frequency bands, and coordinated multi-point (CoMP) transmission, can be employed to opportunistically exploit low-altitude UAV swarms. The authors of \cite{wang2019multip} analyzed the requirements of UAV-aided mmWave communication in 5G ultra-dense networks (UDN), and proposed a novel link-adaptive constellation-division multiple access (CoDMA) mechanism. The authors of \cite{tafintsev2020aerial} reviewed the latest activities of the Third Generation Partnership Project (3GPP) regarding IAB system design, and evaluated the feasibility of IAB for supporting mmWave-UAV BSs and relays.
In addition to the above magazine papers \cite{xiao2016enabli,solomitckii2018techno,zhang2019resear,feng2019spectr,wang2019multip,tafintsev2020aerial}, a survey on 5G mmWave communications for UAV-assisted wireless networks was conducted in \cite{zhang2019asurve}, where the authors reviewed the key technologies from seven different aspects, i.e., antenna techniques, radio propagation channel, multiple access mechanisms, spatial configurations, resource management, security strategies, and performance assessment. However, the survey in \cite{zhang2019asurve} only focused on terrestrial mmWave cellular networks where UAVs serve as BSs or relays to serve ground UEs, while two important scenarios of mmWave-UAV communication, namely aerial UEs that connect to ground cellular networks and ad hoc networks that are constructed by UAV swarms, were not included in \cite{zhang2019asurve}. The networking technologies for mmWave-UAV communication systems, such as network architecture, neighbor discovery, and routing, were not covered in \cite{zhang2019asurve}. Besides, some up-to-date progress from the perspective of channel modeling, conformal antenna arrays, intelligent reflecting surface (IRS)/RIS assisted UAV communications, performance analysis, 3D flexible beam coverage, and seamless handover were not provided in \cite{zhang2019asurve}.
	
	\begin{figure*}[t]
		\centering
		\includegraphics[width=16 cm]{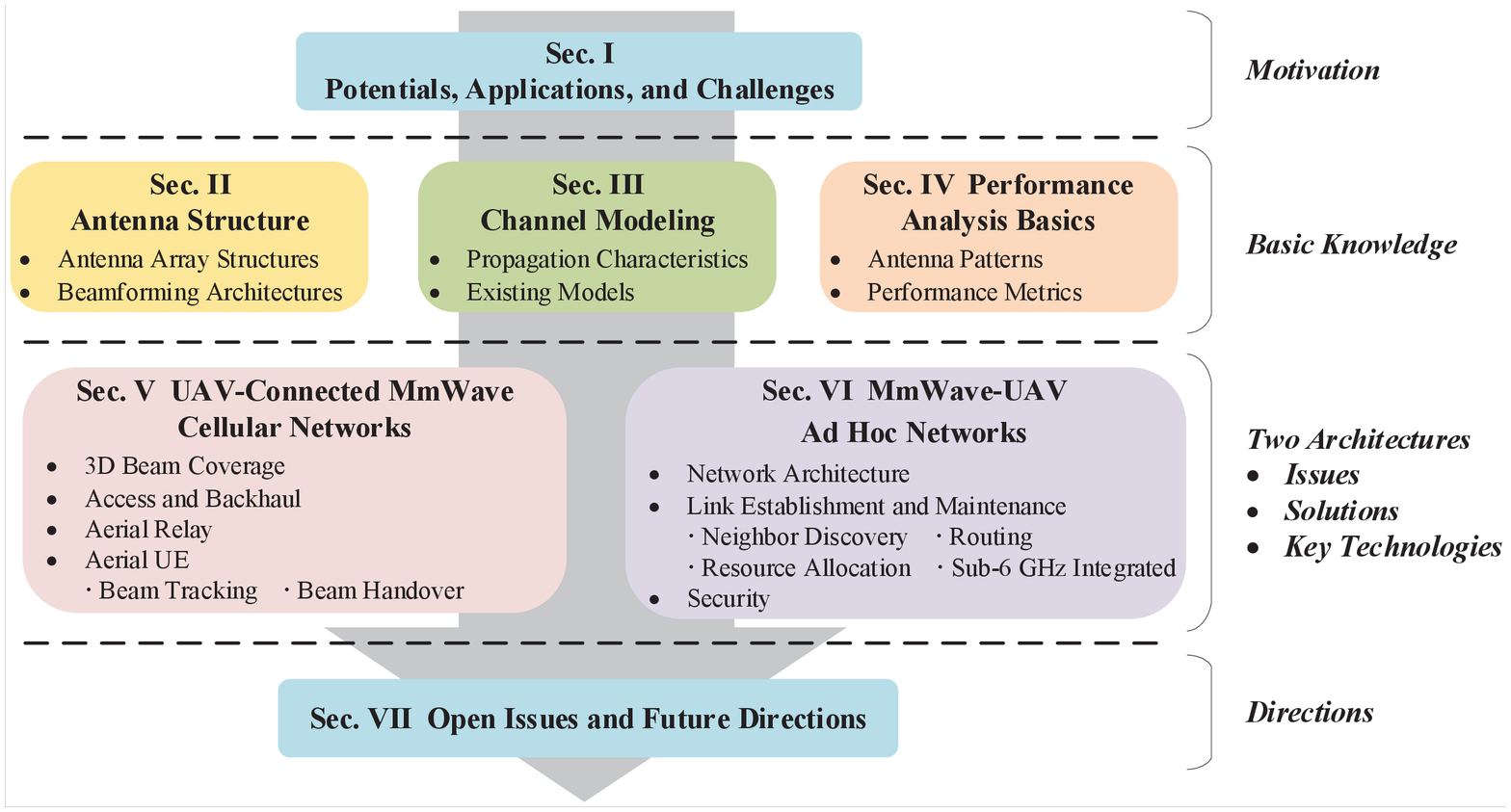}
		\caption{Organization of this paper.}
		\label{fig:organiz}
	\end{figure*}
	
	Compared with the above works, we aim at providing a more comprehensive survey on the basic issues, important technologies, and state-of-the-art progress for \emph{mmWave beamforming} enabled UAV communications and networking\footnote{In \cite{xiao2020uavcom}, we provided an overview on UAV communications employing mmWave beamforming, where the technical potentials, typical scenarios, and key challenges were preliminarily summarized. In contrast, this paper provides a much more comprehensive survey on mmWave beamforming enabled UAV communications and networking. More state-of-the-art works are reviewed from the perspective of antenna structures, channel modeling, and UAV-connected mmWave cellular networks. Furthermore, an outlook on the potential technologies and solutions for mmWave-UAV ad hoc networks is provided.}. Corresponding review on antenna structures and channel modeling is carried out. This paper significantly covers two promising paradigms for mmWave-UAV communications. The first one is UAV-connected mmWave cellular network, where UAVs serve as aerial APs/relays to assist ground UEs, or act as aerial UEs connected to the ground cellular networks. The other one is mmWave-UAV ad hoc networks, where multiple UAVs form an aerial ad hoc mesh network via directional mmWave communication links. Corresponding performance metrics and analysis methodologies are introduced focusing on the interplay of high-altitude characteristic of UAVs and the highly directional beamforming characteristic of mmWave communication systems. The key technologies, potential solutions, and open problems for different application scenarios are provided. The comparison of the contents of the existing survey/tutorial papers and this paper is summarized in Table~\ref{tab:surv}.
	
	\emph{Organization}: Fig. \ref{fig:organiz} provides an overview of the main content of this paper. Specifically, the potentials, applications, and challenges of mmWave beamforming enabled UAV communications and networking outlined in Section I provide the main motivation for this paper. Sections II-IV  survey  the basic background knowledge required for mmWave-enabled UAV communication system design and review the basic mathematical concepts which are indispensable for theoretical and practical performance analysis. In particular, in Section II, we introduce antenna array structures and beamforming architectures. Section III provides an overview on propagation characteristics and channel modeling. Section IV introduces antenna radiation patterns and performance metrics for mmWave-enabled UAV communications. Sections V and VI discuss two typical network architectures for mmWave-UAV communications. Specifically, Section V outlines the key concepts for UAV-connected mmWave cellular networks, including 3D beam coverage, access and backhaul, aerial relays, and aerial UEs. The important issues, solutions, key technologies, and new progress are comprehensively discussed and compared. In Section VI, we discuss potential technologies and solutions for the key issues arising for mmWave-UAV ad hoc networks, including the network architecture, link establishment and maintenance, integration of the sub-6 GHz and mmWave bands, and network security. Open problems and promising research directions for mmWave-UAV communications and networking are summarized in Section VII.
	 Finally, we conclude this paper in Section VIII. For ease of reading, the important acronyms employed in this paper are summarized in Table \ref{tab:acronym}.

	{\linespread{1}
		\begin{table*}[t]\scriptsize
			\caption{Summary of important acronyms}\label{tab:acronym}
			\begin{center}
				\begin{tabular}{p{1.2cm}p{6.8cm}p{0.01cm}p{1.2cm}p{6.8cm}}
					\cline{1-2}\cline{1-2} \cline{4-5}\cline{4-5}
					\textbf{Acronyms} & \textbf{Meaning}                        & \textbf{} & \textbf{Acronyms} & \textbf{Meaning}                           \\  \cline{1-2} \cline{4-5}
					2D    & Two-Dimensional                         &  & IRS    & Intelligent Reflecting Surface                       \\
                    3D    & Three-Dimensional                       &  & ITU &International Telecommunication Union   \\
					3GPP  &Third Generation Partnership Project  & & LAP   & Low-Altitude Platform \\
					4G    & Fourth-Generation                       &  & LTE  &Long Term Evolution \\                                
					5G    & Fifth-Generation                        &  & LoS    & Line-of-Sight        \\
					6G    & Sixth-Generation                        &  & MAB    & Multi-Armed Bandit                                   \\
					A2A   & Air-to-Air                              &  & MIMO   & Multiple-Input Multiple-Output                       \\
					A2G   & Air-to-Ground                           &  & ML     & Machine Learning                                     \\
					A2S   & Air-to-Satellite  &  & MMSE   & Minimum Mean Square Error                            \\
					ABC   & Artificial Bee Colony                   &  & mmWave & Millimeter-wave                                      \\
					ACK   & Acknowledgement                         &  & MPC    & Multi-Path Component                                 \\
					ADC   & Analog-to-Digital Converter             &  & NLoS   & Non-LoS                                              \\
					AE    & Antenna Element                         &  & NOMA   & Non-Orthogonal Multiple Access                       \\
					AF    & Amplify-and-Forward                     &  & NR     & New Radio        \\
					AI    & Artificial Intelligence    &  & NSEE   & Network-wide Secrecy Energy Efficiency    \\
                    AiP   & Antenna-in-Package                      &  & NST    & Network-wide Secrecy Throughput        \\
					AoA   & Angle of Arrival                        &  & OFDMA  & Orthogonal Frequency Division Multiple Access        \\
					AoC   & Antenna-on-Chip                         &  & OFDM   & Orthogonal Frequency Division Multiplexing           \\
					AoD   & Angle of Departure                      &  & OMA    & Orthogonal Multiple Access                           \\
					AP    & Access Point                            &  & PSO    & Particle Swarm Optimization                          \\
					AR    & Augmented Reality                       &  & QoS    & Quality of Service                                   \\
					ASLN  & Area Secure Link Number                 &  & RF     & Radio-Frequency                                      \\
					B5G   & Beyond 5G                               &  & RIS    & Reconfigurable Intelligent Surface                   \\
					BER   & Bit Error Rate                          &  & RSS    & Received Signal Strength                             \\
					BS    & Base Station                            &  & RTS    & Request-To-Send                                      \\
                    CCA   & Cylindrical Conformal Array             &  & SAGIN  & Space-Air-Ground Integrated Network      \\
					CDMA  & Code Division Multiple Access           &  & SDMA   & Spatial Division Multiple Access                     \\
					CF    & Compress-and-Forward       &  & SDN    & Software Defined Networking   \\
                    CMOS  & Complementary Metal Oxide Semiconductor    &  &SIC  & Successive Interference Cancellation  \\
					CoDMA & Constellation-Division Multiple Access  &  & SINR   & Signal-to-Interference-plus-Noise Ratio              \\
					CoMP  & Coordinated Multi-Point                 &  & SIR    & Signal-to-Interference Ratio                         \\
					CSI   & Channel State Information               &  & SNR    & Signal-to-Noise Ratio                                \\
					CTS   & Clear-To-Send                           &  & SWAP   & Size, Weight, and Power                              \\
					CU    & Centralized Unit                        &  & SWIPT  & Simultaneous Wireless Information and Power Transfer \\
					D2D   & Device-to-Device                        &  & TDL    & Tapped Delay Line                                    \\
					DAC   & Digital-to-Analog Converter             &  & TDMA   & Time Division Multiple Access                        \\
					DF    & Decode-and-Forward                      &  & TRN-R  & Receive Training                                     \\
					DLA   & Directed Lens Array                     &  & TRN-T  & Transmit Training                                    \\
					DoF   & Degree of Freedom                       &  & UAV    & Unmanned Aerial Vehicle                              \\
					DU    & Distributed Unit                        &  & UCA    & Uniform Circular Array                               \\
					FANET & Flying Ad Hoc Network                   &  & UDN    & Ultra-Dense Network                                  \\
					FD    & Full-Duplex                             &  & UE     & User Equipment                                       \\
					FDMA  & Frequency Division Multiple Access      &  & ULA    & Uniform Linear Array                                 \\
					GPS & Global Positioning System &  & UPA    & Uniform Planar Array             \\
					GSCM  & Geometry-based Stochastic Channel Model                  &  & URA    & Uniform Rectangular Array           \\
					HAP   & High-Altitude Platform                           &  & URLLC  & Ultra Reliable Low Latency Communications   \\
					HD    & Half-Duplex                             &  & VR     & Virtual Reality                                      \\
					HPBW  & Half-Power Beamwidth                    &  & WPAN   & Wireless Personal Area Network                       \\
					IAB   & Integrated Access and Backhaul          &  & WTS    & Wait-To-Send                                         \\
					IBCS  & Inter-Beam Coordinated Scheduling       &  & ZF     & Zero-Forcing                                         \\
					\cline{1-2} \cline{4-5}
				\end{tabular}
			\end{center}
		\end{table*}
	}

	\section{Antenna Structure}
	Antennas emit/receive electromagnetic waves into/from physical space and are an indispensable component of any wireless communication system. The antenna gain directly impacts the quality of signal transmission. Although the mmWave frequency bands provide unique advantages, such as large bandwidth and spatial sparsity, they also suffer from higher free-space path loss and more severe atmospheric attenuation compared to the sub-6 GHz bands. Therefore, high-gain antennas are essential for mmWave-UAV communications to compensate for the path loss. On the other hand, UAV platforms also impose additional limitations for antenna design because of the SWAP constraints.
	
	In the following, we first provide an overview on directional antennas, which radiate (receive) more power in (from) a specific direction so as to obtain a higher gain than omnidirectional antennas. Then, we review architectures for antenna arrays, which achieve a higher gain by using multiple connected antennas. Conformal antennas and arrays are also introduced, which are designed to conform some prescribed shapes, such as the aircraft body surface. Besides, beamforming technologies based on antenna arrays are reviewed. Furthermore, we introduce the emerging IRSs/RISs which enable passive reflective beamforming. Finally, we discuss the load capacity of UAVs carrying antenna arrays.

	\subsection{Directional Antennas}
	\subsubsection{Aperture Antennas}
	Typical aperture antennas include horn antennas, reflector antennas, and lens antennas. In general, these directional antennas have fixed directional patterns due to their carefully designed shape. In terms of their structure, a horn antenna is a gradually expanding waveguide, which shapes like a horn to direct a beam. A reflector antenna consists of a feed and a reflector, while a lens antenna includes a feed and a lens. By reflection and refraction, the reflector and lens antennas can change the propagation directions of the radio waves and concentrate the energy in a specific direction.
	
	Horn antennas are characterized by moderate directivity, wide band, low cost, and easily implemented. Except for being used as aperture antennas with medium gain, horn antennas are usually utilized as feed sources for large-aperture antennas, such as reflector antennas~\cite{shu2019awideb, shu2019adualc} and lens antennas~\cite{costa2009compac}. For instance, wideband horn antennas operating in the W-band (75-110~GHz) were designed to be used as primary feeds for reflector antennas, offering both high gain and full-duplex (FD) capability for mmWave communications in \cite{shu2019awideb, shu2019adualc}. In~\cite{costa2009compac}, a circular horn antenna was used as the feed for a lens antenna.
	
	Reflector antennas have advantages such as high gain, wide bandwidth, high angular resolution, and low cost. Compared to antenna arrays, reflector antennas are easier to implement since they do not need a complex feed network, and the feed source is simple. Hence, they are widely used in diverse applications, such as radar and satellite communications, especially in scenarios where sufficient space is available and a low-speed beam scanning is required. For example, a Ku-band wideband parabolic antenna is employed in the Global Hawk UAV and can achieve a communication rate of up to 50~Mbps \cite{naftel2011nasagl}.
	
	Lens antennas also achieve high gain, high directivity, and wide bandwidth. Besides, compared to reflector antennas, the feed source of lens antennas is located on the back of the aperture, which eliminates aperture shielding. However, in order to generate narrow beams, the size of the lens needs to be much larger than the wavelength of the radio waves. As a result, lens antennas are mainly used in the high frequency band, such as mmWave frequency bands \cite{costa2009compac,zeng2016millim}, so as to get a manageable antenna size. For example, a compact lens antenna allowing mechanical beam steering in the 60~GHz band was proposed in~\cite{costa2009compac}, which can be readily adjusted for HAP applications. In addition to being fed by a single antenna, lens antennas can also be fed by an array of antennas. Following this structure, mmWave lens MIMO was studied in~\cite{zeng2016millim}, and was shown to achieve significant throughput gains.

	\subsubsection{Integrated Antennas}
	The three types of aperture antennas discussed above, with large gains and good directivity, usually require significant space for deployment. This makes them only suitable for large and medium-sized UAVs with enough space. For example, Global Hawk UAV features a front bulge in order to house a satellite-communication antenna \cite{naftel2011nasagl}. However, large-scale antennas may not be appropriate for small-scale UAVs because of the SWAP constraints. In contrast, for such applications, integrated antennas are a potential solution.
	
	Integrated antennas comprise antenna-on-chip (AoC) and antenna-in-package (AiP) structures~\cite{zhang2019anover}. AoC implements an antenna (or antenna array) together with other circuits on a chip via a semiconductor process~\cite{zhang2009antenn}, while AiP packs an antenna (or antenna array) with a radio into a surface-mounted package~\cite{zhang2019anover}. Since AoC is integrated on a chip, it takes little space and has low cost. However, since the materials and processes of an antenna are constrained by the other elements on the same chip, the performance of AoCs may be degraded. Compared to AoC, AiP employs heterogeneous materials and processes for different functional blocks, and thus achieves a better performance but requires a higher cost. Generally speaking, both AoC and AiP are suitable for integrating antenna arrays to achieve higher gains~\cite{beer2013dbandg,gu2015wbands}. AoC and AiP structures are emerging as the mainstream antennas for various mmWave applications, such as high-capacity communications~\cite{gu2019develo,gu2018antenn}, high-resolution radio imaging~\cite{gu2015wbands}, and automotive radar~\cite{bauer2013a79ghz}. For mmWave-UAV communications, integrated antenna structures are promising technologies to accommodate the SWAP constraints.
	
	\begin{figure*}[t]
		\centering  
		\includegraphics[width=0.95\textwidth, trim=0 0 0 0,clip]{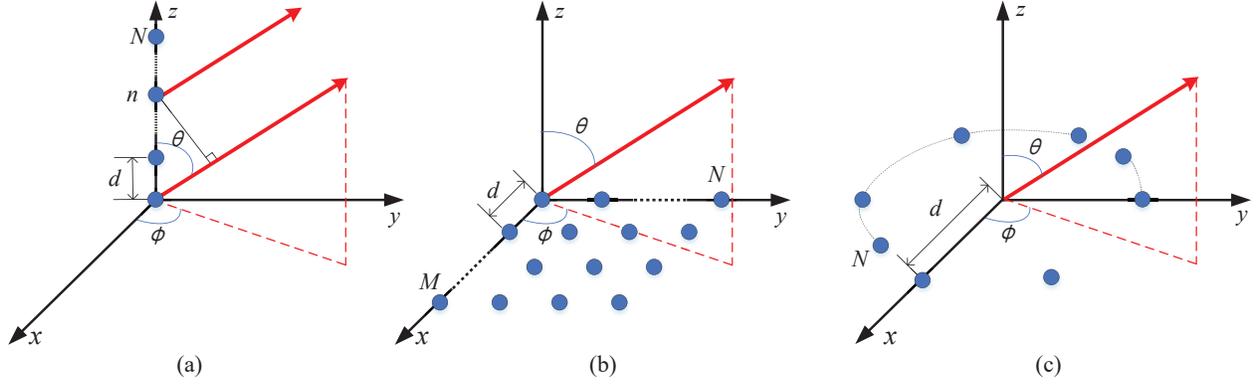}  
		\caption{Geometric structures of three types of antenna arrays: (a) ULA, (b) URA, (c) UCA. }  
		\label{fig:arrays_all}   
	\end{figure*}
	
	\subsection{Antenna Array}
	Antenna arrays are a powerful option to achieve high directional gains by employing multiple connected antenna elements (AEs) to work cooperatively. They will play an important role in mmWave-UAV communications, since the short wavelength of mmWave signals makes it possible to pack a large number of AEs in a small area~\cite{rangan2014millim,xiao2016enabli}. Different from aperture antennas, such as horn antennas, reflector antennas, and lens antennas, which have fixed radiation patterns due to their shapes, the overall radiation pattern of an antenna array is controlled by the type, number, spacing, and geometries of the elements~\cite{hemadeh2018millim}, which affords considerable flexibility. The most common geometries for an antenna array include linear, rectangular, and circular. If the AEs are equidistantly distributed, the corresponding arrays are called uniform linear array (ULA), uniform rectangular array (URA), and uniform circular array (UCA), respectively, as shown in Fig.~\ref{fig:arrays_all}.
	
	\begin{figure*}[t]
		\centering
		\subfigure[$N=8,d=0.5\lambda$]{\includegraphics[width=0.32\textwidth]{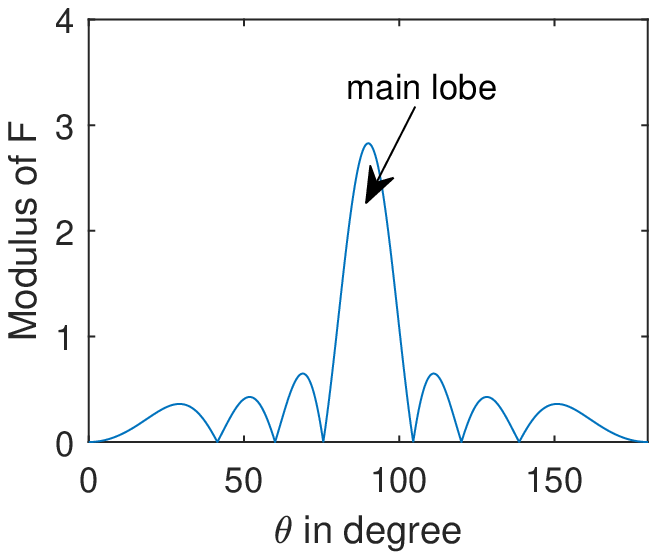}}
		\subfigure[$N=16,d=0.5\lambda$]{\includegraphics[width=0.32\textwidth]{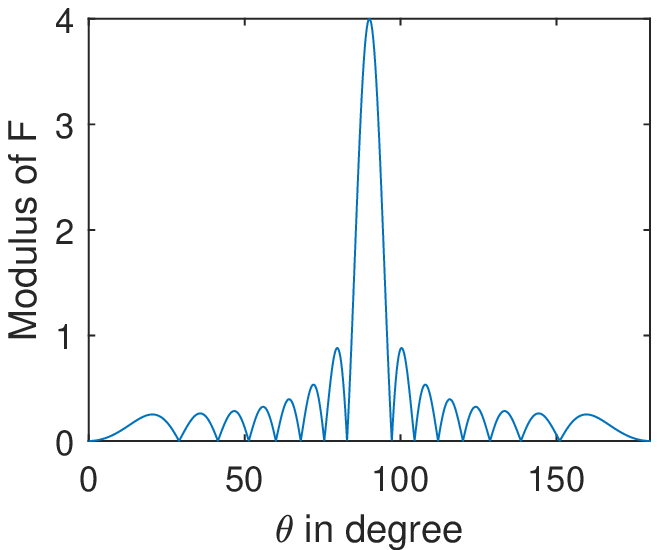}}
		\subfigure[$N=8,d=\lambda$]{\includegraphics[width=0.32\textwidth]{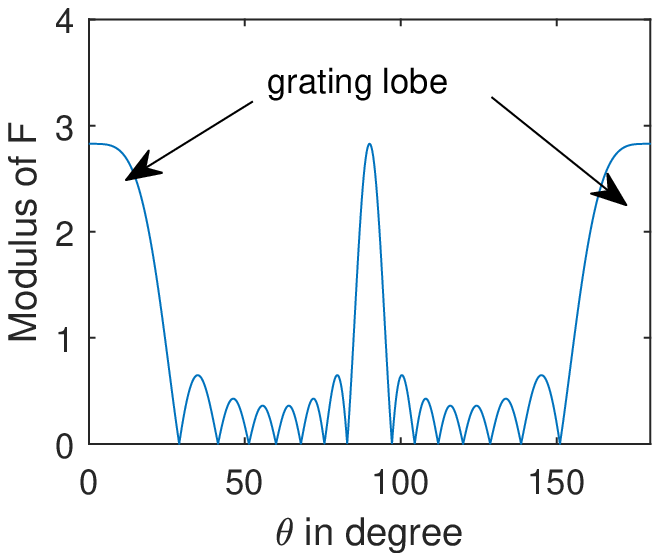}}
		\caption{Array factors under different array configurations.}
		\label{fig:AF}
	\end{figure*}
	
	In the following, we analyze the $N$-element ULA shown in Fig.~\ref{fig:arrays_all}~(a) as an example to demonstrate the basic principles of antenna array theory. The characteristics of URA and UCA can be readily derived using a similar method. We consider a scenario where the distance between the transceiver is much larger than the element spacing $d$, i.e., the far-field hypothesis holds. Then, electromagnetic waves arrive at the receiver in the form of a plane wave. The direction can be specified by azimuth angle $\phi$ and elevation angle $\theta$, and the beam direction vector is given by $\mathbf{\hat{r}}=[\sin\theta \cos\phi, \sin\theta \sin\phi,\cos\theta]^{\mathrm{T}}$. The position vector of the $n$-th AE is given by $\mathbf{p}_n=[0,0,(n-1)d]^{\mathrm{T}},~n\in \{1,2,\cdots,N\}$. Thus, the difference of the wave path of the $n$-th AE and that of the first AE is the inner product of the position vector and the beam direction vector:
	
	\begin{equation}\label{eq:dif_wave}
		\begin{aligned}
			R_n-R_1=-\mathbf{\hat{r}}^{\mathrm{H}}\mathbf{p}_n=-(n-1)d\cos\theta.
		\end{aligned}
	\end{equation}
	
	Furthermore, if the bandwidth of the signal is much less than the carrier frequency, i.e., the signal is narrow-band, the difference between the wave paths leads to a phase difference $-\frac{2\pi}{\lambda}(n-1)d\cos\theta$, where $\lambda$ denotes the carrier wavelength. These phase differences lead to the following array propagation vector for ULAs:
	\begin{equation}\label{eq:steering_vector}
		\begin{aligned}
			\mathbf{a}={\frac{1}{\sqrt{N}}\left[  1,e^{-j\frac{2\pi}{\lambda}d\cos\theta},\cdots,e^{-j\frac{2\pi}{\lambda}(N-1)d\cos\theta} \right]^{\mathrm{T}}},
		\end{aligned}
	\end{equation}
	which is also known as the \emph{steering vector}. Then, the far-field radiation of an isotropic ULA, often called the \emph{array factor}, is the inner product of the antenna weight vector $\mathbf{w}$ and the steering vector $\mathbf{a}$:
	\begin{equation}\label{eq:array_factor}
		\begin{aligned}
			F(\phi,\theta)=\sqrt{N} \mathbf{w}\cdot\mathbf{a},
		\end{aligned}
	\end{equation}
	where $\mathbf{w}\in \mathbb{C}^{N \times 1}$ denotes the excitation (amplitude and phase) applied to each AE.
	
	According to basic geometry, it can be readily to know that the steering vectors of the URA with $M\times N$ AEs, and the UCA with $N$ AEs are respectively given by
	\begin{equation} \label{eq:steering_vector_URA}
		\begin{aligned}
			\mathbf{a}_{\mathrm{URA}}&=\frac{1}{\sqrt{MN}}[  1,\cdots,e^{j\frac{2\pi}{\lambda}d\sin\theta[(m-1)\cos\phi+(n-1)\sin\phi]},\\
			&~~~~~~~~~~~~~\cdots,e^{j\frac{2\pi}{\lambda}d\sin\theta[(M-1)\cos\phi+(N-1)\sin\phi]} ]^{\mathrm{T}}, \\
			\mathbf{a}_{\mathrm{UCA}}&=\frac{1}{\sqrt{N}}[  1,e^{j\frac{2\pi}{\lambda}d\sin\theta \cos(\phi-\frac{2\pi}{N})},\\
			&~~~~~~~~~~~~~~~~~~~~~\cdots,e^{j\frac{2\pi}{\lambda}d\sin\theta \cos[\phi-\frac{2\pi}{N}(N-1)]} ]^{\mathrm{T}}.
		\end{aligned}
	\end{equation}
	
	Setting a uniform weight coefficient for each AE, i.e., $\mathbf{w}=[\frac{1}{\sqrt{N}},\cdots,\frac{1}{\sqrt{N}}]^\mathrm{T}$, the effects of the number of AEs $N$ and the AE spacing $d$ on the array factors of ULAs can be observed in Fig.~\ref{fig:AF}. Comparing Fig.~\ref{fig:AF}~(a) with (b), we observe that an increasing number of AEs results in a narrower main lobe and a larger amplitude, which increases the directivity and array gain. Comparing Fig.~\ref{fig:AF}~(c) with (a), it can be observed that a larger AE spacing also narrows the main lobe, while the amplitude of the array factor does not increase. Besides, high side lobes occur, also called \emph{grating lobes}, centered at $0^\circ$ and $180^\circ$. These grating lobes are harmful as energy is radiated to or received from undesired directions. To avoid grating lobes, the AE spacing should not exceed half a wavelength.

	As can be seen, the design of antenna arrays offers many different DoFs. A variety of antenna arrays have been designed for different applications, where several different types of antennas have been used as AEs, including patches, microstrips, horns, and reflectors. For example, a 28~GHz horn phased array and a dual-band (27/32 GHz) reflector antenna array were designed for 5G applications in~\cite{yc2019a2by2s} and \cite{costanzo2020dualba}, respectively. For UAV platforms, patches and microstrips are often preferred because of their benefits in terms of size, weight, cost, fabrication, and integration~\cite{dweik2014aplana, sun2017circul,siddiq2016micros,seo2019widebe,zheng2018alowpr}.
	Besides, UAVs usually experience jitter and swaying, leading to beam misalignment problems for mmWave directional communication. The simulation results in~\cite{zhang20175gmill} suggested that under such conditions, circular antenna arrays are a better choice compared to other planar array geometries, since circular arrays have a flat gain fluctuation in the main lobe and are thus robust to angle variations. Moreover, the integration of sub-6 GHz and mmWave antenna arrays is becoming a hot topic~\cite{lan2020anaper,ko2020planar}. It is a promising antenna approach for mmWave-UAV communications and networking as it allows the combination of the advantages of the high and low frequency bands. For instance, sub-6 GHz antennas can be utilized to establish control links over the network with high stability, while mmWave antennas can be used to perform directional transmission with high data rate.
	
	\subsection{Conformal Array}
\begin{figure}[t]
		\centering  
		{\includegraphics[width=0.25\textwidth, trim=0 0 0 0,clip]{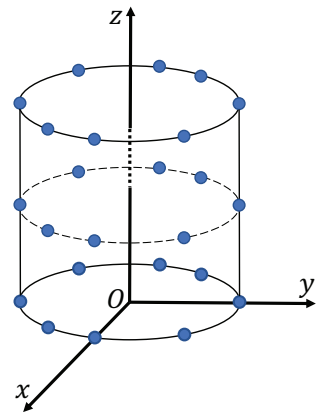}} 
		\caption{The geometric structure of a CCA. }  
		\label{fig:yuanzhu}   
	\end{figure}

\begin{figure}[t]
		\centering  
		{\includegraphics[width=0.48\textwidth, trim=0 0 0 0,clip]{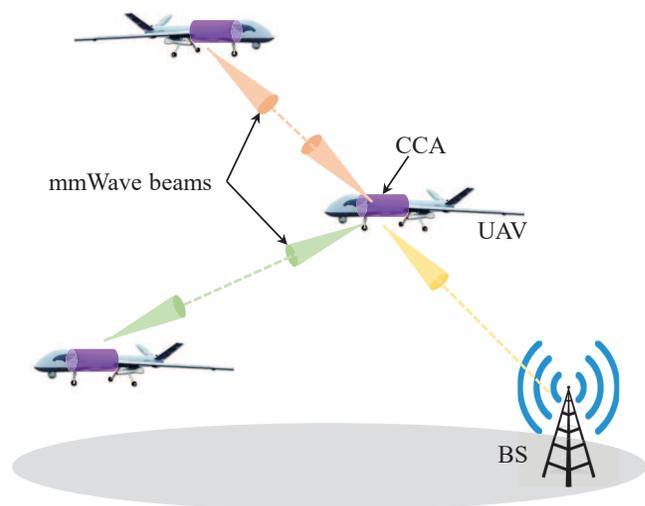}}
		\caption{Illustration of the CCA-enabled mmWave-UAV communication network \cite{zhang2020codebo}. }  
		\label{fig:UAVcca}   
	\end{figure}

	Conformal array is a type of antenna array designed to conform some prescribed shapes, such as aircraft bodies and wings. Conformal array is an attractive option for airborne and space applications because of its wide-angle coverage, low radar cross section, and good aerodynamic properties. For mmWave-UAV communications, conformal arrays have several unique advantages. First, the space, payload, and energy supply on UAVs are limited. With their lightweight and compact design, conformal arrays can be conformed to the surface of the UAVs such that they do not occupy extra space. Therefore, compared to a planar array that requires an additional nacelle, a conformal array has no effect on the aerodynamic characteristics of a UAV and leads to a reduction of the drag and fuel consumption. Second, by properly exploiting the shape and size of the UAV fuselage, there is more surface area available for integrating additional AEs. This is valuable for mmWave communications because large-scale arrays can achieve a larger beam gain. Third, unlike regular arrays, such as ULA, URA, and UCA, which can only offer half of space coverage, conformal arrays introduce more DoFs for geometry design, which allows for a larger spatial coverage. This characteristic opens the possibility of full-space beam scanning for UAV communications, which can reduce communication outages caused by UAV mobility or posture changes. A simple type of conformal array, namely a cylindrical conformal array (CCA), is shown in Fig.~\ref{fig:yuanzhu}. The authors of \cite{zhang2020codebo} proposed CCA-enabled mmWave-UAV networks as shown in Fig. \ref{fig:UAVcca}. Mounted with CCAs in the mmWave frequency band, UAVs can form multiple beams across the full space to connect with neighbor UAVs and ground BSs simultaneously.

	
	Despite these unique advantages, the design of conformal arrays involves many challenges depending on the application and platform. In terms of the choice of materials, the compatibility with the surface, lightweight, and low profile requirements is crucial, especially for space-limited UAV platforms. To this end, flexible materials are practicable alternatives compared to the conventional materials. Due to their low dielectric constant, low loss-tangent, low moisture absorption, corrosion and high-temperature resistance, liquid crystal polymers are attractive options for 5G mmWave conformal array applications \cite{jilani2019millim}. Polyimide film, due to its flexibility, was used as the substrate for an eight-element linear conformal dipole array in~\cite{peng2020widesc}. This conformal array was designed to be mounted on the front wing of a UAV with the wide-scanning capability needed for radar applications. Polysulfone was used as the array-supported layer of a conical conformal array~\cite{xu2019versat}, which is able to facilitate many UAV applications, such as point-to-point communication and target detection.
	
	As for the geometry of conformal antennas, some regular geometric shapes, such as cylinder, cone, and sphere are widely studied~\cite{semkin2016beamsw,gao2019ahighg,xu2019versat,braaten2014phasec,zhang2020codebo}. These geometries have certain advantages for theoretical analysis because of the simple and concise definition of the steering vector. For example, the theory on the relationship between the required phase compensation, the element spacing, and the radius of spherical conformal arrays was given in~\cite{braaten2014phasec}. A beam tracking problem for conformal array-enabled mmWave-UAV networks was studied in~\cite{zhang2020codebo}, based on the theoretical analysis of the radiation pattern of ideal CCAs. In~\cite{abdelhakam2018effici}, a beamforming optimization problem was investigated for a downlink multiuser mmWave MIMO network, where cylinder and cone geometrical configurations were considered and their steering vectors were generated. On the other hand, there are many other geometries of conformal antennas that are chosen to fit the available surfaces for a given application, such as the front wing of a fixed-wing UAV~\cite{peng2020widesc}, the wing of a quadcopter UAV~\cite{balderas2019lowpro}, or missiles and wearable devices~\cite{jilani2019millim}. The resulting irregular shapes complicate the analysis and synthesis of the conformal arrays as the geometry, the form, and the distribution of the AEs have to be considered in the design.
	
	In mmWave communications, wideband spectrum is exploited to support high-data-rate transmission, and thus wideband conformal antennas are needed. Furthermore, an in-band flat antenna gain is also important for antenna design, where the antenna should have a relatively constant gain within the operating bandwidth. Aiming for this, a mmWave conformal array was designed in~\cite{jilani2019millim}, which achieves 9 dBi gain in the Ka-band (26.5-40 GHz) and a peak gain of 11.35 dBi at 35 GHz. A Y-rounded shape conformal antenna was proposed in~\cite{balderas2019lowpro}, which achieved an ultra-wideband performance from 2.9 GHz to 15.9 GHz, and is suitable for beyond-6 GHz UAV applications. Given the increasing demand for more communication capacity, conformal antennas operated at mmWave bands are fast developing. As communication terminals or servers, UAVs mounted with conformal antennas can obtain wide-angle coverage and better aerodynamic performance. As a result, conformal antenna-enabled mmWave-UAV communication is a promising research direction, and mmWave conformal antennas specifically designed for UAV applications are needed.

	\subsection{Beamforming Architectures}
	According to (\ref{eq:array_factor}), the far-field radiation of an antenna array is determined by both the steering vector and the antenna weight vector. The steering vector is predetermined by the configuration of the array, which has been discussed in Section II-B. The antenna weight vector, composed of the phase and amplitude of the weight of each AE, can be electronically controlled to form different radiation patterns. This signal processing technique is called \emph{beamforming}~\cite{van1988beamfo}. With proper beamforming, the beams can be steered into desired directions, which not only improves the received signal power at the target users but also reduces the interference to undesired users. According to the hardware structure of the antenna array, beamforming architectures can be roughly divided into three categories, namely digital beamforming, analog beamforming, and hybrid beamforming.

	\subsubsection{Digital beamforming}
	Fig.~\ref{fig:beamformingAD} (a) illustrates the basic digital beamforming architecture at the transmitter, where each AE is connected to an independent radio-frequency (RF) chain. Beamforming is performed in the baseband via digital signal processing, which yields a high flexibility with sufficient DoFs to implement efficient precoding algorithms. Thus, in theory, digital beamforming achieves a higher performance compared to other architectures~\cite{roh2014millim}. It can accommodate multi-stream transmission, and can distinguish signals simultaneously received from different directions. However, the digital beamforming architecture requires a dedicated RF chain for each AE. The corresponding hardware components, including ADCs, DACs, data converters, and mixers, entail a high hardware complexity and a large energy consumption.

	\subsubsection{Analog beamforming}
\begin{figure}[t]
		\centering  
		\includegraphics[width=\figwidth cm, trim=0 0 0 0,clip]{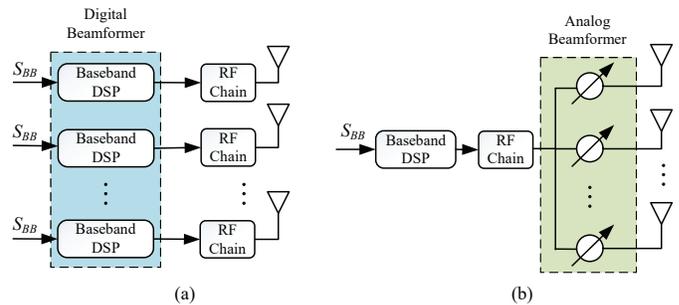}  
		\caption{Beamforming architectures: (a) digital beamforming, (b) analog beamforming. }  
		\label{fig:beamformingAD}   
	\end{figure}

	A basic analog beamforming architecture is shown in Fig.~\ref{fig:beamformingAD} (b). Different from digital beamforming, analog beamforming requires only one RF chain and is implemented by using phase shifters or switches in the analog domain. With analog beamforming enabled by phase shifters, only the phase of the signal can be adjusted at each AE, and thus less DoFs are available.

	\subsubsection{Hybrid beamforming}
\begin{figure}[t]
		\centering  
		\includegraphics[width=\figwidth cm, trim=0 0 0 0 ,clip]{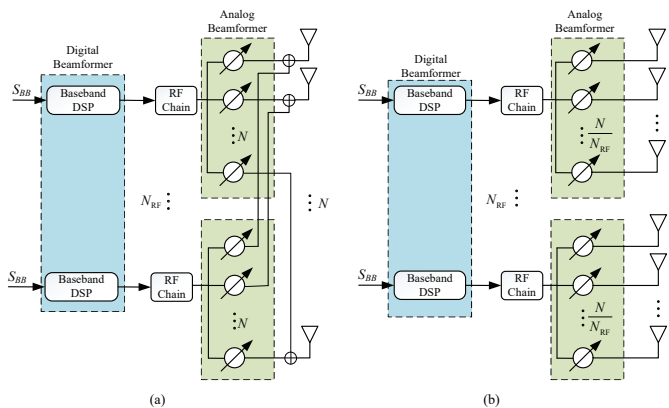}  
		\caption{Hybrid beamforming architectures: (a) fully-connected, (b) partially-connected. }  
		\label{fig:beamformingHybrid}   
	\end{figure}

	Hybrid beamforming is attracting growing interest as a possible solution to combine the advantages of the digital and analog beamforming architectures~\cite{roh2014millim}. This architecture uses a reduced number of RF chains to reduce cost and energy consumption, while enabling multi-stream transmission to meet the overall performance requirements. There are broadly two types of hybrid beamforming architectures, namely fully-connected and partially-connected as shown in Fig.~\ref{fig:beamformingHybrid}. For the fully-connected hybrid beamforming architecture, a small number of RF chains are employed. Let $N_{\mathrm{RF}}$ denote the number of RF chains, we usually have $N_{\mathrm{RF}}\ll N$ for mmWave arrays. Each RF chain connects with all AEs via $N$ phase shifters. The fully-connected structure provides the full beamforming gain for each RF chain but requires $N_{RF}\times N$ RF paths which entail a high complexity. For the partially-connected hybrid beamforming architecture, an antenna array with $N$ AEs is grouped into $N_{\mathrm{RF}}$ sub-arrays. Each sub-array is connected to one RF chain via $N/N_{\mathrm{RF}}$ phase shifters. Thus, the partially-connected structure has lower hardware complexity, as it requires only $N$ RF paths, but results in lower beamforming gain. In other words, there is a tradeoff between hardware complexity and beamforming performance for hybrid beamforming architectures.

	Due to the high hardware complexity, high cost, and high power consumption, fully digital beamforming may not be suitable for practical implementation in mmWave-UAV communications, especially for large numbers of antennas~\cite{roh2014millim,xiao2016enabli}. In contrast, analog and hybrid beamforming structures are preferred for mmWave-UAV communication systems because of their cost and energy efficiency \cite{zhu2020millim,zhao2018channe,zhang2019positi,cheng2019beamst,ren2019machin}. In addition to the above architectures, there are more options. For example, different from conventional analog/hybrid beamforming architecture based on phase shifters, switch-based architecture replaces phase shifters with switches to further reduce the hardware complexity and power consumption, at the expense of communication performance \cite{ahmed2018asurve}. Another alternative architecture is \emph{lens antenna array}~\cite{brady2013beamsp}, where a number of antennas are mounted behind a lens to transmit/receive signals to/from different directions. The antennas can be selectively connected to a small number of RF chains via switches, achieving low-complexity hybrid beamforming. \emph{Passive beamforming} is achieved by IRS/RIS, which consists of lots of passive reflecting elements. Each element is able to independently reflect the incident signals and adjust their amplitudes and phases~\cite{li2020reconf}. In summary, there is a tradeoff between the spectral efficiency and the energy efficiency for the above beamforming architectures, which are compared in Table~\ref{tab:beamforming}.
	
	\begin{table*}[t]
		\caption{Comparison of different beamforming architectures.}
		\label{tab:beamforming}
		\centering
        \begin{spacing}{1.1}
		\begin{tabular}{|c|c|l|}
			\hline
			\textbf{Structure} & \textbf{RF chains} & \multicolumn{1}{c|}{\textbf{characteristics}}                                                                                                                           \\ \toprule[0.5pt]\hline
			Digital            & $N$                  & \begin{tabular}[c]{@{}l@{}}High spectral efficiency, high computational efficiency, and multi-stream transmission;\\ High energy consumption and high hardware cost\end{tabular} \\ \hline
			\begin{tabular}[c]{@{}c@{}}Analog \\(phase shifter) \end{tabular}            & 1                  & \begin{tabular}[c]{@{}l@{}}High energy efficiency and low hardware cost;\\ Performance loss, low flexibility, and single-stream transmission \end{tabular}                    \\ \hline
			\begin{tabular}[c]{@{}c@{}}Hybrid \\(phase shifter) \end{tabular}            & $N_{\mathrm{RF}}\ll N$            & \begin{tabular}[c]{@{}l@{}}Tradeoff between spectral efficiency and energy efficiency;\\ High computational complexity \end{tabular}                                                                                                            \\ \hline
			
			\begin{tabular}[c]{@{}c@{}}Switch-based \\ (analog/hybrid) \end{tabular}         &  $1$ or $N_{\mathrm{RF}}$           &               \begin{tabular}[c]{@{}l@{}}Replace phase shifters with switches;
				\\	Lower hardware complexity and power consumption, but worse performance \end{tabular}                                                      \\ \hline

			\begin{tabular}[c]{@{}c@{}}Lens \\Antenna Array \end{tabular}         & $N_{\mathrm{RF}}\ll N$           &               \begin{tabular}[c]{@{}l@{}}RF chains and antennas are selectively connected by switches;
				\\	Low hardware complexity \end{tabular}                                                      \\ \hline
			IRS/RIS                & 0                  &
			\begin{tabular}[c]{@{}l@{}}Passive beamforming by adjusting amplitudes and phases of reflected signals;\\ Low hardware cost and high energy efficiency  \end{tabular}                                                                                               \\ \hline
		\end{tabular}
        \end{spacing}
	\end{table*}

	\subsection{Intelligent Reflecting Surface}
		\begin{figure*}[th]
		\centering  
		\includegraphics[width=14 cm,trim=0 0 0 0,clip]{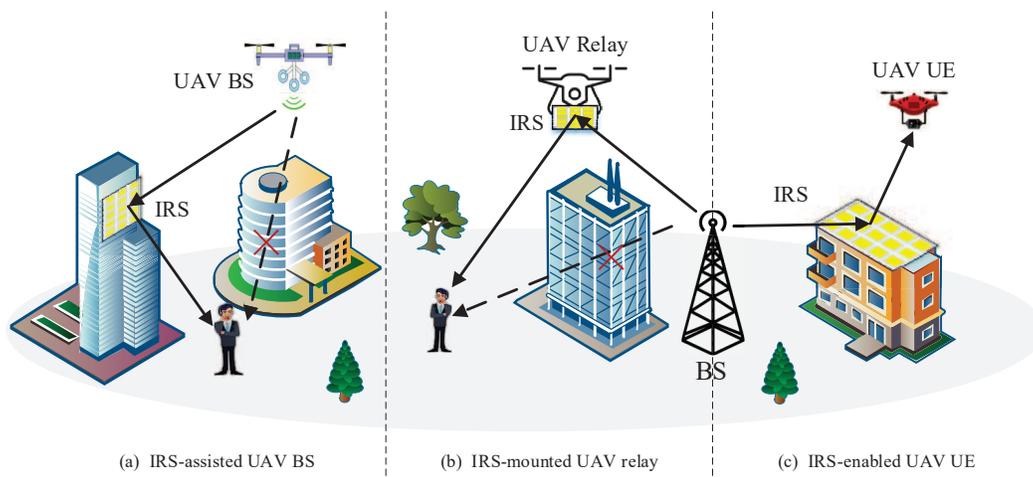}  
		\caption{UAV communication systems assisted by IRS/RIS. }  
		\label{fig:RIS}   
	\end{figure*}
	
	MmWave signals are easily blocked by obstacles, especially in urban and indoor environments. To tackle this issue, IRS/RIS, which is able to reconfigure the propagation environment by providing adjustable reflected paths for signals~\cite{wu2019intell, wu2020toward, abeywickrama2020intell},   has attracted increasing attention. Specifically, an IRS is a planar array which consists of abundant passive reflecting elements. Each element is able to independently reflect the incident signal and induce different reflection amplitude and phase, to steer the reflected signal towards the desired receiver, known as \emph{passive beamforming}~\cite{li2020reconf}. Mathematically, let $x_i$ denote the incident signal of the $i$-th element of the IRS, then the reflected signal is given by $y_i=(I_i e^{j\alpha_i})x_i, i=1,...,M$, where $I_i \in [0,1]$ and $\alpha_i \in [0,2\pi)$ control the amplitude and the phase shift of the reflected signal, respectively. $M$ is the total number of reflecting elements. For a sufficiently large array, IRS is able to achieve \emph{squared power gain} (i.e., $\mathcal{O}(M^2)$) in a single-user system~\cite{wu2019intell}.

	Three typical application scenarios for UAV communication systems enabled by IRS/RIS, namely \emph{IRS-assisted UAV BS}, \emph{IRS-mounted UAV relay}, and \emph{IRS-enabled UAV UE}, are shown in Fig.~\ref{fig:RIS}. First, a UAV may act as an aerial BS, while IRSs are usually placed on building facades to assist reconfiguring wireless propagation channels between the UAV and ground users. The mobility of UAV and the configurability of IRS provide more DoFs for the system design. The joint UAV trajectory and passive beamforming design was investigated in~\cite{li2020reconf}. Considering multiple IRSs, a similar work was conducted in~\cite{wang2020jointt} by utilizing deep reinforcement learning approaches. The results showed that considerable improvement of the data rate could be achieved by employing IRSs for UAV communication systems. Besides, the passive beamforming characteristic of IRS was leveraged in~\cite{fang2021jointo} to facilitate secure UAV communications. Second, the IRS may be equipped on a UAV, which acts as a passive relay. The passive beamforming at the UAV relay can be jointly designed with the active beamforming at the ground BS. In~\cite{shafique2021optimi}, the end-to-end performance is analyzed for different communication modes, in terms of outage probability, ergodic capacity, and energy efficiency. Joint active and passive beamforming design for IRS-enabled UAV relays was studied in~\cite{mohamed2020levera}, aiming to achieve energy-efficient communications. In~\cite{long2020reflec}, secure communications was investigated by optimizing the trajectory of the UAV relay and passive beamforming of the IRS. Third, the UAVs may act as aerial UEs and be served by ground BSs, where the IRSs can be leveraged to improve the signal strength received at UAVs caused by the down-tilt of BS antennas~\cite{ma2020enhanc}. In a word, the combination of mmWave-UAV communications and IRS/RIS can significantly improve the communication capabilities by synergistically leveraging their advantages.
	
\begin{table*}[t]
\begin{center}
	\small
	\caption{Comparison of different antenna structures.}
	\label{Tab:antenna_struture}
\begin{spacing}{1.3}
	\begin{tabular}{|c|c|c|c|}
		\hline
		\textbf{Structure} & \textbf{Description}                                                                           & \textbf{Advantages}                                                                                                                               & \textbf{Disadvantages}                                                                                                                                \\\toprule[0.5pt] \hline
		Aperture antenna   & \begin{tabular}[c]{@{}c@{}}Antenna with an \\ aperture at the end\end{tabular}                   & High gains, good directivity                                                                                                                      & \begin{tabular}[c]{@{}c@{}}Fixed radiation pattern; \\ require significant space\\ for deployment\end{tabular}                            \\ \hline
		Antenna array      & \begin{tabular}[c]{@{}c@{}}Multiple connected AEs \\ which work cooperatively\end{tabular}       & \begin{tabular}[c]{@{}c@{}}Flexible radiation patterns, \\ supports beamforming\end{tabular}                                                    & Half-of-space beam coverage                                                                                                                  \\ \hline
		Conformal array    & \begin{tabular}[c]{@{}c@{}}Antenna array conformed \\ to some curved surface\end{tabular}        & \begin{tabular}[c]{@{}c@{}}Good aerodynamic performance;\\ more DoFs for geometry design, \\ possible full-space beam coverage\end{tabular} & \begin{tabular}[c]{@{}c@{}}Design challenges such as materials;\\ complicated analysis and synthesis\\  due to irregular shapes\end{tabular} \\ \hline
		IRS                & \begin{tabular}[c]{@{}c@{}}Planar array comprised of \\ passive reflecting elements\end{tabular} & \begin{tabular}[c]{@{}c@{}}Reconfigures the propagation \\ environment by passive beamforming\end{tabular}                                         &  \begin{tabular}[c]{@{}c@{}}Theoretical results need to be \\ confirmed in real-world systems\end{tabular}                                                                                                                                            \\ \hline
	\end{tabular}
\end{spacing}
\end{center}
\end{table*}

    \subsection{Carrying Capability of UAVs}
Different UAVs have different SWAP limits. In general, UAVs can be divided into five categories in terms of aerodynamics, i.e., fixed-wing, rotary-wing, monorotor, airship, and flapping wing \cite{chittoor2021arevie,karabulut2021avisio}.
Naturally, the load capacity of UAVs restricts the use of antenna arrays. Rotary-wing UAVs generally have a relatively low load capacity. The average payload value of this category ranges from 0.3 to 2\,kg, and professional carrier UAVs are able to lift 20 to 200\,kg \cite{james2020howmuc}.
The use of antenna arrays in early UAVs was motivated by military radar applications, such as the phased array radar employed by the U.S. Global Hawk, which weighs hundreds of kilograms. With the development of integration technology, antenna arrays are getting lighter and lighter, with weights as low as 66.5\,g for the $16\times6$ element array in \cite{huang2018aultra}.
More recently, communication companies, such as IBM and Qualcomm, have developed phased-array beamformer chips for 5G mmWave dual-polarized MIMO systems in micron-sized SiGe or nanoscale CMOS technologies \cite{sadhu2017aghzel,dunworth2018aghzbu,pang2021acmosd}.
Meanwhile, the emerging patch antenna array and conformal antenna array techniques meet the requirement of mmWave-enabled UAVs of light weight \cite{nachev2020radara, peng2021confor}. In addition, the payload capacity of UAVs has also increased in recent years. 3GPP stipulates that the payload weight of low and high performance UAVs for communications is not less than 6.5\,kg and 11.5\,kg, respectively \cite{3GPPEUAV}. JD's self-developed logistics drone JDX-500 is a civil monorotor UAV and can carry a weight of up to a few hundred kilograms in 2020 \cite{wang2020jdssel}, and HAPs (e.g., unmanned airships) can accommodate even larger payloads \cite{karabulut2021avisio}.
Thus, thanks to the short wavelength of mmWave signals and the rapid development of antenna and UAV manufacturing technologies, nowadays massive numbers of antenna elements as well as high-gain antennas can be deployed in a relatively limited space on a UAV \cite{zhang2020multip}.

\subsection{Summary and Discussion}

	To compensate for the severe path loss in the mmWave frequency bands, directional antennas which achieve high antenna gain are necessary. Besides, UAV platforms demand extra considerations for antenna design. Different antenna structures are compared in Table~\ref{Tab:antenna_struture}. Aperture antennas, such as horn antennas, reflector antennas, and lens antennas, usually achieve high gains and good directivity. However, they also require significant space for deployment. The suitability of a particular antennna design depends on the size of the UAVs. Antenna arrays, which integrate multiple connected AEs working together to obtain high gain, may emerge as the mainstream antenna design for mmWave-UAV communications. Through beamforming, antenna arrays can adjust the beam direction and gain to adjust to different application scenarios \cite{rangan2014millim,hemadeh2018millim,roh2014millim}. Furthermore, by conforming AEs to some curved surfaces,  such as aircraft bodies and wings, conformal arrays can achieve better aerodynamic performance and offer more DoFs for geometry design \cite{zhang2020codebo,jilani2019millim,peng2020widesc}. It is possible for conformal arrays to realize full-space beam scanning. The challenges for conformal arrays concern the antenna design, analysis, and synthesis.
	Another array-structured technology, namely IRSs, has become a focus of research recently in academia \cite{wang2020jointt,fang2021jointo,shafique2021optimi,mohamed2020levera,long2020reflec,ma2020enhanc}. IRSs can reconfigure the propagation environment by reflecting signals. Theoretical analysis and validation via simulations were conducted for different envisioned application scenarios. However, real-world system implementations and experiments are needed to confirm these theoretical results.

	\section{UAV MmWave Channel Modeling}
	Channel modeling is indispensable for wireless communication system design. Accurate A2G and A2A channel models facilitate the performance analysis of UAV-enabled wireless communications in terms of capacity and coverage \cite{holis2008elevat,alhourani2014modeli}. For mmWave enabled UAV communications, the radio propagation characteristics are significantly different from those for classical ground channels. Thus, it is important to study the characteristics of UAV mmWave channels. However, the research on UAV mmWave channel measurement and modeling is still in an initial stage. In the following, the propagation characteristics and channel modeling campaigns for mmWave-UAV communications are surveyed.
	
	\subsection{Propagation Characteristics}
\begin{figure*}[t]
		\centering
		\includegraphics[width=16 cm]{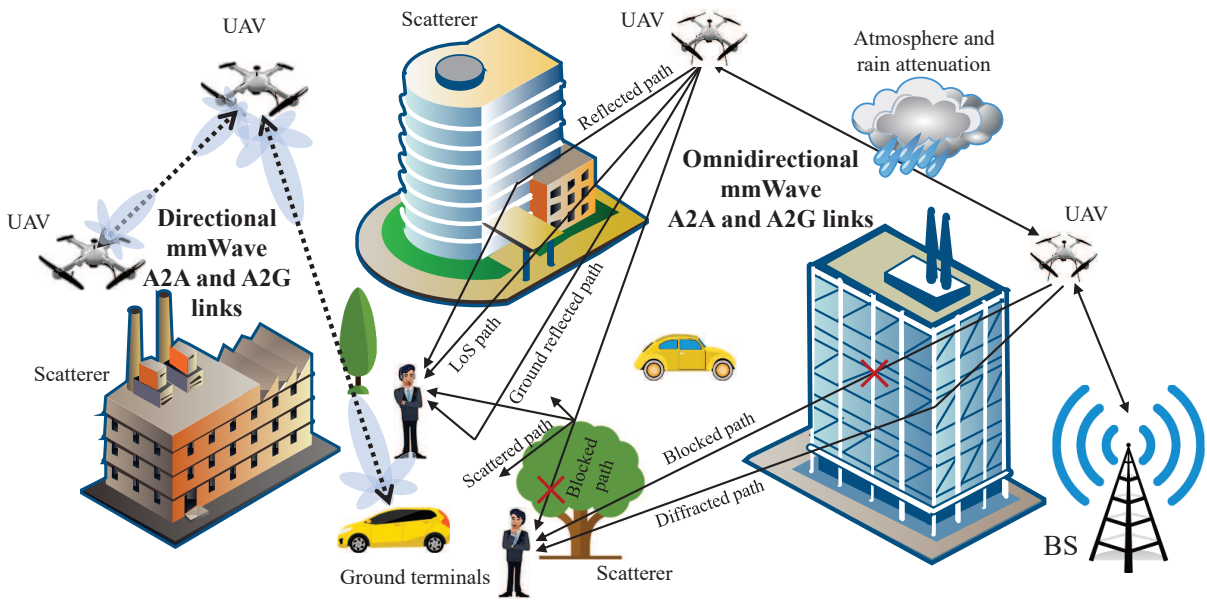}
		\caption{A typical scenario of UAV mmWave channels and the major propagation characteristics.}
		\label{fig:propa}
	\end{figure*}

	A typical mmWave-UAV communication scenario and the corresponding propagation characteristics are illustrated in Fig. \ref{fig:propa}. Compared to the microwave frequency bands, the main characters of the mmWave frequency bands include a short wavelength, large bandwidth, large penetration loss, and strong atmospheric attenuation. Moreover, due to the mobility of the UAV, the main differences between UAV and terrestrial communications include temporal variations of the non-stationary channels, dynamic change between LoS and NLoS environments, and inherent airframe shadowing and fluctuations. The salient propagation characteristics of UAV mmWave channels are summarized in the following.
	
	\subsubsection{Path Loss}
	
	The transmission range is one of the bottleneck problems faced by mmWave-UAV communication systems. According to Friis transmission formula, the free-space path loss is given by \cite{friis1946anoteo, rappaport2013millim}
	\begin{equation}\label{FPL}
		\beta(\bar{d})=\left(\frac{4 \pi \bar{d}}{\lambda}\right)^{2}=\beta_{0} \bar{d}^{2}
	\end{equation}
	where $\lambda$ is the carrier wavelength, $\bar{d}$ is the distance between the transmitter
	and the receiver, and $\beta_{0} \triangleq \left(4 \pi/\lambda\right)^{2}$ is the free-space path loss at the reference distance of 1\,m. For an ideal scenario without signal obstruction and reflection, e.g., typical ideal A2A environment, the free-space path loss model can be employed \cite{cuvelier2018mmwave, ye2019secure, zhou2019beamma}. For certain scenarios, such as rural areas with few blockages and scatterers and UAVs flying at high altitudes to maintain LoS links, the free-space path loss model provides a reasonable approximation and has been widely used in the existing works \cite{zeng2016throug, zeng2017energy, wu2018jointt}. If the antenna gain is assumed to be constant over frequency, the path loss of mmWave signals with their extremely short wavelengths is much higher than that of microwave signals.
	
	In addition, mmWave signals are more vulnerable to the attenuation caused by atmospheric absorption, including oxygen absorption, water-vapor absorption, and rain attenuation \cite{qingqing2006rainat, zhang2015rainfa,banday2019effect}. Atmospheric absorption has a peak around 60 GHz \cite{xiao2017millim, rappaport2017overvi}, while rain and hail cause substantial attenuation at frequencies above 10\,GHz \cite{xiao2017millim, rappaport2017overvi, itu2015propag}. Indeed, the attenuation at mmWave frequencies varies with the time of the day and the season of the year, because of the changes in temperature and humidity \cite{hosseini2016analyz, busari2018millim}. According to the measurement campaigns in \cite{shakhatreh2021modeli, semkin2021lightw, kachroo2021emulat}, different values of the path loss exponents are obtained for different frequency bands in different scenarios.
	\subsubsection{Penetration Loss and Blockage Effects}
	When a signal propagates through building walls, trees or human bodies, there are penetration losses \cite{aragon2017radiop,zhao2013ghzmil, meng2010invest}. 3GPP \cite{3gpp2019studyo} and International Telecommunication Union (ITU) \cite{itu2017guidel} provide formulas for calculating the penetration losses for different materials. In general, the penetration loss increases with the carrier frequency \cite{3gpp2019studyo, itu2017guidel}. Compared to rural areas, penetration losses are more prevalent in dense environments such as urban areas \cite{3gpp2018studyo}. In addition, the penetration loss of foliage is a key factor that cannot be ignored \cite{rahim2015millim, rappaport2015ghzwid}. Common foliage can lead to losses of dozens of dB, even for low-frequency mmWave signals \cite{marcus2005millim, meng2010invest, hemadeh2018millim}. Besides, the penetration loss of the human body varies with the transceiver distance, carrier frequency, and human body depth \cite{aftabimomo2019effect}. For example, an average sized male with a body depth of 0.28\,m can cause as much as 30-40\,dB of the penetration loss at 73\,GHz \cite{maccartney2016millim}.

	When an obstacle exists between transmitter and receiver, the LoS path is blocked. For terrestrial communications, ITU and 3GPP provide probabilistic LoS channel models and obstacle models based on the 3D geometric method and various types of environments \cite{itu2017guidel, 3gpp2018studyo,3gpp2019studyo}. A widely used A2G probabilistic LoS model is given in \cite{alhourani2014modeli} and \cite{alhourani2014optima}, and was derived by using the statistical parameters provided by ITU. The LoS probability is modeled as a logistic function of the elevation angle $\theta$ as follows \cite{alhourani2014optima, alhourani2014modeli}
	\begin{equation}\label{LoSP}
		P_{\mathrm{LoS}}(\theta)=\frac{1}{1+a \exp (-b(\theta-a))},
	\end{equation}
	where $a$ and $b$ are modeling parameters that depend on the environment. In fact, the measurement results in \cite{bergh2016lteint, matolak2017airgro, lin2018thesky} have shown that A2G communication channels are mainly dominated by LoS links even if a UAV is located at a moderate altitude. For example, for a UAV operating at an altitude of 120\,m, the LoS probability of A2G links in a rural environment exceeds 95\% \cite{lin2018thesky}. Moreover, if there are obstacles between the transceivers, the LoS path can be rapidly restored by flexibly adjusting the 3D position of the UAV.
	\subsubsection{Scattering Characteristics}
	Due to the short wavelength of mmWave signals and the high altitude of UAVs, A2G mmWave signal propagation mechanisms differ from the conventional terrestrial propagation. The MPCs of mmWave channels are mainly caused by reflections from ground scatterers, including the earth surface, buildings and human bodies \cite{khawaja2019asurve}. In general, it is difficult for mmWave signals to diffract due to their narrow first Fresnel zone \cite{marcus2005millim, rangan2014millim, khawaja2017uavair}. Since the scattering of mmWave signals is not significant, the number of effective MPCs in mmWave communications is very limited in practice. Besides, the scattering characteristics and MPCs are highly dependent on the operating environment. Specifically, in rural areas, in addition to the LoS component, the path reflected from the ground is usually the strongest component \cite{khawaja2019asurve}. In dense environments, depending on the size and density of the scatterers, the reflection paths introduce rapid fluctuations of the received signal strength (RSS) \cite{khawaja2017uavair}. 
	It is worth noting that for a UAV at a high altitude, the signals scattered from buildings may not be relevant \cite{khawaja2017uavair}. Thus, increasing the altitude of the UAV may reduce the root mean square delay spread of the channels and mitigate the impact of scatterers on the RSS.
	\subsubsection{Doppler Effect}
	The motions of both the aerial nodes and the ground nodes introduce Doppler shifts, which result in carrier frequency offset, inter-carrier interference, and limited channel coherence time. The doppler shift is proportional to the carrier frequency and the mobile velocity, and is also influenced by the angular dispersion \cite{hemadeh2018millim, rangan2014millim}. In mmWave-UAV communication systems, severe Doppler shifts are caused by the high carrier frequency and high mobility. Besides, different MPCs may have largely different Doppler frequencies. However, if the UAV is located at a sufficiently high altitude and is far away from the ground node, the MPCs are expected to have a very similar Doppler shift as they will all arrive from similar angles at the UAV. Then, large Doppler shifts can be well mitigated via frequency synchronization \cite{khawaja2019asurve}. Moreover, the sparsity of mmWave channels and the directivity of the antennas can further reduce the MPC angular spread \cite{durgin2000theory, rappaport2013millim, lorca2017onover}. For mmWave-UAV communications with orthogonal frequency division multiplexing (OFDM), the inter-carrier interference caused by the Doppler effect may be alleviated by ensuring a sufficient subcarrier spacing \cite{zhou2019beamma}.
	\subsubsection{Airframe Shadowing and Fluctuation}
	Airframe shadowing and hovering fluctuation are unique to the UAV communications. In A2A and A2G communications, the LoS paths may be blocked due to the UAV structure design, on-board antenna placement, and UAV flight status \cite{khuwaja2018asurve}. Moreover, the short-wavelength signals may be more easily blocked and reflected by the metallic aircraft body \cite{yan2019acompr}. In fact, the UAV fuselage is a potential scatterer, which needs to be considered in the modeling of UAV mmWave channels. The effect of airframe shadowing cannot be eliminated by exploiting the spatial diversity at the ground node. In addition, there is no significant correlation between the airframe shadowing loss and the shadowing duration in A2G environments \cite{sun2017airgro}. Specifically, the airframe shadowing loss can be modeled as a function of the aircraft roll angle, while the shadowing duration is mainly affected by the flight speed \cite{sun2017airgro}.
	
	Fluctuations of the positions of the on-board antennas may be caused by the engine vibrations and wind turbulence. 
	For example, utilizing a robotic arm to simulate the UAV motion caused by wind gusts, the average Doppler spread was measured to be around -20\,Hz to +20\,Hz in an anechoic chamber for transceiver distances from 1.1 to 7.2 meters and a carrier frequency of 28\,GHz \cite{kachroo2021emulat}.
	Although installing a precise stabilizer can suppress the fluctuations of the UAV fuselage, a UAV may be limited by the strict SWAP constraints and it is difficult to accomplish perfect mechanical control. In mmWave-UAV communication systems, although high directional antenna gains can compensate the high path loss, the vibrations of the transceivers deteriorate the channel quality because of the narrow beamwidth. 
	The position of the UAV also influences the degree of AoA/AoD fluctuation at the UAV side \cite{wang2021jitter}.
	Due to the stochastic of UAV fluctuations, it is challenging to realize precise beam alignment \cite{dabiri2018channe, dabiri2020optima}. The resulting mismatch between the directional antennas between transceivers significantly impacts reliability, channel capacity, bit error rate (BER), and many other system performance metrics for mmWave-UAV communications \cite{dabiri2020analyt}. 
	One potential approach is to utilize AoA and AoD estimates to guide the beam alignment. In particular, beam training schemes assisted by UAV navigation information and compressed sensing may enhance the AoA/AoD estimation accuracy \cite{wang2021jitter}, while this scheme will increase the training time and may be not applicable for the mobile UAV scenario. For the dynamic scenario, a suitable way is carefully designing and optimizing the antenna patterns to find a favorable tradeoff between the beam width and beam gain, in order to reduce the probability of the potential sharp decline of the received power \cite{xiao2020uavcom}. In addition to the performance loss caused by antenna mismatch, the channel coherence time in the mmWave frequency bands is in the order of microsecond due to UAV jittering \cite{banagar2020impact}, which increases the difficulty of channel tracking and phase estimation.
	\subsection{Existing Channel Models}
	We first review mmWave channel models and then review UAV mmWave channel models. A basic mmWave channel model consists of large-scale fading (mainly including the distance-dependent path loss and blockage effect), small-scale fading (mainly resulting from the constructive and destructive interference of the MPCs between transceivers), and spatial and temporal characteristics. Due to the unique transmission characteristics, the number of MPCs for mmWave channels is much less than the antenna number at transceivers, which are different from the sub-6\,GHz channels with relatively rich MPCs.
	
\begin{figure}[t]
		\centering
		\includegraphics[width=8.5 cm]{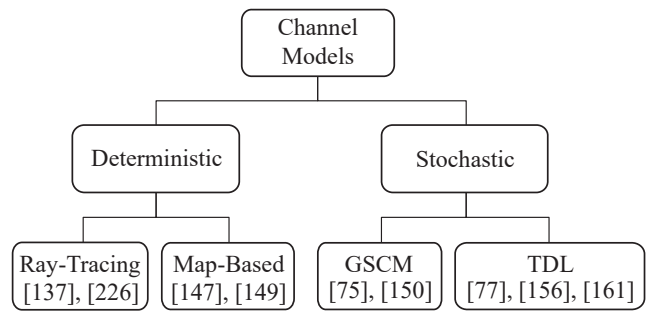}
		\caption{Classification of channel models in existing mmWave-UAV communication works.}
		\label{fig:classification}
	\end{figure}

	In the existing works, channel models are mainly classified into deterministic and stochastic channel models as shown in Fig. \ref{fig:classification}. Deterministic channel models, such as ray-tracing and map-based channel models, try to model the actual propagation characteristics of electromagnetic waves. These models rely on propagation measurements and information collected in databases regarding the environment. The ray-tracing modeling approach was first used for mmWave-UAV  A2G channel modeling in \cite{khawaja2017uavair} and the actual performance of the two ray propagation model in various environments was tested. Moreover, 3GPP \cite{3gpp2018studyo} provided a map-based hybrid method to model the mmWave channel.
	In particular, combining ray-tracing and digital map based methods to model UAV mmWave channels and to compute the exact channel parameters, the authors in \cite{jiang2020mapbas, zhu2020effect, zhu2021mapbas} studied the channel characteristics for different scenarios and the influence of the reconstruction accuracy of the digital map database. 
	
	Stochastic channel models, such as geometry-based stochastic channel models (GSCMs) and tapped delay line (TDL) models, utilize statistical distribution models and empirical parameters to mathematically analyze the channel characteristics with a relatively low computational complexity. The GSCM approach evaluates the spatial-temporal channel characteristics by simulating a virtual 3D environment confined to specific geometrical shapes and was applied to mmWave-UAV MIMO channels \cite{zhao2018channe, michailidis2020threed}. 3GPP \cite{3gpp2018studyo} also proposed a TDL channel modeling framework, which can accommodate fading statistics of the MPCs derived from the channel impulse response. This modeling approach has a relatively low computational complexity and the fading parameters of each tap can be empirically obtained from statistic measurement data. Besides, the Saleh-Valenzuela channel model \cite{saleh1987astati} which is a TDL model, has been widely used in the existing mmWave-UAV communication systems. 
	
	The time-varying complex impulse response of the Saleh-Valenzuela model is given as follows \cite{saleh1987astati}
	\begin{equation}
		h(t) = \sum\limits_{l=1}^L {{a_l}} {e^{ - j{\psi _l}(t)}}\delta \left( {t - {\tau _l}} \right),
	\end{equation}
	where $a_{l}$, $\psi _l$ and $\tau _l$ denote the time-varying amplitude, phase, and delay of the $l$-th MPC, respectively. For UAV communications, the fading can be modeled by the Nakagami distribution which can capture various channel fading conditions and provides a good fit with experimentally measured data \cite{goddemeier2015invest}.
	
	For a mmWave-MIMO system with $N_{\rm{T}}$ transmit and $N_{\rm{R}}$ receive antennas, the time-varying channel response in the frequency domain is given by
	\begin{equation}\label{timevary}
		\begin{aligned}
			{\bf{H}}\left( {t,f} \right) = &\sqrt {\frac{{{N_{\rm{T}}}{N_{\rm{R}}}}}{L}} \sum\limits_{l = 1}^L {{\alpha _l}} {e^{j2\pi ({\upsilon _l}t - {\tau _l}f)}} \times\\
			&~~~~~~~~~~~~~~{{\bf{a}}_{r}}\left( {{\theta _{r,l}},{\phi _{r,l}}} \right){\bf{a}}_{t}^{\mathrm{H}}\left( {{\theta _{t,l}},{\phi _{t,l}}} \right),
		\end{aligned}
	\end{equation}
	where $L$ is the total number of MPCs.
	For each MPC $l$, $\alpha _l$ denotes the complex gain, which includes the large-scale fading and small-scale fading. ${\theta _{r,l}}$, ${\phi _{r,l}}$, ${\theta _{t,l}}$, and ${\phi _{t,l}}$ represent the elevation angle of arrival (AoA), azimuth AoA, elevation angle of departure (AoD), and azimuth AoD, respectively. Parameters $\tau _l$ and $\upsilon _l$ are the delay and Doppler shift of the $l$-th MPC, respectively. The steering vectors ${\bf{a}}_{r}$ at the receiver and ${\bf{a}}_{t}$ at the transmitter are determined by the geometry of the arrays and have been defined in \eqref{eq:steering_vector} and \eqref{eq:steering_vector_URA}.
	Supposing that the channel is sufficiently slow-varying over the signal duration of interest, i.e., the Doppler shifts are small, (\ref{timevary})
	can be simplified as follows
	\begin{equation}\label{slowvaring}
		\begin{aligned}
			{\bf{H}}\left( f \right) = &\sqrt {\frac{{{N_{\rm{T}}}{N_{\rm{R}}}}}{L}} \sum\limits_{l = 1}^L {{\alpha _l}} {e^{ - j2\pi {\tau _l}f}}
			{{\bf{a}}_r}\left( {{\theta _{r,l}},{\phi _{r,l}}} \right){\bf{a}}_t^{\mathrm{H}}\left( {{\theta _{t,l}},{\phi _{t,l}}} \right).
		\end{aligned}
	\end{equation}
	If the bandwidth is sufficiently small, the narrow-band discrete channel model is obtained as follows
	\begin{equation}\label{narrowband}
		\begin{aligned}
			{\bf{H}} = &\sqrt {\frac{{{N_{\rm{T}}}{N_{\rm{R}}}}}{L}} \sum\limits_{l = 1}^L  {{\alpha _l}}{{\bf{a}}_{r}}\left( {{\theta _{r,l}},{\phi _{r,l}}} \right)
			{\bf{a}}_{t}^{\mathrm{H}}\left( {{\theta _{t,l}},{\phi _{t,l}}} \right),
		\end{aligned}
	\end{equation}
	which is also known as the extended Saleh-Valenzuela model.
	
	Although there are many channel modeling works on ground mmWave communications \cite{3gpp2019studyo, itu2015propag ,itu2017guidel, rangan2014millim, rappaport2017overvi, xiao2017millim} and sub-6\,GHz enabled UAV networks \cite{alhourani2014modeli, alhourani2014optima, dabiri2020analyt, yi2019modeli,yi2019aunifi, yi2020cluste}, UAV mmWave channel modeling is still in an initial stage. 
	Most existing works on the analysis of mmWave-UAV communication performance \cite{rodriguezfernandez2018positi, zhao2018channe, zhang2019positi, cheng2019beamst, ke2019positi, cuvelier2018mmwave, zhou2019beamma, yi2019modeli,yi2019aunifi, yi2020cluste, rupasinghe2018compar, rupasinghe2019anglef, rupasinghe2019nonort, li2020millim} adopt the channel models in \eqref{timevary}\,--\,\eqref{narrowband} or make necessary simplification for the application scenario and analytical requirements. The standardization of channel models for UAV communications above the 6\,GHz frequency bands is in its infancy, and the modeling parameters proposed by 3GPP \cite{3gpp2019studyo} and ITU \cite{itu2015propag, itu2017guidel} are commonly used. In the following, we briefly discuss recent works on analytical channel modeling for mmWave-UAV communication systems. A summary is provided in Table \ref{Tab:channel_model}.

	\begin{table*}[t]\footnotesize
		\begin{center}
			\caption{Parameters in existing A2A and A2G mmWave-UAV communication channel models.}
			\label{Tab:channel_model}
            \begin{spacing}{1.3}
			\begin{tabular}{|c|c|c|c|c|c|c|c|c|c|}
				\hline
				\textbf{References} &\textbf{Scenario} &\textbf{Antenna} &\textbf{Path loss} &\textbf{Atmosphere} &\textbf{Time-varying} &\textbf{Blockage} &\textbf{MPCs} &\textbf{Doppler effect} &\textbf{UAV fluctuations}\\
				\toprule[0.5pt]\hline
				\cite{zhang2019positi, ke2019positi} &A2A &UPA &$\surd$ &\hspace*{\fill}  &\hspace*{\fill}  &\hspace*{\fill} &\hspace*{\fill} &\hspace*{\fill} &\hspace*{\fill}\\
				\hline
				\cite{rodriguezfernandez2018positi} &A2A &UPA &$\surd$ &\hspace*{\fill} &$\surd$ &\hspace*{\fill}  &\hspace*{\fill} &\hspace*{\fill} &\hspace*{\fill}\\
				\hline
				\cite{cuvelier2018mmwave} &A2A &UPA &$\surd$ &$\surd$ &\hspace*{\fill} &\hspace*{\fill}  &\hspace*{\fill} &\hspace*{\fill} &$\surd$\\
				\hline
				\cite{zhou2019beamma} &A2A &UPA &$\surd$ &\hspace*{\fill}  &\hspace*{\fill}  &\hspace*{\fill} &\hspace*{\fill} &$\surd$ &\hspace*{\fill}\\
				\hline
				\cite{dabiri2020analyt} &A2A &ULA &$\surd$ &\hspace*{\fill}  &\hspace*{\fill}  &\hspace*{\fill} &\hspace*{\fill} &\hspace*{\fill} &$\surd$\\
				\hline
				\cite{xiao2020unmann} &A2G &ULA, UPA &$\surd$ &\hspace*{\fill}  &\hspace*{\fill}  &$\surd$ &$\surd$ &\hspace*{\fill} &\hspace*{\fill}\\
				\hline
				\cite{cheng2019beamst, li2020millim} &A2G &DLA &$\surd$ &\hspace*{\fill}  &\hspace*{\fill} &\hspace*{\fill} &\hspace*{\fill} &\hspace*{\fill} &\hspace*{\fill}\\
				\hline
				\cite{rupasinghe2018compar, rupasinghe2019anglef, rupasinghe2019nonort} &A2G &ULA &$\surd$ &\hspace*{\fill}  &\hspace*{\fill}  &\hspace*{\fill} &\hspace*{\fill} &\hspace*{\fill} &\hspace*{\fill}\\
				\hline
				\cite{yi2019modeli,yi2019aunifi, yi2020cluste} &A2G &UPA &$\surd$ &\hspace*{\fill}  &\hspace*{\fill} &$\surd$ &$\surd$ &\hspace*{\fill} &\hspace*{\fill}\\
				\hline
				\cite{michailidis2020threed} &A2G &URA &$\surd$ &\hspace*{\fill}  &$\surd$  &$\surd$ &$\surd$ &$\surd$ &\hspace*{\fill}\\
				\hline
				\cite{zhang2021distri} &A2G &ULA &$\surd$ &\hspace*{\fill}  &$\surd$  &$\surd$ &\hspace*{\fill}  &\hspace*{\fill} &\hspace*{\fill} \\
				\hline
			\end{tabular}
            \end{spacing}
		\end{center}
	\end{table*}
	
	\subsubsection{A2A MmWave Channel Modeling}
	A2A channels are typically time-varying due to the high mobility of the UAVs. When UAVs operate at appropriate altitudes, the probability that an LoS path exists is very high. Besides, since in the air there are few scatterers except the airframes, the number of MPCs for an A2A mmWave channel is small. Actually, the power gain of the LoS path is expected to be much larger than that of the NLoS paths. Hence, the most basic A2A mmWave channel models only consider the LoS path. In addition, many analytical channel models are assumed to be quasi-static in each appropriate time slot which is much smaller than the channel coherent time \cite{zhang2019positi, ke2019positi}.
	
	In \cite{cuvelier2018mmwave}, the authors first showed that the channel coherence time is fatally short compared to the time slot for communication, even in the extreme case with very high velocities, high frequencies, and narrow beams of UAV-to-UAV mmWave communication systems. Then, the authors considered a static channel model, in which many practical factors were taken into account, including atmospheric absorption, precipitation, and small-scale fading caused by small fluctuations of the UAVs' positions.
	In highly dynamic scenarios, channel estimation and tracking are essential due to the time-varying characteristic. To this end, the authors in \cite{rodriguezfernandez2018positi} considered an A2A mmWave multi-user MIMO communication network employing uniform planar arrays (UPAs) and hybrid beamforming structures. The channel model in \eqref{timevary} was adopted to study the multi-user channel estimation and tracking with prior information on the UAV position. A frequency-selective channel estimation algorithm was proposed for pure LoS channels and could be extended to the environment with MPCs.
	
	When adopting OFDM, the Doppler effect is more significant and should be considered for channel modeling. The authors in \cite{zhou2019beamma} evaluated the inter-carrier interference caused by the Doppler shift in mmWave-UAV mesh networks, employing a switch-based analog beam pattern at the transceivers \cite{xue2017beamsp}. It was shown that the impact of the Doppler spread becomes negligible by making the sufficient subcarrier spacing sufficiently large. For example, the subcarrier spacing is 5.15625\,MHz in the IEEE 802.11ay based mmWave networks \cite{ieee2018ieeest}. A radial velocity of 10\,m/s between the transceiver causes a maximum Doppler shift of 2000\,Hz, and thus the power of the inter-carrier interference is negligible.
	
	Since multi-rotor hovering UAVs may suffer from random vibrations, the authors in \cite{dabiri2020analyt} proposed a segment ULA gain model and derived analytical expressions of the probability distribution function and the cumulative distribution function of the end-to-end signal-to-noise ratio (SNR) in closed form. By evaluating the outage probability as a function of the vibration angle and the number of AEs, it was shown that UAVs with a high directional gain are more vulnerable to orientation fluctuations in the high SNR regime.

	\subsubsection{A2G MmWave Channel Modeling}
	In A2G mmWave communications, blockage effects are one of the key factors, which is especially important in dense urban scenarios because of the high penetration losses of mmWave signals. In the absence of any prior information regarding the obstacles, the randomness of having LoS or NLoS conditions should be taken into account for channel modeling. ITU \cite{itu2017guidel} and 3GPP \cite{3gpp2019studyo} have provided the LoS probability models and the blockage models in different terrestrial environments. As we have discussed before, a typical probabilistic LoS model which is widely used for A2G propagation environments \cite{xiao2020unmann, wang2019covera}, is given by \eqref{LoSP}. Besides, due to the abundant scatterers on the ground, MPCs should be considered for A2G channels \cite{xiao2020unmann, wang2019covera}.
	
	Early studies on mmWave-UAV communications focused on the performance effects of the UAV's altitude \cite{cheng2019beamst,li2020millim, rupasinghe2018compar, rupasinghe2019anglef, rupasinghe2019nonort}, and thus the basic channel model with only a LoS path was adopted. From a stochastic view, the authors in \cite{yi2019modeli,yi2019aunifi, yi2020cluste} utilized a typical 3D blockage model of ITU and  a practical close-in free-space path loss model proposed by New York University \cite{rappaport2015wideba} to analyze the UAV coverage problem in the urban environment. In particular, the authors showed that the effect of NLoS transmissions was negligible in the low-density blockage environment. By Accounting for the human body blockage and adopting a terrestrial mmWave channel model, the authors in \cite{gapeyenko2018effect} derived the optimal altitude of the UAV-BS and solved the 3D placement problem of the UAV. It was shown that the density of human blockers has a significant impact on the communication performance.
	
	For accurate channel modeling, it is essential to locate the positions of the scatterers in the vicinity of the ground terminals. The authors in \cite{michailidis2020threed} modeled mmWave massive MIMO aerial fading channels using a geometry-based approach. By employing 3D regular-shaped GSCM and spherical wavefront assumption, the birth-death processes were developed to describe the spatial-temporal cluster evolution feature of non-stationary channels. Whereafter, the authors derived the time-variant transfer function, the space-time-frequency correlation function, the Doppler power spectral density, and the standard deviation of the Doppler frequency, and developed a sum-of-sinusoids based simulation model to verify the theoretical analysis. However, the argument of this channel model with real-world measurement data needs to be further explored. To address this issue, the authors in \cite{zhang2021distri} developed a data-driven channel estimation framework based on a multi-UAV cooperative deep learning approach. Specifically, each UAV firstly applies a conditional generative adversarial network framework to model the channel distribution for a pre-determined codebook and collected channel information. Then, the generator of each UAV generates channel samples (containing the location, channel gain, and AoA/AoD information) and the discriminator distinguishes the fake data based on the real measurement data during the training process. In particular, multiple UAVs may share the generated channel samples using OFDMA over sub-6~GHz frequencies in a distributed hop-by-hop manner. Simulation results show that compared to non-cooperative strategies, the proposed learning approach yields a higher modeling accuracy and a larger average data rate in the downlink.
	
	\subsection{Summary and Discussion}
Compared to terrestrial mmWave communications, the propagation characteristics in mmWave-UAV communications are very unique because of obstacles, large Doppler shifts, aircraft shadowing, and UAV fluctuation. In particular, in real-world applications, the moving UAV may cause transceivers to become blocked by the fuselage itself \cite{khuwaja2018asurve}. Moreover, even a slight jitter can have a significant impact on the performance of the highly directional mmWave beam \cite{kachroo2021emulat}. These unique properties introduce great challenges for channel modeling for mmWave-UAV communications. Existing works on mmWave-UAV communications focus mainly on theoretical system performance analysis and usually use statistical channel models which are based on terrestrial measurement results \cite{rodriguezfernandez2018positi,ke2019positi, yi2019modeli,yi2019aunifi, yi2020cluste, rupasinghe2018compar, rupasinghe2019anglef, rupasinghe2019nonort, li2020millim}, which may not be accurate enough for mmWave-UAV communications. Therefore, a wide range of channel measurement campaigns are urgently needed to acquire the real channel parameters to facilitate the corresponding standardization work and to support statistical channel models for mmWave-UAV communications. In particular, the combination of ray tracing and 3D mapping provides high accuracy for the channel measurement \cite{jiang2020mapbas, zhu2020effect, zhu2021mapbas}. Moreover, the real UAV fluctuations have to be accounted for for channel modeling to ensure the robustness of the system performance to weak jitter. In addition, in order to obtain the practical channel information in the real dynamic scenario, it is necessary to develop low-complexity channel estimation methods. Particularly, channel estimation can be carried out cooperatively based on learning methods to improve performance for multi-UAV scenario \cite{zhang2021distri}, while the convergence and time efficiency need to be ensured.

\section{Performance Analysis Basics}
	To unveil the potentials of mmWave-UAV communications, system-level and network-level performance analysis has to be conducted. Recently, several works have reported results for performance analysis and performance evaluation for networks where mmWave-UAV APs serve ground UEs \cite{yi2020cluste,kovalchukov2018analyz,jan2019qosbas,wang2019covera,liu2019mmwave,fontanesi2020outage,khosravi2018perfor}. In these works, stochastic geometry is the most commonly used analysis tool to capture the behavior of the network and to generate the distributions of the ground UEs, buildings, and UAVs. Different from terrestrial networks, the modeling of the aerial AP systems has to take into account the effect of the high altitude of the UAVs, where the existing results for the two-dimensional (2D) plane have to be generalized to 3D space. The system performance is highly dependent on the flight altitude of the UAV because it impacts both the probability that an LoS path exists and the transmission distance. In addition to the 3D location of the UAVs, the system performance also depends on other aspects, such as the type of antenna array, beamforming architecture, channel environment, frequency bands, and antenna radiation patterns. Some of these aspects have already been discussed in Sections II and III, and thus, we will focus on the antenna radiation patterns and key performance indicators for mmWave-UAV communications in this section.
	
	\subsection{Antenna Radiation Pattern}
	Different from the omnidirectional antennas in the sub-6~GHz frequency bands, mmWave antennas have a highly directional characteristic, which makes the performance analysis for mmWave-UAV communication systems more difficult. For directional antennas with fixed radiation patterns, the function of the beam gain with respect to the elevation and azimuth angles is difficult to be obtained in closed form. This is because the antenna radiation pattern is usually influenced by the environment and has an irregular shape. For a typical half-wavelength spacing ULA, if we select a steering vector for beamforming, the beam pattern is the well-known Fej\'{e}r kernel function given by
	\begin{equation} \label{actual_pattern}
		\begin{aligned}
			&G(N,\theta_{\mathrm{tilt}},\theta)\\
			=&|\mathbf{a}^{\mathrm{H}}(N,\theta_{\mathrm{tilt}})\mathbf{a}(N,\theta)|^2=\frac{1}{N^2}\left|\sum \limits _{n=0}^{N-1} e^{j n\pi(\cos\theta_{\mathrm{tilt}}-\cos\theta)}\right|^2 \\
			=&\frac{\sin^2[\frac{\pi N}{2} (\cos\theta_{\mathrm{tilt}}-\cos\theta)]}{N^2 \sin^2[\frac{\pi}{2} (\cos\theta_{\mathrm{tilt}}-\cos\theta)]}
			\triangleq \frac{\sin^2[\frac{\pi N}{2} \Delta\Omega]}{N^2 \sin^2[\frac{\pi}{2} \Delta\Omega]},
		\end{aligned}
	\end{equation}
	where $N$ is the antenna size. $\theta_{\mathrm{tilt}}$ and $\theta$ are the tilt of the beam boresight and the elevation AoD, respectively. As can be observed, the array gain depends on the difference between the cosines of two angles, i.e., $\Delta\Omega$. Due to the fact that the sine functions appear in both the numerator and denominator, it is not easy for further analysis of the system performance in general. Hence, to make the analytical form more tractable, several approximate antenna patterns have been proposed for mmWave communications \cite{yu2017covera,yi2020cluste,wang2019covera,li2019closed}. Next, we introduce these antenna patterns to facilitate the performance analysis for mmWave-UAV communication systems.
	
	\emph{Flat Pattern:} This is the most common simplified antenna pattern which assumes flat gains in the mainlobe and sidelobe, respectively. If we assume the antenna pattern for horizontal region is omnidirectional, the antenna gain with respect to the elevation angle is given by
	\begin{equation} \label{flat_pattern}
		G_{\mathrm{flat}}(\theta)=\left\{
		\begin{aligned}
			&G,~\theta \in [\theta_{\mathrm{tilt}}-\frac{\theta_{\mathrm{3dB}}}{2},\theta_{\mathrm{tilt}}+\frac{\theta_{\mathrm{3dB}}}{2}],\\
			&g, ~\text{otherwise,}
		\end{aligned}\right.
	\end{equation}
	where $\theta_{\mathrm{3dB}}$ represents the half-power beamwidth (HPBW) in vertical direction. $G$ and $g$ are the antenna gains in the mainlobe and sidelobe, respectively. The flat pattern shown in \eqref{flat_pattern} has a simple form and it is more tractable for further analysis. However, this excessive approximation may result in a low accuracy for the analytical results. Especially for the interference-dominated systems, the flat pattern has a relatively low reference significance because the periodical characteristic and the null points in the sidelobe of the original radiation pattern are not captured in this model.
	
	\emph{Sinc Pattern:} An approximation of the Fej\'{e}r kernel antenna pattern can be obtained by substituting the sine function $\sin[\frac{\pi}{2} \Delta\Omega]$ with an affine function $\frac{\pi}{2} \Delta\Omega$ in the denominator. The sinc pattern is given by
	\begin{equation} \label{sinc_pattern}
		\begin{aligned}
			G_{\mathrm{sinc}}(N,\Delta\Omega)&=\frac{\sin^2[\frac{\pi N}{2} \Delta\Omega]}{ \frac{N^2\pi^2}{4} \Delta\Omega^2}.
		\end{aligned}
	\end{equation}
	Due to the fact that when $x$ approaches to 0, the inequality $\sin^2 x \leq x^2$ approaches to the equality, a tight lower bound on the antenna pattern of \eqref{actual_pattern} is defined in \eqref{sinc_pattern}. Especially for the smaller $\Delta\Omega$, the sinc pattern can accurately match the actual antenna pattern. Since the numerators of \eqref{actual_pattern} and \eqref{sinc_pattern} are identical, the null space of the Fej\'{e}r kernel is included in the sinc pattern. Moreover, the denominator of the sinc pattern is an affine function, which is more tractable compared to the Fej\'{e}r kernel pattern.
	
	\emph{Cosine Pattern:} Another approximation of the Fej\'{e}r kernel antenna pattern is the cosine pattern given by
	\begin{equation} \label{cosine_pattern}
		G_{\mathrm{cos}}(N,\Delta\Omega)=\left\{
		\begin{aligned}
			&\cos^2\left(\frac{\pi N}{4} \Delta\Omega\right),~|\Delta\Omega| \leq \frac{2}{N},\\
			&0,~\text{otherwise,}
		\end{aligned}\right.
	\end{equation}
	which properly approximates the mainlobe of the Fej\'{e}r kernel for $|\Delta\Omega| \leq \frac{2}{N}$ by using an elementary function, and simply neglects the sidelobe. Compared to the flat pattern, the cosine pattern reflects the change of the antenna gain with respect to the AoD within the mainlobe. Hence, it can be viewed as a tradeoff between the tractability and the accuracy of the antenna pattern.
	
	\emph{3GPP Pattern:} 3GPP defines a 3D directional model for radiation power pattern of a single AE as follows \cite{3gpp2018studyo}
	\begin{equation} \label{3GPP_pattern}
		G_{\mathrm{3GPP}}(N,\theta)=10^{\frac{-\min\left\{12\left(\frac{\theta-\theta_{\mathrm{tilt}}}{\theta_{\mathrm{3dB}}}\right)^2,G_{\mathrm{SLL}}\right\}}{10}},
	\end{equation}
	where $G_{\mathrm{SLL}}$ (in dB) is the antenna gain of the sidelobe level, and 3GPP suggests a typical value of 30 dB. In fact, the 3GPP pattern utilizes an elementary function to capture the characteristic of the mainlobe, while uses a small constant to define the antenna gain of the sidelobe.
	
	\begin{figure}[t]
		\begin{center}
			\includegraphics[width=\figwidth cm]{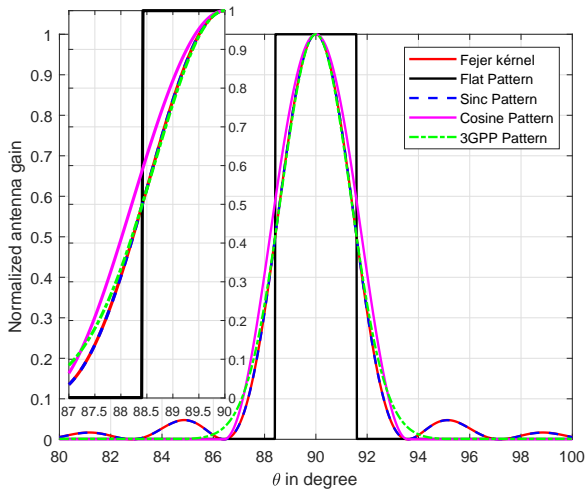}
			\caption{Comparison of the antenna patterns of different approximation strategies for $N=32$ and $\theta_{\mathrm{tilt}}=90^{\circ}$.}
			\label{Fig:antenna_pattern}
		\end{center}
	\end{figure}	
	
	In Fig. \ref{Fig:antenna_pattern}, we show a comparison of the antenna patterns for the approximate strategies mentioned above. To make these schemes comparable, we set the number of antennas as $N=32$. The beam boresight is set as $\theta_{\mathrm{tilt}}=90^{\circ}$. Accordingly, the HPBW can be obtained as $\theta_{\mathrm{3dB}}=3.17^{\circ}$ \cite{balanis2016antenn}. As can be observed, sinc pattern achieves the best performance for approaching the Fej\'{e}r kernel antenna pattern, where the mainlobe and sidelobe are both closely approximated. The cosine pattern and 3GPP pattern both approximate the mainlobe very well.
	
	As we know, the steering vector for a URA/UPA can be decomposed into the Kronecker product of the steering vectors for two ULAs along orthogonal directions. Hence, these strategies above can be easily generalized to the 3D antenna patterns that are characterized by both the elevation AoD in the vertical plane and the azimuth AoD in the horizontal plane. For example, by employing a steering vector for beamforming, the effective antenna gain of a UPA can be represented by the product of two Fej\'{e}r kernel functions \cite{yi2020cluste}.	
	
	\subsection{Performance Metric}	
\begin{figure}[t]
		\begin{center}
			\includegraphics[width=\figwidth cm]{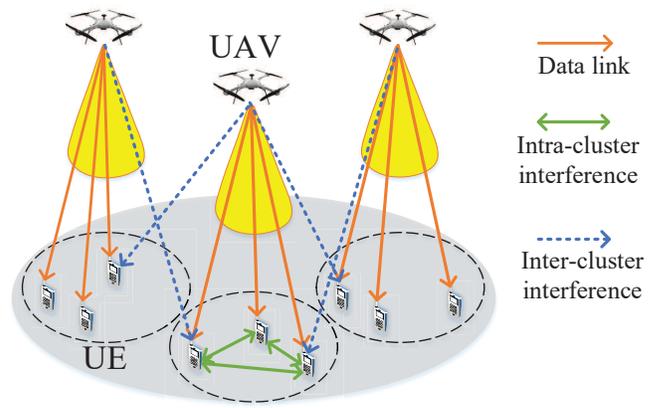}
			\caption{Architecture of the UAV-assisted mmWave communication network.}
			\label{Fig:system_AP}
		\end{center}
	\end{figure}

	Similar to the terrestrial networks, the performance metrics of mmWave communication networks assisted by UAV APs include the signal-to-interference-plus-noise ratio (SINR), spectrum efficiency, energy efficiency, outage probability, communication throughput, delay, etc. Moreover, due to the high altitude and mobility of the UAVs, some new performance metrics should be defined and evaluated to distinguish the ability of the UAV APs from that of the ground BSs. Next, we show the definition of the performance metrics of UAV-assisted mmWave cellular networks. The system model is shown in Fig. \ref{Fig:system_AP}, where $M$ UAV APs are deployed to serve $K$ ground UEs.
	
	\emph{SINR:} We assume that each UE is served by one UAV at the same time, and the UE set served by the $m$-th UAV is denoted by $\mathcal{K}_{m}$. For the downlink/uplink transmission between a UE $k \in \mathcal{K}_{m}$ and its corresponding serving UAV $m$, the SINR can be expressed as
	\begin{equation} \label{SINR}
		\gamma_{m,k}= \frac{|h_{m,k}|^{2}G_{m,k}^{\mathrm{UAV}}G_{m,k}^{\mathrm{UE}}P_{\mathrm{t}}}{I_{\mathrm{intra}}+I_{\mathrm{inter}}+WN_{0}},
	\end{equation}
	where $h_{m,k}$ is the channel gain between UAV $m$ and UE $k$, which embodies the impact of large-scale path loss, shadowing, small-scale fading, multi-path effect, and delay distribution. The channel gain depends on the positions of the UAV and UE, and the propagation environment. $G_{m,k}^{\mathrm{UAV}}$ and $G_{m,k}^{\mathrm{UE}}$ denote the antenna gains at the UAV and UE sides, respectively. Since the mmWave antennas have directional characteristics, the antenna gains depend on the AoDs/AoAs of the available paths of the channel between the UAV and UE. Especially when antenna arrays are employed, the beamforming vectors are highly coupled with the channel matrices. $P_{\mathrm{t}}$ is the transmit power of the target signal, which corresponds to the UAV and UE sides for downlink and uplink transmissions, respectively. $W$ and $N_{0}$ are the bandwidth and the power spectrum density of the additive white Gaussian noise. $I_{\mathrm{intra}}$ and $I_{\mathrm{inter}}$ represent the powers of the intra-cluster interference caused by the signals for the UEs served by the same UAV and the inter-cluster interference from the signals for the UEs served by other UAVs, respectively. The expressions of the intra-cluster and inter-cluster can be derived similar to that of the target signal. From \eqref{SINR}, we find that the positions and antenna patterns of the UAVs will not only influence the power of the target signals, but also influence the power of the interference, which makes the performance analysis and optimization for UAV-assisted mmWave networks more challenging.
	
	To properly simplify the performance analysis, mmWave-UAV communication systems can be classified as a noise-limited system or an interference-limited system. As the bandwidth increases, the noise power becomes dominant compared to the interference. Especially when the density of the buildings is high, the LoS paths are more likely to be blocked and the average signal power decreases. In such a case, the system is noise-limited and the SINR can be approximated to SNR. On the contrary, when the bandwidth is small and the density of the UAVs is high, the system becomes interference-limited and the SINR can be approximated to signal-to-interference ratio (SIR). It was shown in \cite{yi2020cluste} that the system performance will rapidly deteriorate as the density of UAVs increases because it introduces severe interference. In such a case, efficient interference management techniques are required for mmWave-UAV communication networks.
	
	\emph{Coverage probability:} For a predefined threshold $\bar{\gamma}_{k}$, the coverage probability for user $k$ served by UAV $m$ is given by
	\begin{equation} \label{cover_pro}
		P_{m,k}^{\mathrm{cover}}=\mathrm{Pr}(\gamma_{m,k} \geq \bar{\gamma}_{k}),
	\end{equation}
	which is the probability that the SINR of user $k$ is no smaller than the threshold. Hence, the factors that influence the SINR, such as the position of the UAV, the antenna gain, the transmit power, and the interference, are relevant to the coverage probability. Besides, the randomness of the UE distribution and small-scale fading should be considered. Especially for a network where both the UAV and ground BSs can provide communication service, a \emph{multi-cell coverage probability} is defined as the function of the coverage probabilities of the UAV and ground BS \cite{liu2019mmwave}.
	
	Note that the coverage probability reflects the performance of the link between a UAV and a single UE. To evaluate the coverage ability of a UAV AP, the \emph{coverage density} can be defined based on the coverage probabilities of the multiple UEs. Different from the ground BSs which only consider the 2D coverage along the horizontal plane, UAV APs should take the 3D coverage into account. The \emph{3D coverage density} for a UAV AP can be expressed as
	\begin{equation} \label{cover_den}
		D_{m}^{\mathrm{cover}}=\frac{\mathbb{E} [|\mathcal{K}_{m}|]}{V_{m}},
	\end{equation}
	where $\mathcal{K}_{m}$ is the UE set where the SINR of the UE is no smaller than the threshold. $V_{m}$ (in meter$^3$) represents the volume of the region covered by UAV $m$. The expectation operation on the number of covered UEs can be averaged over the time (serving period) and space (trajectory of the UAV).
	
	\emph{Traffic capacity:} The instantaneous capacity for the link between a UAV and a UE for Gaussian signals can be expressed as $C_{m,k}=W\log_{2}(1+\gamma_{m,k})$ measured in bps. Thus, the \emph{average traffic capacity} for UE $k$ is given by
	\begin{equation} \label{traffic_cap}
		\bar{C}_{m,k}=\mathbb{E}\left[W\log_{2}(1+\gamma_{m,k})\right].
	\end{equation}
	Note that the equation in \eqref{traffic_cap} shows the date traffic for a single UE. The \emph{total traffic capacity} of the whole network can be measured by the summation of the average traffic capacity of all UEs in this area, i.e., $C^{\mathrm{total}}=\sum \limits _{m=1}^{M} \sum \limits _{k \in \mathcal{K}_{m}}\bar{C}_{m,k}$. The application of UAVs and HAPs promotes the communication service from the ground to the integrated air-ground space. The service ability of the aerial APs can be measured by the \emph{local data traffic capacity}, which is defined as the ratio between the total traffic capacity and the volume of the covered region, i.e.,
	\begin{equation} \label{local_traffic_cap}
		C^{\mathrm{local}}=\frac{\sum \limits _{m=1}^{M} \sum \limits _{k \in \mathcal{K}_{m}} \mathbb{E}\left[W\log_{2}(1+\gamma_{m,k})\right]}{\sum \limits _{m=1}^{M} V_{m}}.
	\end{equation}
	
	\emph{Energy efficiency:} The energy efficiency, measured in bits per joule, is the amount of information bits that can be successfully transmitted per unit energy consumed. For UAV APs, the energy consumption mainly includes two parts, i.e., navigation-related and communication-related power consumptions. Thus, the energy efficiency of a UAV AP can be expressed as
	\begin{equation} \label{EE}
		\mathrm{EE}_{m}=\frac{\sum \limits _{k \in \mathcal{K}_{m}}\bar{C}_{m,k}}{P_{\mathrm{nav}}+P_{\mathrm{com}}},
	\end{equation}
	where the navigation-related power consumption $P_{\mathrm{nav}}$ includes the energy consumed by UAV hovering and propulsion, which are detailed in \cite{filippone2006flight,difranco2015energy,alhabob2021energy}. The communication-related power consumption $P_{\mathrm{com}}$ includes the transmit power of the signals and the power consumption of the RF and baseband modules. For a network, the energy efficiency can be defined as the ratio of the total traffic capacity and the total power consumption. For mmWave-UAV APs, the energy consumption is usually large. Thus, the energy efficiency is an important performance metric to be evaluated and optimized, especially for high-mobility UAVs \cite{zeng2016throug,zeng2019energy}.
	
	\emph{Secrecy throughput:} UAVs may be deployed in extreme environments to provide communication services, such as battlefields and frontiers. In these scenarios, the electromagnetic environments are very complex. There may exist antagonistic eavesdroppers and jammers. Thus, the secrecy and privacy of UAV communication are important. The secrecy throughput is defined as the effective average transmission rate of the confidential message, which is given by \cite{ju2018safegu,zhu2018secrec,wu2018securi}
	\begin{equation} \label{secrecy_thr}
		\mathrm{S}_{m,k}= \mathbb{E}[\log_{2}(1+\gamma_{m,k})-\log_{2}(1+\gamma_{\mathrm{eav}})]^{+},
	\end{equation}
	where $\gamma_{\mathrm{eav}}$ is the equivalent SINR of the wiretap channel between the serving UAV and the eavesdroppers. The multiple eavesdroppers, which may be located on the ground or in the air, may have colluding or non-colluding ability. For the former one, the eavesdroppers can share their received messages and the equivalent SINR is given by the summation of the SINR for all eavesdroppers. For the latter one, the eavesdroppers cannot share their received messages, and thus the equivalent SINR is the maximum of the SINR among the eavesdroppers. It was shown that the performance of the secrecy throughput can not always improve as the transmit power of the UAV AP increases \cite{zhu2018secrec}. This is because both the achievable rates at the UE and the eavesdroppers increase with the transmit power. Instead, the secrecy throughput can be enhanced by optimizing the UAV positioning and beamforming.

	\begin{table*}[t]\small
	\begin{center}
		\caption{Comparison of different antenna patterns with a typical half-wavelength spacing ULA.}
		\label{Tab:antenna-pattern}
		\begin{spacing}{1.8}
			\begin{tabular}{|c|c|c|c|c|c|}
				\hline
				\textbf{Name of antenna patterns} &\textbf{Fej\'{e}r kernel pattern} &\textbf{Flat pattern} &\textbf{Sinc pattern} &\textbf{Cosine pattern} &\textbf{3GPP pattern} \\
				\toprule[0.5pt]\hline
				\textbf{Accuracy} &Perfect &Very low &Very high &Medium &High\\
				\hline
				\textbf{Tractability} &Very low &Very high &Low &High &Medium \\
				\hline
			\end{tabular}
		\end{spacing}
	\end{center}
\end{table*}

	\subsection{Summary and Discussion}
In Table \ref{Tab:antenna-pattern}, we have compared different antenna radiation patterns in terms of accuracy and tractability. For theoretical performance analysis, approximate antenna patterns are usually used for mathematical tractability. Compared to other models, the cosine pattern and the 3GPP pattern achieve a tradeoff between tractability and accuracy. In addition, key performance indicators pertinent to mmWave-UAV communications have been presented in this section. In particular, the optimization of system parameters, such as 3D position, transmit power, and beamforming vectors, play an important role for performance improvement.

	\section{UAV-Connected MmWave Cellular Networks}
	UAV-connected mmWave cellular networks can be classified into two categories. One is UAV-assisted wireless communication, where the UAVs are deployed as aerial APs or relays to enhance the service ability of the traditional ground cellular network. The other is UAVs acting as aerial UEs connected to the ground cellular network.  In emergency and hotspot regions with urgent communication needs, UAVs can be rapidly deployed to serve as aerial APs and establish temporary data links with ground users. This is one of the most typical communication scenarios for UAV-assisted cellular networks. There have been numerous studies on this topic in the past few years, but most of them focus on UAV communication systems operating in the low frequency bands with omnidirectional antennas \cite{zeng2019access,cao2018airbor,mozaffari2019atutor}. Due to the sparse channel characteristic and the directional beamforming, UAV communication in the mmWave frequency band presents different challenges and requires different solutions compared to systems operating in the low frequency band.

	In the following, four typical application scenarios for UAV-connected mmWave cellular networks are discussed, including beam coverage extension, multiple access, relaying, and mobility enhancement, accompanied by the corresponding challenges and key technologies.
		Firstly, 3D beam coverage based on mmWave beamforming is discussed, which constitutes one of the bottlenecks of 5G BSs. The emerging and flexible multi-beam coverage techniques in mmWave-UAV communication will increase the coverage ability significantly. Secondly, promising multiple access techniques that exploit the spatial characteristics are introduced, and the great potential, challenges, and solutions of mmWave assisted IAB are elaborated. Subsequently, the recent developments regarding mmWave UAV relays are comprehensively reviewed. In particular, IRS-assisted UAV relays will play an important role in the intelligent communication environments envisioned for 6G. Finally, aiming at the mobility of UAVs, we review the beam tracking procedures in 802.11 ad, and then survey and compare the recent progress on beam tracking for mmWave-UAV communications. Meanwhile,  solutions that apply existing techniques to UAV beam handovers are also discussed.

	\subsection{3D Beam Coverage}
	For a ground BS in cellular network, multiple antenna panels are usually equipped to generate fixed beam patterns. Each antenna panel can form a directional beam to cover a sector. In general, the beamwidth in the vertical plane is small because both the BSs and UEs are distributed on the ground. In fact, 2D beams in the horizontal plane are used at the ground BSs. However, for a UAV AP, a fixed 2D beam cannot satisfy the requirement of the coverage for multiple ground UEs. Since a UAV usually operates at a high altitude, 3D beams are required to realize the coverage from the air to the ground, where both the azimuth and elevation angles should be taken into account \cite{cheng2019beamst,bao2019cellco,tafintsev2018improv}. Besides, due to the mobility of UAVs, the flexibility of the beam should be controlled to adapt to the varying of the environment. In the following, we introduce the 3D beam coverage strategies for UAV APs.
	
	\subsubsection{Single Beam Coverage}
	Although the beamforming techniques for mmWave massive MIMO, such as fully digital beamforming and hybrid analog-digital beamforming, have been investigated for many years. The application of these solutions for UAV communication still has a long way to go because of the hardware bottleneck of the mmWave components and the SWAP constraints for UAV platforms. In contrast, the directional antenna is a more mature technology and has been widely used in industries \cite{bao2019cellco}. A directional antenna can shape a single beam with a fixed radiation pattern. Equipping a directional antenna at a UAV, the direction of the beam can be adjusted by using a mechanical adjustment module, such as tripod head and servo system. In addition to the directional antenna, phased array can also shape a direction beam. Compared to the fully digital beamforming and hybrid analog-digital beamforming structures, phased array employs pure analog beamforming, which has a low hardware cost and power consumption. Different from the directional antenna, a phased array can change the direction of the beam by using the electrical adjustment, i.e., changing the phase of the signal for each antenna branch. Compared to the mechanical adjustment, the electrical adjustment has a lower latency and higher efficiency.

	As we have analyzed before, if we employ a steering vector as the beamforming vector, the beam pattern of a phased array is similar to that of the directional antenna. A common characteristic of the directional antenna and a small antenna array employing the steering-vector based beamforming is that they can only generate a single-directional beam in space at one time. If the number of UEs is small and the distribution of the UEs is concentrated, a single beam can cover the UEs. In this case, only the azimuth and elevation angles of the beam need to be optimized, which has a low computational complexity in general \cite{vaezy2020beamfo,tafintsev2018improv,zhang2020uavbea}. When the number of the UEs becomes large or the UEs locate dispersedly, a single narrow beam may not be able to cover all the UEs simultaneously. In such a case, the UEs which are located outside the mainlobe of the beam may face bad channel conditions. To solve this problem, an intuitive way is to dynamically adjust the direction of the beam and serve different UEs in different time slots. The advantage of the single beam coverage is that it requires a low hardware cost, low power consumption, and low computational complexity. However, the coverage efficiency is limited because of the single narrow beam.

	\subsubsection{Multi-Beam Coverage}
	An efficient way to improve the coverage efficiency is to increase the number of beams. For example, multiple directional antennas can be equipped at the UAV and each antenna steers to a specific direction. By optimizing the steering beams of the antennas, the coverage performance of the UAV AP can be effectively improved \cite{li2019closed}. In the following, we will introduce the different schemes for generating multi-beam patterns, and the comparison is shown in Table \ref{Tab:multi-beam}.
	
	For large antenna arrays, the hybrid analog-digital beamforming structure is promising to be used for UAV APs. Although there are plenty of research works on mmWave hybrid beamforming in the past few years, the application of hybrid beamforming for UAV platforms has few studies. Different from the phased array, the hybrid structure employs a small number of RF chains connected with a large number of antennas via phase shifters or switches \cite{ayach2014spatia,xiao2017millim}. In fact, the hybrid beamforming achieves a tradeoff between the high hardware-cost digital beamforming and the low-efficiency analog beamforming. In general, each RF chain can generate a single analog beam. The interference between the multiple beams can be mitigated by digital beamforming, based on ZF, minimum mean square error (MMSE), and other convex optimization techniques.
	
	\begin{figure}[t]
		\begin{center}
			\includegraphics[width=\figwidth cm]{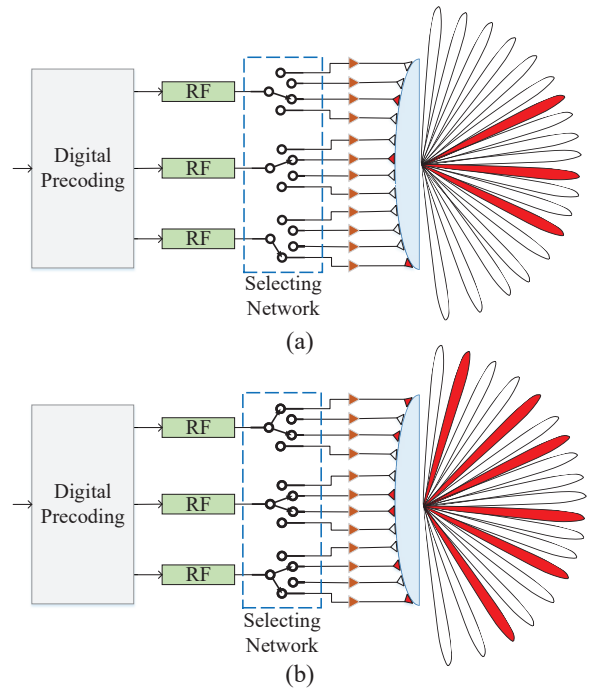}
			\caption{Architecture of the lens antenna array: (a) traditional switch-based selecting network, (b) switch and splitter/mixer based multi-beam forming network.}
			\label{Fig:lens_array}
		\end{center}
	\end{figure}
	
	In addition to the phased-array-based structure, lens antenna array is another typical structure to achieve multi-beam coverage. As shown in Fig. \ref{Fig:lens_array} (a), a lens is equipped on the front end to change the propagation characteristic of the mmWave channels, which are known as beamspace MIMO \cite{zeng2016millim,xie2019onthep}. Then, the received signals at different antennas have different gains, which depend on the AoDs/AoAs of the signals. The analog beamforming portion is realized by employing a switch-inverter network, where a small number of RF chains can select antennas that have high antenna gains. Recently, some research works have studied the application of mmWave lens array for UAV BSs to improve the communication service for UEs \cite{cheng2019beamst,ren2019machin}, where the beam selection is one of the most important issues for UE coverage in the analog domain. Compared to phase-shifter-based hybrid beamforming structure, the lens array has a low hardware cost and power consumption. It is easier to generate multi-beam to improve the coverage of the UEs. However, due to the discrete deployment of the antennas, there exists power leakage for lens arrays \cite{xie2019onthep}. The selected beams may not perfectly match the AoDs/AoAs because of the low flexibility of analog beamforming, and the interference between multiple beams cannot be mitigated in the analog domain.
	
\begin{figure}[t]
		\begin{center}
			\includegraphics[width=\figwidth cm]{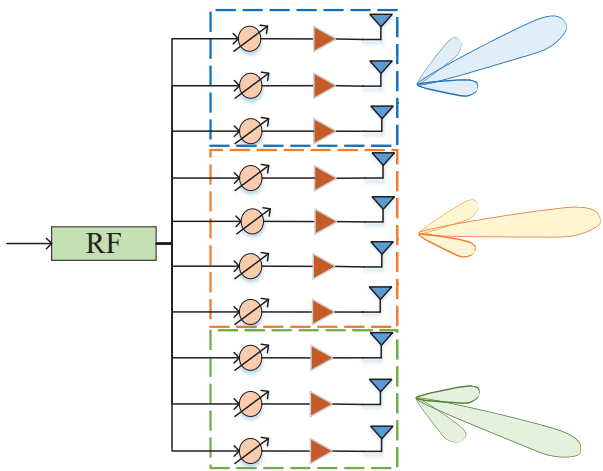}
			\caption{Architecture of the phased antenna array: multi-beam forming in the analog domain based on the sub-array technique.}
			\label{Fig:phased_array}
		\end{center}
	\end{figure}

	It is worth noting that for the above hybrid beamforming structures based on both the phase shifter and lens array, multiple beams are usually generated by different RF chains and they are separate in the space domain. Hence, the number of analog beams is generally no larger than that of the RF chains. In other words, these multi-beam strategies straightway increase the number of beams by improving the hardware facility. However, a UAV platform has SWAP constraints, the number of RF chains may not be so large to generate enough beams covering all ground UEs. To solve this problem, multi-beam forming techniques in the analog domain can be used \cite{wei2019multib,zhu2019millimB5G}. Specifically, one RF chain can obtain a beam that achieves high antenna gain along different directions. For the lens antenna array, as shown in Fig. \ref{Fig:lens_array} (b), one RF chain can be connected with multiple antennas via power splitter/mixer and switch \cite{xie2019onthep}, and thus the UEs along directions of the selected beams all have high antenna gains. Similar to this idea, for the phased antenna array, as shown in Fig. \ref{Fig:phased_array}, the large array can be divided to multiple virtual sub-arrays, where each sub-array steers to a specific direction \cite{wei2019multib,zhu2019millimB5G}. Note that the two multi-beam forming techniques inevitably result in power division, which means that the signal from the RF chain has to be divided to deliver to different branches. Hence, there exists a loss of the beam gain for the above two multi-beam forming solutions.
	
	In addition to the techniques based on improving the physical structure, mathematical optimization methods can also be used to achieve multi-beam forming in the analog domain. For an $N$-element array, we use $\mathbf{a}_{i}$ and $\mathbf{w}$ to denote the steering vector of the $i$-th desired beam and the beamforming vector of the phased antenna array, respectively. Then, the array gain of the $i$-th beam is given by $G_{i}=|\mathbf{a}_{i}^{\mathrm{H}}\mathbf{w}|^2$, $1 \leq i \leq I$. To obtain multiple beams, an optimization problem can be formulated as follows \cite{xiao2018jointp,zhu2018jointp}
	\begin{equation} \label{multi_beam}
		\begin{aligned}
			\mathop{\mathrm{Maximize}}\limits_{\mathbf{w},\alpha}~~~~~ &~~\alpha\\
			\mathrm{Subject~ to}~~~~~ &\left|\mathbf{a}_{i}^{\mathrm{H}}\mathbf{w}\right|^2 \geq \alpha , ~ 1 \leq i \leq I,\\
			&\left|\left[ \mathbf{w}\right]_{n}\right| = \frac{1}{\sqrt{N}}, ~ 1 \leq n \leq N,
		\end{aligned}
	\end{equation}
	where $\alpha$ is a slack variable to improve the beam gain along the desired directions. Note that Problem \eqref{multi_beam} is a non-convex problem because of the basic non-convex form of the constraints. To solve this problem, proper relaxations can be used to make the original problem convex. For example, the first constraint can be relaxed to be a linear one, such as $\mathrm{Re}(\mathbf{a}_{i}^{\mathrm{H}}\mathbf{w}e^{-j\phi_{i}})\geq \sqrt{\alpha}$, where $\phi_{i}$ denotes the phase of $\mathbf{a}_{i}^{\mathrm{H}}\mathbf{w}$. The strict constant-modulus constraint can be relaxed to $\left|\left[ \mathbf{w}\right]_{n}\right| \leq \frac{1}{\sqrt{N}}$, which is a convex constraint. However, the impact of the relaxation on the optimality should be evaluated. In general, the optimal beamforming vector for Problem \eqref{multi_beam} after modulus relaxation has only a small number of elements that do not satisfy the constant-modulus constraint \cite{xiao2018jointp,zhu2018jointp}. This is an interesting conclusion for ease of handling the constant-modulus constraint when optimizing the analog beamforming. Besides, some heuristic algorithms can also be used to solve the multi-beam forming and coverage problem, such as the particle swarm optimization (PSO) and artificial bee colony algorithms \cite{zhu2019jointt,xiao2020unmann}. These strategies may obtain a near-optimal solution for multi-beam coverage, however, at the expense of a high computational complexity.
	
	\begin{table*}[t]\footnotesize
		\begin{center}
			\caption{Comparison of different schemes for generating multi-beam patterns.}
			\label{Tab:multi-beam}
			\begin{spacing}{1.3} \begin{tabular}{|m{0.07\textwidth}<{\centering}|m{0.06\textwidth}<{\centering}|m{0.08\textwidth}<{\centering}|m{0.09\textwidth}<{\centering}|m{0.06\textwidth}<{\centering}|m{0.08\textwidth}<{\centering}|m{0.10\textwidth}<{\centering}|m{0.26\textwidth}<{\centering}|}
					\hline
					\textbf{Reference} &\textbf{Type of antenna array} &\textbf{Multi-beam generating mode} &\textbf{Direction adjustment mechanism} &\textbf{Hardware cost} &\textbf{Power consumption} &\textbf{Computational complexity} &\textbf{Limitations}\\
					\toprule[0.5pt]\hline
					\cite{xiao2017millim,ayach2014spatia} &Phased antenna array &Multiple RF chains &Many phase shifters and switches &Very high &Very high &Low &Number of RF chains is limited by UAV SWAP constraints\\
					\hline
					\cite{zeng2016millim, cheng2019beamst, ren2019machin} &Lens antenna array &Multiple RF chains &Switch-inverter network &High &High &Low &Number of RF chains is limited by UAV SWAP constraints,  power leakage exists\\
					\hline
					\cite{wei2019multib, zhu2019millimB5G} &Phased antenna array &Single RF chain &Multiple partitioned virtual sub-arrays &Medium &Medium &Medium &Loss in beam gain\\
					\hline
					\cite{xie2019onthep} &Lens antenna array &Single RF chain &Power splitter/mixer and switches &Low &Low &Medium &Power leakage and beam gain loss\\
					\hline
					\cite{xiao2020unmann, xiao2018jointp,zhu2018jointp, zhu2019jointt} &Phased antenna array &Single RF chain &Mathematical optimization methods &Low &Low &High &High computational complexity may lead to low time efficiency\\
					\hline
				\end{tabular}
			\end{spacing}
		\end{center}
	\end{table*}
	
	\subsubsection{Flexible Beam Coverage}
	Note that the mmWave directional beams have particular resolutions in the angle domain, i.e., the beamwidth. As the number of ground UEs is very large, it may not always be possible to cover all UEs by generating a large number of beams because these beams may cause interference to each other when they have overlaps. In such a case, the single or multiple directional beams cannot perfectly cover all UEs. One possible way to improve the coverage is to use a flexible beam, where the shape of the 3D beam can be adaptively adjusted according to the distribution of the UEs for realizing a full coverage of the target region. Due to the mobility of the UAV, the relative position between the UAV and the target region may change frequently. Thus, a flexible 3D beam coverage is necessary for UAV APs. It is worth noting that the flexible beam coverage is hard to be accomplished by directional antennas or lens arrays because of their fixed beam patterns in the space domain. In contrast, the phase-shifter based analog beamforming and hybrid beamforming structures are more likely to obtain flexible beams because more DoFs are available for beamforming optimization. Next, we introduce some schemes for flexible beam coverage.
	
	A low-complexity 3D beam coverage strategy based on sub-array techniques was proposed in \cite{zhu20193dbeam}, where a UPA is equipped at the UAV to cover the target region. In general, the beam coverage is distinguished in the angle domain for antenna arrays. Given the positions of the UEs and UAV, the range of the AoDs/AoAs can be easily calculated according to basic geometry. Then, the envelope of the target region is obtained in the angle domain and a broad beam is designed to cover this envelope. The basic idea to generate a broad beam is to divide a large array into multiple sub-arrays and schedule the sub-arrays steering to adjacent directions with particular interval. On one hand, the beam can be broadening for each sub-array because the beamwidth is approximately inversely proportional to the array size. On the other hand, the beam can be broadening by combining multiple sub-beams of the sub-arrays. After determining the minimum number of required sub-arrays, a beamforming vector can be obtained in closed form to fully cover the target region and avoid the leakage of the antenna gain \cite{zhu20193dbeam}.
	
	Another possible way to obtain a flexible beam is to use the codebook \cite{wang2019joint3}. It was shown that any arbitrary beam pattern can be acquired by using a two-step codeword design \cite{chen2020twoste}. In this strategy, ideal codewords are first designed to approach the desired beam pattern by ignoring the hardware constraints, such as the resolution of the phase shifter and the number of RF chains. Both the amplitude and phase of the target beam pattern are taken into account to make full use of the DoF for beamforming. It was shown that a closed-form solution can be obtained by alternately optimizing each element of the ideal codeword. Then, considering the hardware constraints, practical codewords are designed to approach the ideal codewords \cite{chen2020twoste}.
	
	Riemannian manifold optimization is a useful tool for analog beamforming design to handle the constant-modulus constraint due to the hardware limit \cite{yu2016altern,fan2019flatbe}. Given any desired beam pattern, a beamforming problem can be formulated to minimize the mean square error between the desired and designed beam patterns. The power and/or modulus constraints restrict the beamforming problem into a Riemannian manifold. Thus, the beamforming problem can be seen as an unconstrained problem in a Riemannian manifold, where the Riemannian gradient can be found for the alternating optimization of the beamforming vector. It was shown that a flat beam pattern can be obtained by using the Riemannian optimization technique \cite{fan2019flatbe}.
	
	\subsubsection{ML-Based Beam Coverage}
	ML-based beam coverage can break the performance limits of the traditional beamforming techniques for mmWave communications \cite{gao2017machin,li2020millim,ren2019machin}. For mmWave-UAV APs, the ML strategies for beam coverage have advantages in two aspects. On one hand, the mmWave beamforming usually suffers from hardware constraints, which makes the optimal solution for beam coverage difficult to obtain. All the beamforming schemes mentioned above, including the single beam coverage, multi-beam coverage, and flexible coverage, can only find a sub-optimal solution in general. ML techniques are promising to further reduce the performance gap to the optimum. On the other hand, the coverage of UAV APs can be enhanced by adjusting their positions. For example, a UAV may move to a position where the channel conditions of the UEs are better \cite{tafintsev2018improv}. However, it is impossible for a UAV to travel the whole airspace to acquire the channel information. Thus, it is quite difficult to evaluate and judge which position is the optimal one to achieve the best communication performance. In such a case, ML methods may be utilized to make a sensible decision based on partial environmental information \cite{tafintsev2018improv}.
	
	Since the digital beamforming is easier to be designed by using the model-driven optimization methods \cite{roh2014millim,cao2019orthog,abdelhakam2018effici}, most of the existing works for ML-based beamforming in mmWave MIMO systems are dedicated to analog beam steering or forming, such as cross-entropy optimization \cite{gao2017machin,li2020millim,ren2019machin}, reinforcement learning \cite{vaezy2020beamfo}, deep reinforcement learning, and meta learning. When using these ML-based methods for beamforming at UAV APs, one problem is the overhead of the training. It may require a long period for sharing and training the parameters at different nodes of the network. One may use the offline learning scheme for parameter training, and employ the trained model for online decisions. However, when the environment of the UAV APs changes frequently, the offline learning method may not perform satisfactorily. Besides, there are also some works on the flight control and path planning for UAV APs to improve the communication service \cite{wang2017autono,yin2019intell,yang2019threed,zhang2019cellul}. However, the characteristics of mmWave channels and beamforming are not involved. The ML-inspired beam coverage for mmWave-UAV APs, considering both the beamforming and trajectory control, is an interesting topic for future study.

    \subsubsection{Summary and Discussion}
If the number of the UEs is large or the UEs are widely dispersed, a single narrow beam may not be able to cover all UEs at the same time, which limits the coverage efficiency. Therefore, it is critical to increase the number of beams, and we have compared different methods for generating multi-beam patterns as shown in Table \ref{Tab:multi-beam}. In particular, considering the SWAP constraints, using a single RF chain combined with mathematical optimization or a lens array seems promising for generating multiple beams  \cite{xie2019onthep,xiao2018jointp,zhu2018jointp,zhu2019jointt}. Furthermore, compared to the fixed beam patterns generated by lens arrays, phase-shifter based analog and hybrid beamforming architectures offer more DoFs, and can be exploited to generate flexible beams to mitigate beam overlapping and incomplete coverage \cite{zhu20193dbeam}. In addition, ML techniques are promising to further improve performance compared to other beam coverage approaches \cite{gao2017machin,li2020millim,ren2019machin}, but may introduce severe training overhead. However, ML-based methods for improving the communication coverage have not been investigated for beamforming enabled mmWave-UAV communications yet. Besides, integrating trajectory planning into the design problem is another interesting topic.

	\subsection{Access and Backhaul}
	Multiple access and backhaul are necessary for UAV APs to support the ground UEs and connect to the core network. In this subsection, we will introduce the key technologies for access and backhaul separately, and show the IAB technology for supporting mmWave-UAV APs.
	
	\subsubsection{Multiple Access Technologies}
	The multiple access technologies in the low frequency bands have been widely investigated. For example, orthogonal multiple access (OMA) technologies, such as frequency division multiple access (FDMA), time division multiple access (TDMA), code division multiple access (CDMA), and orthogonal frequency division multiple access (OFDMA) have been successfully used in mobile communication networks. These strategies may also be used in UAV-assisted mmWave communication network, but may not perform well compared to the terrestrial network in low frequency bands. For example, OFDMA technologies have been widely used in 4G (Fourth-Generation) and 5G networks. However, the Doppler effect is more severe for high-frequency carriers and high-mobility systems. The Doppler shift and spread may offset the multiplexing gain of OFDMA, and interrupt the link because of the strong inter-carrier interference. Besides, the directional transmission and the loading limit of UAV platform should be considered for the multiplexing technique design of mmWave-UAV APs \cite{wang2019multip}. In the next, some promising multiple-access technologies that exploit the spatial characteristics will be discussed, including SDMA, non-orthogonal multiple access (NOMA), and CoDMA.

	\emph{SDMA:} Thanks to the sparsity of mmWave channels and the directivity of mmWave beams, much more multiplexing gain in the spatial domain can be leveraged for mmWave-UAV APs \cite{xiao2020uavcom,roy1991spatia,cao2020dualas,sun2015beamdi,zhu2020optimi}. SDMA technology, which is also known as beam division multiple access (BDMA), was first proposed in \cite{roy1991spatia} for increasing the capacity of the communication link between a BS and multiple remote UEs. Specifically, according to the positions or channels of the UEs, the BS can generate different beams to cover the UEs. Then, the UEs located in different beams can transmit their signals in the same time-frequency-code domain while distinguishing them in the spatial domain. Let us consider a phase-shifter-based hybrid beamforming structure. The analog beamforming is first designed according to the AoDs/AoAs of the UE channels, where an independent RF chain is used to shape a beam steering to one UE. Then, by utilizing the ZF-based digital beamforming, the interference between different beams can be canceled. However, when two neighboring UEs have highly correlated channels, the interference between the corresponding beams will become dominant. In such a scenario, the signals cannot be easily separated at the receiver, where ZF-based digital beamforming results in severe power attenuation of the signals or amplification of the noise. Even if two UEs are sufficiently isolated, there still exists interference because of the power leakage between the two channels/beams, which is caused by the MPCs of the channels and the existence of the sidelobe in the beam patterns. A possible way to address this problem is to schedule the UEs according to their channels. The UEs with low-correlation channels or non-overlap beams can be scheduled together and served by the UAV AP simultaneously \cite{sun2015beamdi}.
	
	As we have analyzed in the beam coverage stage, the number of directional beams is generally no larger than that of the RF chains. Especially for a UAV platform, the number of RF chains is small. As a result, the number of UEs served by the UAV cannot exceed the number of RF chains \cite{xiao2016enabli}. To break this limit, SDMA can be combined with other OMA technologies. The UEs are first clustered into different groups, where each group of UEs is covered by one beam. The UEs in different beams perform SDMA while the UEs in the same beam perform OMA strategies, such as TDMA, CDMA, and OFDMA. With this implementation, the high-correlation of the mmWave channels becomes a potential to make full use of the array gain because a narrow beam can be reused to cover multiple UEs. Here comes to another problem, beam coverage. For example, if the array size is large at the UAV, and the UEs are distributed dispersedly, it may be difficult to cover the UEs even after the UE clustering. To tackle this problem, there are two possible solutions. First, we may schedule the UEs according to their distribution. The UAV can navigate in the air with a predetermined trajectory, and serve different UEs according to the scheduling result. In each time slot, SDMA or hybrid SDMA and OMA strategies can be used to serve a number of UEs. Thus, the joint design of user scheduling, beamforming, and UAV trajectory should be considered. This problem was preliminarily investigated in \cite{li2019closed}, where multiple directional antennas with fixed beam patterns are employed at the UAV AP. An iterative algorithm was proposed to alternately optimize the user scheduling, beam steering, and UAV trajectory. The second way to solve the above problem is to employ the multi-beam coverage and/or flexible beam coverage. Specifically, for each RF chain, we may generate multi-beam or flexible beam in the analog domain to fully cover the UEs in each group. Then, the UEs in the same group can be served by using OMA techniques. Note that the interference between different beams/groups may increase because of the beam broadening. Thus, interference management is an important topic for further investigation.
	
	\emph{NOMA:} NOMA is an emerging technology to support massive access in wireless communication networks \cite{ding2017asurve,dai2018asurve}. The transmitter superimposes different signals in the same orthogonal resource block while distinguishing them via different power levels. At the receiver, successive interference cancellation (SIC) technology is used to decode the signals in a successive manner. In general, for downlink NOMA, the signals of the UEs with worse channel conditions are likely to be decoded in prior \cite{ding2017asurve,zhu2019optima}. For uplink NOMA, the signals of UEs with better channel conditions and lower data rate requirements are more likely to be decoded first \cite{zhang2020optima}. Different from the low frequency bands, the application of NOMA for mmWave communication has some unique advantages \cite{ding2017random,cui2018optima,zhu2019millimB5G}. First, the propagation attenuation of mmWave signals is high, which enlarges the channel difference of the UEs and increases the multiplexing gain compared to OMA. Second, the directional propagation characteristic makes the mmWave channels of UEs highly corrected, and thus the UEs within the same beam are ideal to perform NOMA to make full use of array gain. Third, the number of RF chains in mmWave systems, especially for UAV platforms, is limited. 
By using the sub-array technique, a multi-beam mmWave-MIMO-NOMA scheme was proposed in \cite{wei2019multib}. The joint analog beamforming and power allocation for two-user mmWave-NOMA systems was investigated in \cite{xiao2018jointp} and \cite{zhu2018jointp}, where the multi-beam forming techniques are used. Based on the random-beamforming strategy, the authors in \cite{cui2018optima} studied the optimal user scheduling and power allocation for mmWave-NOMA systems by utilizing the game theory and branch-and-bound algorithm. Employing hybrid beamforming structures, the user grouping, beamforming, and power allocation were optimized to maximize the achievable sum rate of mmWave-NOMA systems \cite{dai2019hybrid,zhu2019millim}.
	
	The above works show that mmWave-NOMA outperforms mmWave-OMA in terms of both the spectrum efficiency and energy efficiency. Applying NOMA for mmWave-UAV APs is a promising solution to support more number of UEs \cite{tafintsev2020aerial,rupasinghe2019nonort,rupasinghe2019anglef}. To reduce the training overhead, limited-rate feedback strategies, such as angle and distance feedback, can be adopted for mmWave-UAV AP when performing NOMA \cite{rupasinghe2019nonort,rupasinghe2019anglef}. It was shown that the angle feedback is more important than the distance feedback when the UEs are distributed in a large region. In addition to the sum-rate and outage performance, the application of mmWave-NOMA for UAV APs can also enhance the physical layer security \cite{rupasinghe2018enhanc}. This is because the highly directional transmission in the mmWave frequency bands and the superimposed signals in NOMA transmission bring new potentials to improve security against eavesdroppers. In \cite{sun2020secure}, a simultaneous wireless information and power transfer (SWIPT) assisted framework was developed to study the security, reliability, and coverage performance of downlink mmWave UAV networks, where both the OMA and NOMA strategies are considered. The results showed that NOMA outperforms OMA for low transmit power and high codeword transmission rate.
	Moreover, it has been demonstrated that combined with time sharing, NOMA can efficiently increase the number of UEs and improve the fairness of downlink transmission \cite{masaracchia2020energy}.
Therefore, considering the particularity of mmWave communications, the application of NOMA and the coexistence of NOMA and OMA for mmWave-UAV communication systems are worthwhile to be further studied.
	
	\emph{CoDMA:} A new multiple access scheme, namely CoDMA, was proposed for UAV-aided 5G mmWave UDN \cite{wang2019multip}. The key idea of CoDMA is to use flexible mapping constellations to implement self-adaptive data rate with respect to the link quality. For conventional signal constellations, such as quadrature phase shift keying and quadrature amplitude modulation, the diagram of the modulated signals is fixed. A denser constellation diagram corresponds to a higher order modulation and achieves a higher data rate, and vice versa. However, the channel conditions rapidly change because of the mobility of both the UAV and UEs. In such a case, the achievable data rate is restrained by the fixed constellation diagram. Thus, the utilization of rateless-code enabled flexible constellation can break this limit and let multiple UEs communicate with the UAV AP simultaneously. We take the downlink transmission for example. First, the UAV encodes the messages of multiple UEs into groups of rateless symbols, where a constellation encoder is used to realize the constellation domain multiplexing. Then, the rateless symbols are broadcasted to the UEs with a bit rate much larger than the channel capacity. At the receiver, the UEs try to extract the useful information in the rateless symbols. After decoding their own message successfully, the UEs will feed back acknowledgements (ACKs) to the UAV. When the UAV receives ACKs from all UEs, the next round transmission is scheduled. It was shown that in combination with proper user grouping and beamwidth control, CoDMA can adapt to the dynamic mmWave-UAV network and significantly improve the system throughput \cite{wang2019multip}. 
	However, CoDMA may cause significant delay because of the large feedback overhead. This issue was not discussed in \cite{wang2019multip} and the other related works. To address the delay problem, a deadline may be set. Specifically, if a UAV does not receive the ACKs from some UEs before the deadline expires, then retransmission from the UAV to the UEs is launched. Moreover, UEs whose ACKs are not received for several scheduling time slots may be regarded as out-of-range UEs and should be removed from the connected cluster of UEs. With the above approach, the latency and overhead of the links between the UAV and the UEs can be reduced. Nevertheless, more research is needed to address the delay problem and to verify the feasibility of CoDMA for mmWave-UAV communication systems.
	
\begin{table*}[t]\scriptsize
		\begin{center}
			\caption{Summary of representative works on multiple access technologies for mmWave-UAV communications.}
			\label{Tab:multi-access}
			\begin{spacing}{1.2} \begin{tabular}{|m{0.05\textwidth}<{\centering}|m{0.06\textwidth}<{\centering}|m{0.07\textwidth}<{\centering}|m{0.055\textwidth}<{\centering}|m{0.41\textwidth}<{\centering}|m{0.21\textwidth}<{\centering}|}
					\hline
					\textbf{Reference} &\textbf{Multiple access mode} &\textbf{Number of UAV BSs} &\textbf{Antenna} &\textbf{Design objective} &\textbf{Main techniques}\\
					\toprule[0.5pt]\hline
					\cite{xiao2020unmann} &SDMA &Single UAV  &ULA/UPA &Maximizes the achievable sum rate by jointly optimizing the beamforming vector and the 2D position of the UAV subject to a minimum rate constraint for each user, and a constant-modulus constraint &Grid search method and artificial bee colony algorithm\\
					\hline
					\cite{zhu2020optimi} &SDMA &Multiple UAVs &UPA &Maximizes the achievable sum rate of UEs by jointly optimizing the 2D position of the UAVs, UE assignment, and hybrid beamforming subject to a minimum rate
						constraint for each user and a constant-modulus constraint &K-means method, water-filling algorithm, successive convex optimization, greedy algorithm, triangle inequality, and alternating optimization\\
					\hline
					\cite{rupasinghe2019anglef} &NOMA &Single UAV &ULA &Rigorously analyzes the outage probability and sum rate performance of user ordering strategies considering different types of limited feedback involving either distance or angle information &Fej\'{e}r Kej\'{e}l function and stochastic geometry\\
					\hline
					\cite{rupasinghe2019nonort} &NOMA &Single UAV &ULA &Proposes a beam scanning approach to identify the optimal area to be radiated within the region where users are distributed, rigorously derives outage probabilities and respective sum rates for a distance feedback mechanism, and optimizes the operational altitude of the UAV BS &Stochastic geometry and closed-form outage probability expressions for partial CSI\\
					\hline
					\cite{rupasinghe2018enhanc} &NOMA &Single UAV &ULA &Optimizes the protected zone shape to maximize the achievable NOMA sum secrecy rates under potential eavesdropper attacks outside of the user area &Stochastic geometry\\
					\hline
					\cite{sun2020secure} &NOMA &Single UAV &Switched-ULA &Investigates the security, reliability, and
						energy coverage performance in the presence of passive eavesdroppers on the ground considering user security requirements, transmit power and allocation factor, antenna number, and user density &Adaptive genetic simulated annealing algorithm\\
					\hline
					\cite{wang2019multip} &SDMA \& CoDMA &Single UAV  &Not mentioned &Presents a new multiple
						access technique that enables concurrent inter-beam and intra-beam user transmission for mmWave-UAV communications based on
						flexible constellation coding. &Hash function, linear congruential generator, maximum likelihood based tree searching, heuristic grouping and decoding, and correlated symbol sequences\\
					\hline
				\end{tabular}
        \end{spacing}
		\end{center}
	\end{table*}
	
	\subsubsection{Backhaul Technologies}
	Unlike the ground BSs, UAVs as aerial APs cannot establish wired backhaul links over the optical fiber. Thus, wireless backhaul is necessary for UAV APs. To support the ultra high date-rate requirement of mobile UEs, dozens or hundreds of gigabit-per-second rate is required for a UAV backhaul link. Compared to the low-frequency systems, mmWave-band backhaul for UAV APs have the following advantages. First, the large continuous mmWave bands can provide potential GHz bandwidth, which is impossible in the heavy-traffic microwave frequency bands below 6 GHz. Second, the high propagation loss of the signal in the mmWave bands is beneficial to reduce the interference between different cells and improve the reuse of mmWave frequency. This is very important for UAV platforms which usually suffer from severe interference. Third, thanks to the small wavelength of mmWave signals, a large number of antennas can be packed in a small area of UAV and ground macro-BS and acquire considerable transmit and receive array gains simultaneously.
	
	The authors in \cite{gao2015mmwave} discussed the feasibility of mmWave massive-MIMO-based wireless backhaul for 5G UDN. Hybrid analog-digital beamforming structures based on phase-shifter network are used, where one macro-BS can simultaneously support multiple micro-BSs. The key challenges and potential solutions for channel estimation and low-rank precoding/combing were presented. In \cite{shi2019modeli}, a tractable stochastic geometry model was proposed to evaluate the performance of point-to-multipoint assisted mmWave backhaul networks. It was shown that the link distance, LoS interference, and transmit power have a significant impact on the tradeoff between the energy efficiency and spectrum efficiency. To break the bottleneck of backhaul capacity in mmWave communication systems, a deep reinforcement learning approach was proposed to capture and predict the system dynamics caused by channel varying. Then, the backhaul resource can be efficiently utilized by the UEs. The performance analysis for UAV backhaul link was carried out in \cite{galkin2018backha}, where the LTE and mmWave backhaul are both evaluated in an urban environment. Flexible and reconfigurable backhaul architectures have been desirable for 3GPP New Radio. Motivated by this, in \cite{gapeyenko2018flexib}, a flexible and reliable UAV-assisted backhaul framework was proposed for 5G mmWave cellular network, where the dynamic blockage caused by moving humans and possible link rerouting were captured under a 3D multi-path propagation model.

	\subsubsection{IAB}
	Since both the access and backhaul links require wireless transmission, IAB is an important topic for mmWave-UAV APs to improve the reconfigurability of the network. Specifically, a UAV AP can operate backhaul through the access link of a ground macro-BS or another UAV AP. Then, it is possible to support single-hop or multi-hop transmission for a UAV AP to connect to the core network. Note that IAB can utilize any frequency band in which the access link operates. Hence, both the in-band and out-band backhaul strategies are feasible for IAB. For in-band IAB, a higher spectrum efficiency can be achieved because of the frequency reuse, while the interference management is an important issue to be resolved.
	
\begin{figure*}[t]
		\begin{center}
			\includegraphics[width=14 cm]{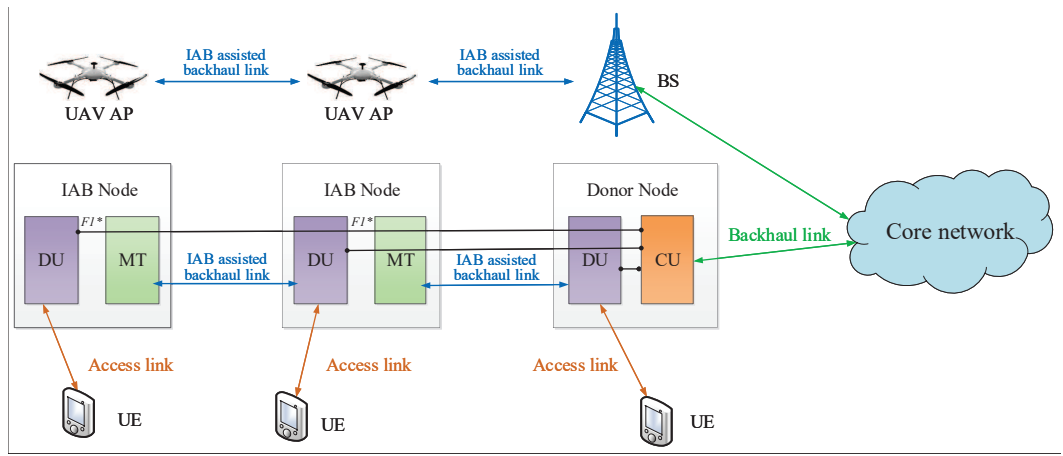}
			\caption{Potential architecture of the IAB network for UAV APs.}
			\label{Fig:IAB}
		\end{center}
	\end{figure*}

	As shown in Fig. \ref{Fig:IAB}, 3GPP introduces a basic architecture of an IAB network \cite{3GPPIAB}, which is potentially to be used for UAV APs. Based on the centralized unit (CU)/distributed unit (DU) split architecture, a CU, including the radio resource control and packet data convergence protocols, is equipped at the donor node. Multiple DUs, including the radio link control and multiple access control protocols, are distributed at the donor node and IAB nodes. The connection between the CU and DUs is standardized F1 interface. The donor node can establish a normal backhaul link to the core network via wired or wireless transmission. Similar to the conventional BS, each IAB node can establish access links with the UEs, while it is treated as a terminal when connecting with the donor node for backhauling. The IAB node includes a mobile-termination part, which can provide connectivity between the IAB node and the DU of the donor node. For an IAB node far away from the donor node, a multi-hop backhaul link can be established by accessing the IAB node to the DU of other available IAB nodes.
	
	It is foreseeable that mmWave frequency bands will be the most promising spectrum for the backhaul link in the IAB network. There are a number of works that investigate mmWave IAB for 5G network. In \cite{saha2018bandwi}, an analytical framework for IAB-enabled 5G mmWave cellular network was established to evaluate the coverage probability of downlink transmission. It was shown that the bandwidth partitioning between the access and backhaul links has an important influence on the network performance. Furthermore, in \cite{saha2019millim}, a stochastic geometry-based analytical framework was developed for a mmWave heterogeneous network, where the macro-BSs have fiber access to the core network and the micro-BSs are wirelessly backhauled by the macro-BSs over mmWave links. The results showed that the capacity of wireless backhaul is a bottleneck for IAB network to improve the UE experience. There are also some preliminary studies employing IAB for mmWave-UAV APs. Inspired by the 3GPP IAB architecture, the authors in \cite{gerasimenko2019protot} designed a prototyping for directional UAV-based wireless communication system, where the practical parameters, such as weather conditions, UAV speed, weight, power consumption, and mmWave antenna configuration, were taken into account. It was shown that high-directivity antenna array can help the UAV realize long-distance backhaul and reduce interference. In \cite{perez2019raytra}, the potential coverage gains of using UAVs in IAB mmWave cellular were analyzed. In particular, ray-tracing methodology was employed to study the propagation characteristics of outdoor mmWave channels. Then, based on the coverage ray-tracing maps at access and backhaul links, the 3D deployment of UAVs can enhance the coverage performance. In \cite{tafintsev2020aerial}, the performance evaluation of UAV-aided radio systems enabled by mmWave IAB was provided. The results demonstrated that leveraging the UAV mobility can improve the performance for serving the moving UEs even if the capacity of the backhaul link is limited. From the above research work, we can find that to break the limit of capacity of the wireless backhaul link, large-antenna-array enabled mmWave beamforming and the adaptive position adjustment of UAV have potentials to enhance the performance of IAB enabled mmWave-UAV network.
	
	\subsubsection{Summary and Discussion}
Specifically, we have summarized the representative works on multiple access technologies for mmWave-UAV communications, as shown in Table \ref{Tab:multi-access}. Three multiple access technologies are particularly promising for mmWave-UAV communications. First, SDMA utilizes multi-beam transmission for broader coverage. However, when two neighboring UEs have highly correlated channels or power leakage, the interference between the two beams will become severe. Thus, UEs with low-correlation channels or beams without overlap can communicate concurrently with the UAV AP via SDMA, while UEs with highly correlated channels or overlapping beams may communicate with the UAV AP based on other OMA techniques, which requires the joint design of UE scheduling, UAV trajectory/deployment optimization, and beamforming. Second, different from traditional OMA methods, NOMA distinguishes signals via different power levels and uses SIC to decode signals. The optimal SIC decode orders for uplink and downlink NOMA were investigated in \cite{zhu2019optima} and \cite{zhang2020optima}, respectively. The combination of NOMA and mmWave communications outperforms mmWave-OMA in terms of the spectrum efficiency and energy efficiency. Besides, NOMA allows UAVs with limited numbers of RF chains to serve multiple UEs. However, there do not seem to be any results available for 3D location optimization for UAVs for the combination of NOMA and SDMA. Therefore, in order to fully exploit UAV deployment flexibility, the application of NOMA combined with SDMA in mmWave-UAV communications and the coexistence with OMA are intersting topics for future research. Third, CoDMA utilizes flexible mapping constellations and adaptively optimizes the data rate according to the link quality of each UE covered by a single beam \cite{wang2019multip}. In particular, CoDMA is compatible with SDMA, which solves the issue of one beam serving one UE. However, the resulting feedback overhead and latency need to be quantified in further studies.

Due to its flexibility, large bandwidth and directionality, mmWave-UAV communications has been considered for backhaul links and mmWave-UAV IAB is a promising architecture for 5G NR and next generation communication networks \cite{3GPPIAB,saha2018bandwi,saha2019millim,gerasimenko2019protot,perez2019raytra}. However, the capacity of wireless backhaul is a bottleneck for IAB networks. Therefore, it is crucial to study the optimal bandwidth partition between the access link and the backhaul link in future works for enhancing the network performance. Besides, mmWave beamforming and deployment/trajectory optimization for UAVs are also worthy of investigation to mitigate the capacity limit of the wireless backhaul links.

	\subsection{Aerial Relay}
	Mobile UAVs can serve as aerial relays to support the transmission between two nodes with poor channel conditions. The application scenarios include the relaying transmission between a ground BS and UEs, D2D communications \cite{wang2018spectr}, assisting long-range backhaul \cite{gapeyenko2018flexib,tafintsev2020aerial,perez2019raytra}, and communication between two terrestrial mobile BSs in emergency situations \cite{cao2018airbor}. Compared to the ground relays, UAVs have better maneuverability and can be rapidly deployed to hot-spot regions or disaster areas to establish a temporary relaying link. Moreover, the high-altitude of UAV makes it possible to establish LoS links with both the source and terminal nodes. For UAV relays, the interference is usually severe because the channel model in the air roughly follows that of free-space attenuation \cite{zeng2019access,khuwaja2018asurve,zeng2016wirele}, especially when omnidirectional antennas are used in the low frequency bands. In contrast, the highly directional mmWave transmission can help reduce the dominant interference for UAV relays. In this subsection, we will introduce the relaying schemes, duplex modes, and some new progress on studying mmWave-UAV relays.
	
	\subsubsection{Relaying Scheme}
	For relaying, there are mainly three categories of relays, namely, amplify-and-forward (AF) \cite{tutuncuoglu2015throug,lee2014ontheo,akhtar2014onthes}, decode-and-forward (DF) \cite{huang2014optima,zhang2015fulldu}, and compress-and-forward (CF) relays \cite{lin2013mimotw,wu2013ontheo}.
	An AF relay directly forwards the received signal from the source node without decoding any symbol. The signal processing at the relay is relatively simple and it even does not introduce additional delay. However, the AF strategy also amplifies the noise and residual self-interference received at the FD relay. The destination node suffers from both the noise of itself and the amplified noise plus residual self-interference from the relay. For a DF relay, the signal from the source node is decoded and then encoded to forward to the destination node. Thus, the DF relay is more reliable compared to the AF relay. Although a DF relay does not forward the noise if it decodes correctly, due to the increasing complexity of signal processing, DF relays usually introduce an additional delay and increase power consumption. In contrast, a CF relay compresses the received signals before forwarding them to the destination node. Hence, the CF strategy provides an effective tradeoff between the complexity of signal processing and communication performance. In particular, when the channel quality between the source and the relay is poor, the communication performance of CF relaying is usually better than that of AF and DF relaying schemes.
	
	\begin{table*}[t]\scriptsize
		\begin{center}
			\caption{Summary of representative works on IRS-assisted UAV relays.}
			\label{Tab:IRS-relay}
			\begin{spacing}{1.2} \begin{tabular}{|m{0.06\textwidth}<{\centering}|m{0.15\textwidth}<{\centering}|m{0.15\textwidth}<{\centering}|m{0.25\textwidth}<{\centering}|m{0.26\textwidth}<{\centering}|}
					\hline
					\textbf{Reference} &\textbf{Scenario} &\textbf{Antenna} &\textbf{Design objective}  &\textbf{Main findings}\\
					\toprule[0.5pt]\hline
					\cite{li2020reconf} &An IRS on a building between  a mobile UAV and a static UE in downlink &A ULA at the IRS; a single omnidirectional antenna at the UAV and UE &Jointly designs UAV trajectory and passive IRS beamforming to maximize the average achievable rate &The assistance of the IRS is beneficial to substantially improve the communication quality of UAV-enabled networks \\
					\hline
					\cite{long2020reflec} &A mobile UAV relay equipped with an IRS between the BS and a group of UEs in uplink &A ULA at the IRS; a single omnidirectional antenna at the BS and UEs &Jointly optimizes the UAV's trajectory, the IRS phase shifts, user association, and transmit power to maximize the secure energy efficiency &The secrecy energy efficiency and achievable rate decrease as the UAV altitude increases \\
					\hline
					\cite{yang2020onthep} &An IRS on a building between  a UAV relay and a source node; DF relaying and HD mode &A ULA at the IRS; a single omnidirectional antenna at the source node, UAV relay, and destination node &Analysis of the outage probability, average BER, and average capacity in IRS-assisted dual-hop UAV communication system &The optimal position of the UAV relay which minimizes the outage probability is close to the destination node \\
					\hline
					\cite{ranjha2020urllcf} &An IRS on a building between  a UAV relay and a destination node; DF relaying and HD mode &A ULA at the IRS; a single omnidirectional antenna at the source node, UAV relay, and destination node &Jointly designs the passive beamforming, blocklength, and UAV position to minimize the decoding error rate &The decoding error rate decreases logarithmically with increasing number of IRS elements, and decreases with increasing blocklength \\
					\hline
					\cite{lu2020aerial} &An aerial IRS between  a source node and the target area &A ULA/UPA at the aerial IRS; a UPA at the source &Jointly optimizes the transmit beamforming for the source node as well as the placement and 3D passive beamforming to maximize the worst-case SNR in the target area &The optimal horizontal aerial IRS position only depends on the ratio between the source-destination distance and the altitude of the aerial IRS; it achieves an optimal balance between minimizing the angular span and the concatenated path loss \\
					\hline
				\end{tabular}
			\end{spacing}
		\end{center}
	\end{table*}
	
	In addition to the AF, DF, and CF relaying schemes, IRS/RIS-assisted UAV relay has also been studied recently. An IRS consists of multiple configurable elements and can change the propagation environment via reflecting the incident signals. For sparse mmWave channels which are sensitive to blockage, IRSs can effectively improve the channel qualities between transmitters and receivers. Different from the traditional AF and DF relays, the passive elements of IRS are more energy-efficient and cost-effective. By adjusting the phases of the incident signals, i.e., passive beamforming, the signal power at the receiver can be enhanced. Since the signals are forwarded via reflection, no transmit power is required and no noise is induced at the IRS. The integration of IRS for UAV relay can provide more potentials to improve the quality of the target links and weaken the quality of the interference links \cite{li2020reconf,yang2020onthep,ranjha2020urllcf,long2020reflec,lu2020aerial}. The IRS can be deployed on the ground buildings to assist the communication between a UAV relay and ground users. Besides, the IRS can also be equipped on a UAV relay to enhance the communication between a BS and multiple users on the ground \cite{lu2020aerial}. For both scenarios, the placement/trajectory of the UAV relay and the passive beamforming of the IRS can be jointly exploited and optimized for the improvement of the system performance \cite{lu2020aerial}. For example, an IRS-assisted UAV relaying system was investigated in \cite{yang2020onthep}, where an IRS is installed on a building to assist the link between a source node and a UAV relay, and the DF relaying protocol is employed at the UAV to serve the destination node. The outage probability, average BER, and average capacity were derived for the considered system \cite{yang2020onthep}, and it was shown that the performance highly depends on the horizontal position and altitude of the UAV relay. The authors in \cite{ranjha2020urllcf} considered an ultra reliable low latency communications (URLLC) scenario facilitated by UAV relay and IRS, where the UAV positioning, passive beamforming of the IRS, and the blocklength were jointly optimized for minimization of the total decoding error rate over the link between the ground transmitter and the ground user. In \cite{long2020reflec}, the joint trajectory, user association, power allocation, and passive beamforming optimization was studied for maximization of the fair secrecy energy efficiency in an IRS on-board UAV relay system. In Table \ref{Tab:IRS-relay}, we summarize these represent works on IRS-assisted UAV relaying communications.
	
	\subsubsection{Duplex Mode}
	
	\begin{figure}[t]
		\begin{center}
			\includegraphics[width=\figwidth cm]{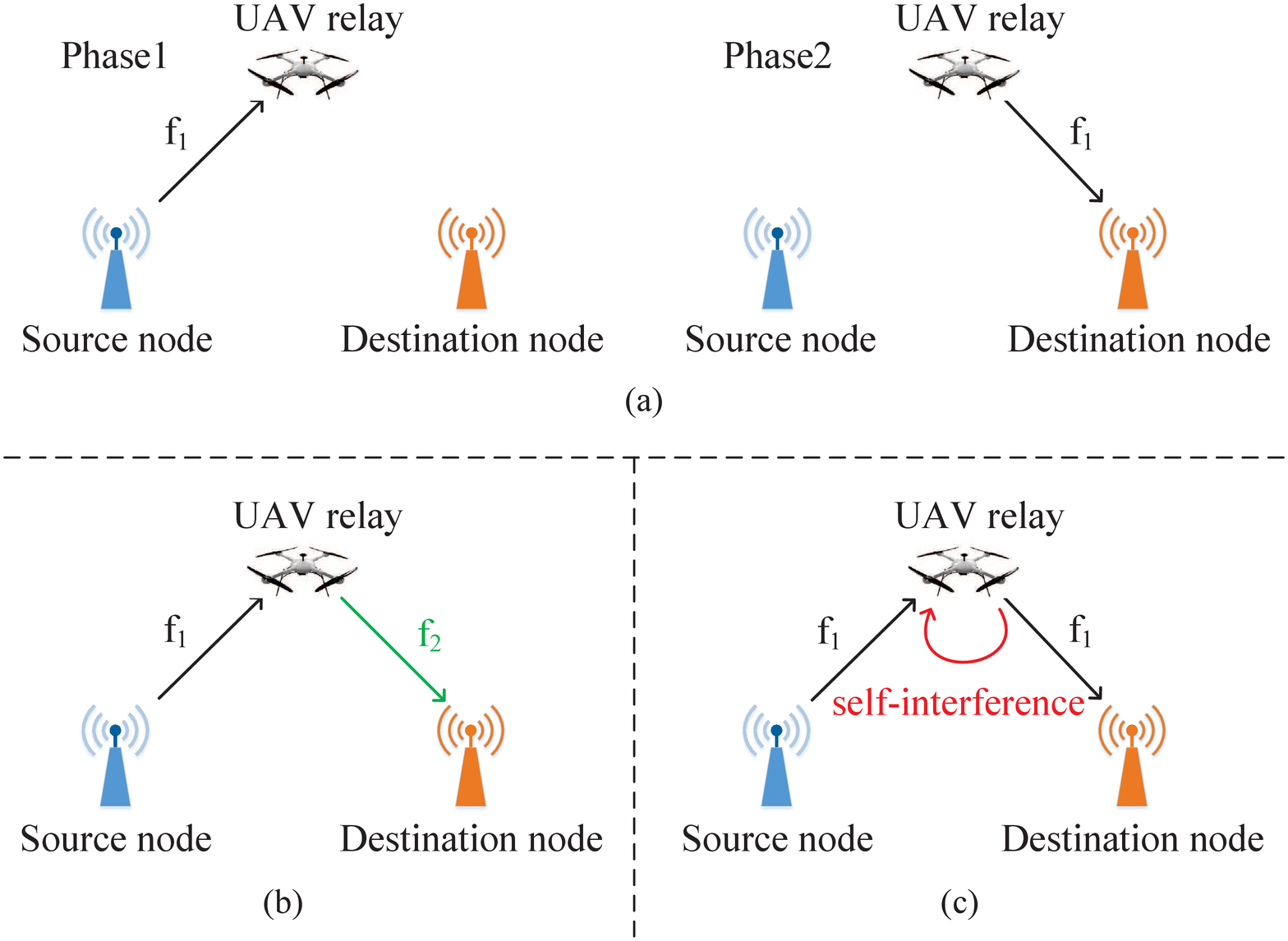}
			\caption{Illustration of different duplex modes for UAV relays: (a) HD relay, (b) out-band FD relay, (c) in-band FD relay.}
			\label{Fig:relay_duplex}
		\end{center}
	\end{figure}
	
	According to the duplex mode, a relay can be roughly classified into half-duplex (HD), out-band FD, and in-band FD relay. As shown in Fig. \ref{Fig:relay_duplex} (a), an HD relay should accomplish the data forwarding with two phases. In the first phase, the source node transmits the signal to the relay. In the second phase, the relay forwards the signal to the destination node. Since the two links operate at different time slots in an alternating manner, synchronization is an important issue to be addressed. Besides, this implementation results in additional delay. For an out-band FD relay, the links from the source node to the relay and from the relay to the destination node can operate simultaneously, where two frequency bands are required as shown in Fig. \ref{Fig:relay_duplex} (b). Note that the HD and out-band FD relays achieve the same spectrum efficiency because of the half-multiplexing in time or frequency domain. In contrast, an in-band FD relay can improve the spectrum efficiency. If the self-interference is ideally eliminated for an in-band FD relay, a twice spectrum efficiency can be achieved compared to the HD or out-band FD relays. Hence, the self-interference cancelation is one of the most important issues for in-band FD relays. It was shown that by isolating the transmit/receive antenna and adopting the interference cancellation techniques in the propagation-analog-digital domain, the self-interference can be well mitigated \cite{riihonen2011mitiga,duarte2014design,everett2014passiv,liu2015inband,duarte2012experi}. In addition to these technologies, the directional beamforming technique can be leveraged for self-interference cancelation for mmWave communication systems with in-band FD UAV relays \cite{zhu2019millim,satyanarayana2019hybrid,zhang2019onprec}.

	\subsubsection{Recent Progress}
	There are some preliminary research works on UAV relays employing mmWave technologies \cite{yi2019aunifi,kong2017autono,ghazzai2018trajec,ma2019secure,sun2019secure,zhang2019reflec,zhu2020millim}. In \cite{yi2019aunifi}, a unified spatial framework for UAV-aided networks was proposed. A UAV is deployed to an area with no communication service for collecting the messages from the UEs. Then, the UAV flies to the nearby BS and sends these messages via the access link. The results revealed that the altitude of UAV and the antenna scale both have significant influence on the coverage performance of the UAV-relay-aided network. The authors in \cite{kong2017autono} proposed a novel method, namely, AutoRelay, to realize the autonomous deployment for a UAV relay in mmWave communication systems. Specifically, the UAV samples the link qualities to the transmitter and receiver in real time when approaching to the target position. The sampled information is utilized to estimate the link qualities in candidate space by leveraging compressive sensing theory. The UAV can calculate and update the optimal position in the candidate space according to the estimation results. In such a manner, the UAV relay can autonomously find the optimal position during the flight, which improves the accuracy and efficiency for UAV deployment. Reference \cite{ghazzai2018trajec} developed a generic
	framework to assign UAV relays to support transceivers, where the trajectories of UAVs were optimized to minimize the service time, and both the typical microwave and the mmWave bands were considered. The secure transmission for UAV-assisted relay in mmWave communication networks is also an interesting topic, which can be found in \cite{ma2019secure,sun2019secure}. Note that in the above works \cite{yi2019aunifi,kong2017autono,ghazzai2018trajec,ma2019secure,sun2019secure}, fixed antenna patterns with simplified expressions are employed to facilitate the performance analysis and optimization for UAV-relay assisted mmWave communication systems. The potentials of mmWave beamforming for interference mitigation and enhanced transmission can be further exploited. In \cite{zhu2020millim}, an FD UAV relay was deployed to assist the transmission between two ground nodes in mmWave networks, where the UAV positioning, beamforming, and power control were jointly optimized to maximize the achievable rate of the system. It was shown that the self-interference at the FD UAV relay can be effectively suppressed via analog beamforming. This inspires us to leverage the DoF in the space domain, i.e., directional beamforming, to handle the strong interference for mmWave-UAV relays.
	
\begin{table*}[t]\scriptsize
		\begin{center}
			\caption{Summary of recent progress in aerial relay assisted mmWave-UAV communications.}
			\label{Tab:aerial-relay}
			\begin{spacing}{1.2} \begin{tabular}{|m{0.05\textwidth}<{\centering}|m{0.05\textwidth}<{\centering}|m{0.04\textwidth}<{\centering}|m{0.15\textwidth}<{\centering}|m{0.26\textwidth}<{\centering}|m{0.30\textwidth}<{\centering}|}
				\hline
				\textbf{Reference} &\textbf{Relaying scheme} &\textbf{Duplex mode} &\textbf{Antenna} &\textbf{Design objective and main techniques} &\textbf{Main findings}\\ \toprule[0.5pt]
				\hline
				\cite{yi2019aunifi} &DF &HD &UPA &Derives several closed-form expressions for the coverage probability in UAV relaying networks with multiple clustered users &There exists an optimal UAV altitude that maximizes the coverage probability; the effects of thermal noise and NLOS transmission can be made negligible in such a system \\
				\hline
				\cite{kong2017autono} &DF &HD &Unstated &Solves a relay deployment problem based on compressive sensing &The proposed scheme achieves high communication quality, but at the expence of a larger delay\\
				\hline
				\cite{ghazzai2018trajec} &DF &HD &An omnidirectional antenna for microwave communications and a directive antenna for mmWave communications &Determines the association between transceivers and UAVs, as well as the trajectories of UAVs to minimize the total service time of the network &There is a tradeoff between the flying time and the communication time in such a scenario; it is important to optimize the number of the UAVs to serve the dispersed nodes\\
				\hline
				\cite{ma2019secure} &DF &HD &UPA &Derives the secrecy outage probability for an opportunistic relay selection scheme, and degrade the qualities of the eavesdropping channels  via cooperative jamming &There exists an optimal attitude of the UAVs, an optimal jamming power, and an optimal density of jamming UAVs for achieving the best secrecy outage probability\\
				\hline
				\cite{sun2019secure} &DF &HD &UPA &Optimizes the source/UAV transmit power, power splitting ratio, and UAV's location to maximize a secrecy rate lower bound &The average secrecy rate at high mmWave frequencies outperforms that at low mmWave frequencies for high eavesdropping node densities\\
				\hline
				\cite{zhu2020millim} &DF &FD &UPA &Jointly designs UAV position, analog beamforming, and power control to maximize the achievable rate between source node and destination node &The proposed method can closely approach a performance upper bound; the proposed alternating interference suppression algorithm shows high robustness with respect to beam misalignment\\
				\hline
			\end{tabular}
        \end{spacing}
		\end{center}
	\end{table*}

	\subsubsection{Summary and Discussion}
	We have summarized the relevant works on IRS-assisted UAV relay communications and aerial relay assisted mmWave-UAV communications, as shown in Tables \ref{Tab:IRS-relay} and \ref{Tab:aerial-relay}, respectively. Specifically, IRS-assisted UAV relays have the potential to significantly improve the channel quality. IRS-based passive beamforming can enhance the signal power and reduce interference. Deployed on the surfaces of buildings, IRSs can assist the UAV relays to communicate with ground obscured users. On the other hand, IRSs can be deployed on UAV relays to assist terrestrial cellular communications. In particular, the joint optimization of UAV 3D trajectory/deployment and passive beamforming can improve the outage probability, average BER, and average capacity \cite{li2020reconf,long2020reflec,yang2020onthep,ranjha2020urllcf,lu2020aerial}. However, obviously UAV fluctuations may potentially have a negative impact on both the transmitter and receiver of IRS-based UAV relays, but existing studies on IRS assisted mmWave-UAV relays do not consider robustness. For aerial relay assisted mmWave communications, most works focus on DF relaying and the HD mode, as shown in Table \ref{Tab:aerial-relay}. Nevertheless, it is of interest to study other relaying schemes and relaying modes for mmWave-UAV communications. Particularly, in-band FD UAV relays can double the spectral efficiency for ideal self-interference cancelation. Therefore, it is important to isolate the transmitting and receiving antennas. Besides,  self-interference can be effectively mitigated by optimizing analog beamforming, which leverages the DoFs in the space domain \cite{zhu2020millim}.
	
	\subsection{Aerial UE}
	In this subsection, we focus on UAVs as aerial UEs connected with the ground BSs in the mmWave frequency band. Although the technologies for supporting aerial UEs have been investigated before, most of the efforts are dedicated to sub-6 GHz technologies \cite{zeng2019access,zeng2019cellul,zhang2019cooper,sundqvist2015onprec,gonzalez2011assess,afonso2016cellul}. 3GPP TR 36.777 shows the requirement of supporting aerial UEs via enhanced LTE network and potential technologies, such as downlink/uplink interference detection, interference mitigation, mobility performance, and aerial UE identification. More sophisticated UAV-ground interference mitigation/cancellation techniques have been proposed in \cite{mei2019cellul,liu2019multib,mei2020cooper}. For aerial UEs, the requirement of the uplink transmission is high because they need to send the application-related data, such as sensor data, images, and videos, to the BSs. 3GPP suggests a data rate of 50 Mbps for uplink transmission from an aerial UE to a terrestrial BS. As some emerging application scenarios for UAVs arise, for example, VR, AR, and hologram, the data requirement for aerial UEs will explosively increase. The use of mmWave frequency bands is promising for supporting the requirements of high data rate and low latency. In \cite{xia2019millim}, a simulation study was conducted to evaluate the feasibility of public safety UAV connectivity through a 5G link at 28 GHz, considering the channel modeling, blockage, and beamforming that follow the 3GPP protocol framework. The experimental results showed that 5G mmWave communication could support a UAV UE to achieve a data rate up to 1 Gbps and a latency not exceeding 1 ms when the BS is located near to the mission area. In \cite{wang2019beyond}, an analytical model was developed to evaluate the potentials of URLLC for mmWave cellular-connected UAV network at 28 GHz. The simulation results demonstrated that for a certain range of message size and block length, a packet error probability of $10^{-5}$ and a latency below 1 ms are achievable. In \cite{ismayilov2018adapti}, an adaptive beam-frequency allocation algorithm was proposed, considering the position uncertainty inherent to high mobility of UEs, such as UAVs, high speed trains and vehicles, in 5G mmWave networks. Connecting UAVs to 4G/5G cellular networks can efficiently reuse the existing resource of the terrestrial infrastructures. In the following, we will show the important issues of supporting aerial UEs enabled by mmWave beamforming, including the beam tracking and handover.
	
	\subsubsection{Beam Tracking}
	Directional beams are important for mmWave communication systems to compensate the high path loss of the signals. Especially for serving aerial UEs which suffer from severe interference, large-scale antenna arrays and 3D beamforming techniques are required at the ground BSs \cite{zeng2019cellul}. When an aerial UE is fast moving in the air, the channel between the ground BS and the aerial UE changes rapidly. Different from the terrestrial UEs, a UAV has a much higher velocity. If the beam training and channel estimation are operated frequently, the pilot overhead will become unacceptable. Thus, efficient beam tracking strategies are required for aerial UEs.
	
	\begin{figure}[t]
		\begin{center}
			\includegraphics[width=\figwidth cm, trim=0 0 0 0,clip]{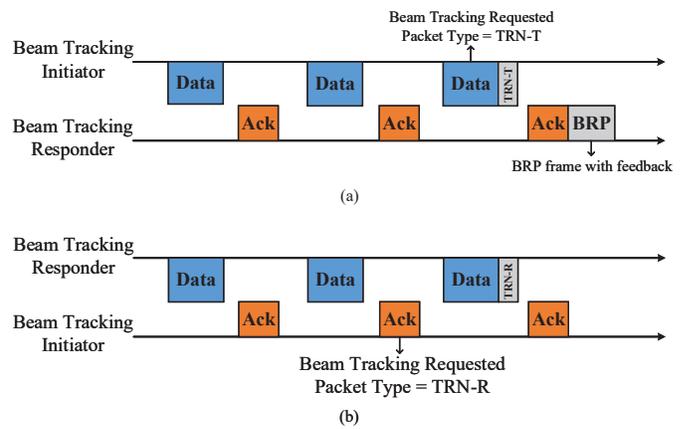}
			\caption{Examples of beam tracking procedures: (a) transmit beam tracking, (b) receive beam tracking.}
			\label{Fig:track_frame}
		\end{center}
	\end{figure}
	
	The IEEE 802.11ad defines a protocol frame for beam tracking in the mmWave frequency bands \cite{nitsche2014ieee80211ad}. When the link quality deteriorates to a certain threshold, the beam tracking initiator will send a beam tracking request to the responder, and a beam refinement process will start. For the transmit beam tracking shown in Fig. \ref{Fig:track_frame} (a), the indicator appends transmit training (TRN-T) units to data frames to perform beam tracking. Then, the responder will transmit an ACK and a beam refinement protocol frame to feed back the tracking results. For the receive beam tracking shown in Fig. \ref{Fig:track_frame} (b), once the responder receives the beam tracking request from the indicator, the receive training (TRN-R) units will be used to perform beam tracking. The optimal beam should be selected to improve the communication quality. Based on the frame defined in IEEE 802.11 ad, two enhanced beam tracking methods were developed in IEEE 802.11 ay \cite{ghasempour2017ieee82011ay,zhou2018ieeeay}. For the beam tracking scheme in the analog domain, a similar method to IEEE 802.11 ad is utilized to cope with the rotation, movement, and blockage via tracking the AoDs/AoAs of the channels. The digital beam tracking is performed under fixed analog beams, and it can cope with the blockage for the hybrid beamforming structures via tracking the baseband effective channels. Compared to the beam tracking method in IEEE 802.11 ad, IEEE 802.11 ay can support multi-channel operation, and thus multiple beams can be tracked simultaneously.

	Different from the terrestrial UEs, a typical feature for a UAV is that it is easy for monitoring and management. Thus, the position and trajectory of a UAV can be sent to the ground BS to assist beam tracking. Note that such information can be transmitted to the BS via the command \& control link with a low data rate. When an LoS path exists between the UAV and BS, the position-aided beam tracking is much more efficient \cite{zhao2018beamtr,zhang2019positi,ke2019positi}. The Kalman filter that needs navigation information has also been used to estimate the movement of UAV and track the channel angle, which can be found in \cite{zhao2018channe} and \cite{ge2019unscen}. However, there are still slight attitude errors from Kalman filter based data fusion, which causes performance losses due to beam misalignment \cite{zhao2018beamtr}. Therefore, the position-altitude prediction is an efficient way to reduce the system overhead and improve performance, where ML-based methods can be used to predict the continuous trajectory of UAV by sampling a few points. When the position information is unavailable, some general solutions can be applied for UAV beam tracking. The authors in \cite{yang2019beamtr} proposed to use the beam coherence time to determine the minimum frequency for beam training, and developed joint beam training and angular velocity estimation to calculate the instantaneous beam coherence time. In \cite{zhang2020uavbea}, based on the UPA structure, the beam tracking strategies were proposed for scenarios of both the random and inertial mobility of UEs. For the former, the angle bounds in the angle domain were derived according to the beam training errors and the maximum velocities of UEs. Then, the candidate narrow beams are obtained for comparison. For the latter, the variation of speed and velocity of the UE is utilized to estimate the beam directions in different time slots. The authors in \cite{huang20203dbeam} proposed a 3D beam tracking scheme for cellular-connected UAV, where the azimuth and elevation angles of the UAV were predicted by utilizing the pilot information, and the dynamic pilot insertion was proposed based on a closed-loop feedback control.
	Based on the measurements for different pairs of analog beamforming vectors, the authors in \cite{chiang2021machin} developed an online/offline Q-learning approach to select the best beam pairs corresponding to the maximum received power.
In these works, the mmWave channels are assumed to be dominated by the LoS paths. However, it is still possible that the LoS path is blocked by ground buildings, mountains or fuselages \cite{bao2017blocka}. In such a case, the performance of the data link between the aerial UE and BS severely deteriorates. How to maintain a stable link under the impact of blockage is still a challenging problem, while handover can be a potential solution.

\begin{table*}[t]\footnotesize
		\begin{center}
			\caption{Summary of representative works on beam tracking for mmWave-UAV communications.}
			\label{Tab:beam-tracking}
			\begin{spacing}{1.2} \begin{tabular}{|m{0.06\textwidth}<{\centering}|m{0.05\textwidth}<{\centering}|m{0.05\textwidth}<{\centering}|m{0.42\textwidth}<{\centering}|m{0.3\textwidth}<{\centering}|}
				\hline
				\textbf{Reference} &\textbf{Scenario} &\textbf{Antenna} &\textbf{Main contents and techniques} &\textbf{Main pros and cons} \\
				\toprule[0.5pt]\hline
				\cite{zhao2018channe} &A2G, one-to-one &URA &Navigation assisted Kalman filter to update UAV movement information, and few pilots to derive channel gain information &Small position and velocity errors; large amount of information provided by navigation sensors \\
				\hline
				\cite{zhao2018beamtr} &A2S, one-to-one &UPA &Rough mechanical adjustment using Kalman filter based data fusion, electrical adjustment based on instantaneous receive power, and pilots to derive gain information &High beam alignment accuracy; partial navigation information, temporal hysteresis in early stage\\
				\hline
				\cite{zhang2019positi} &A2A, one-to-one &UPA &Exchanging motion state information and predicting future position-attitude information of other UAV by Gaussian process based learning algorithm using historical information &High prediction accuracy; partial navigation information, temporal hysteresis, no discussion of time efficiency\\
				\hline
				\cite{ke2019positi} &A2A, one-to-many &UPA &Similar information exchange and prediction method as [76], utilizes a channel evolution parameter to monitor channel state &High spectrum efficiency, guarantees time efficiency; partial navigation information, potential large pilot overhead\\
				\hline
				\cite{yang2019beamtr} &A2G, one-to-one &ULA &Estimating speed of angle variation based on received signal samples, codebook based beamwidth design to adapt to flying speed and range &Adaptable training frequency and beam width, no navigation information; potentially large pilot overhead\\
				\hline
				\cite{zhang2020uavbea} &A2G, one-to-one &UPA &Codebook design with insufficient training based on maximizing the coverage probability sequentially, and bounding angular variation range by using statistical motion parameters &Considering insufficient training, and low overhead; only considers 2D user movement, low accuracy compared to motion prediction\\
				\hline
				\cite{huang20203dbeam} &A2G, one-to-one &UPA &Closed-loop feedback control based dynamic pilot insertion according to angular speed of UAV &Significantly reduces average pilot overhead; beamforming gain drops occasionally\\
				\hline
				\cite{chiang2021machin} &A2A, one-to-many &URA &Online/offline Q-learning based prediction to select analog beams, and SINR based digital beamforming optimization &Strikes a balance between the system resilience and efficiency, high robustness; potential training efficiency\\
				\hline
			\end{tabular}
        \end{spacing}
		\end{center}
	\end{table*}

	\begin{figure}[t]
		\begin{center}
			\includegraphics[width=\figwidth cm, trim=0 0 0 0,clip]{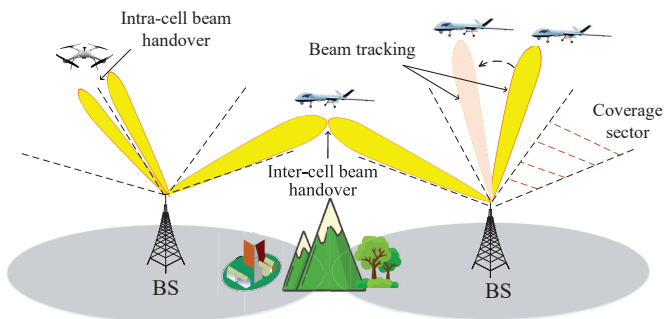}
			\caption{Illustration of the beam tracking and handover for aerial UEs in mmWave cellular network.}
			\label{Fig:aerial_UE}
		\end{center}
	\end{figure}
	
	\subsubsection{Beam Handover}
	Due to the high mobility of an aerial UE, it may pass through different coverage areas of the ground BSs. To guarantee the seamless connection, efficient beam handover strategies are required. The beam handover for a UAV connected with mmWave cellular networks can be divided into two scenarios, i.e., the inter-cell beam handover and the intra-cell beam handover \footnote{ Intra-cell beam handover can be thought of as another mode of beam tracking.}, as shown in Fig. \ref{Fig:aerial_UE}. The former occurs when a UAV is located at the cell edge of the serving BS \cite{tuong2013handov,wang2019alowco,liu2018initia}. The latter occurs when a UAV moves away from the coverage area of one beam and has to be switched to another beam of the serving BS \cite{ren2019arobus,oh2014aneffi}.
	
	We start from the inter-cell beam handover. The first question for cell handover is to judge whether or when to perform handover. For sub-6 GHz networks employing omnidirectional or quasi-omnidirectional antennas, the handover can be determined by comparing the powers of the received signals between the serving BS and the candidate BS. However, when we use directional transmission in the mmWave frequency bands, the received signal power depends on both the channel condition and the beamforming gain. Since the link between the UAV and the candidate BS is not established before handover, it is ``blind'' for the candidate BS to detect the potential aerial UE. It is impossible for the candidate BS to generate a beam accurately pointing to the potential UE because the BS does not know which direction the UE comes from. To solve this problem, an intuitive approach is to let the candidate BS generate an omnidirectional or quasi omnidirectional beam to detect the UE. Then, the powers of the test signals are compared for the determination of handover, where the antenna gain should be normalized when comparing the signal powers of the serving and candidate BSs \cite{tuong2013handov}. This scheme has a low complexity, however, at the expense of low antenna gain. Another possible way is to let the candidate BS scan the whole space by using high directional beams. The beam which achieves the highest signal power will be used for comparison \cite{tuong2013handov}. This method is simple, and the selected beam can be directly used for serving the UE if the handover is declared. However, it requires a high time overhead, especially when large-scale array and 3D beamforming are adopted. To improve the efficiency of the inter-cell beam handover, a promising method is to resort to the cooperation between the serving and candidate BSs. For example, the overlapped coverage area of two neighbouring BSs can be obtained at a candidate BS if the beam pointing of the serving BS is shared. Based on this information, the range of the AoD/AoA between the candidate BS and the UE can be estimated, which efficiently reduces the search space for beam switching \cite{wang2019alowco}. Note that the cooperation between the neighbouring BSs can also significantly improve the accuracy of UE positioning, which can be utilized for cell handover. Employing a mmWave positioning system, the positions of UEs and APs could be obtained at the BS \cite{palacios2019leapl}. Then, the handover and beam selection could be optimized based on the position information. The authors in \cite{yan2019machin} proposed a Kernel-based ML algorithm to predict the positions of the vehicles, where the channel state information (CSI) of sub-6 GHz bands was utilized. Then, by leveraging the historical handover data, K-nearest neighbor ML algorithms were employed to predict handover decisions, where the time consumed for target selection and beam training is greatly reduced.
	
	To solve the intra-cell beam handover problem, the inter-beam coordinated scheduling (IBCS) is an efficient solution \cite{oh2014aneffi,oh2015anenha}. Specifically, for a UE performing beam handover, a BS can allocate radio resources to the UE at the same position in the subframe of the serving and candidate beams. The resource allocation is determined by priority scheduling policy according to the channel qualities of the two beams. Inspired by this idea, a novel jointly-optimized and dynamic scheme was proposed based on multiple beams cooperation, which could reduce the handover failure rate for high speed train \cite{ren2019arobus}. Note that the above inter-beam or intra-beam handover strategies are designed for terrestrial UEs. More specialized research is required to support the beam handover for aerial UEs.
	
	\subsubsection{Summary and Discussion}
For aerial UEs communicating with terrestrial BSs, uplink transmission requires high rates to send application-related data to the ground BS. Various performance tests showed that mmWave-UAV communications can provide high reliability, low delay, and large capacity \cite{xia2019millim,wang2019beyond,ismayilov2018adapti}. In this section, we focused on the challenge of UAV mobility and offered potential solutions for beam tracking and beam handover to solve the problem of mobility. In particular, we have summarized the relevant works on beam tracking for mmWave-UAV communications, as shown in Table \ref{Tab:beam-tracking}. The severe Doppler effect in high-speed UAV communications causes the short beam coherence time, and further influences the frequency of beam training \cite{yang2019beamtr}. If beam training and channel estimation have to be conducted frequently, the pilot overhead becomes unacceptable in mmWave-UAV communications. The position information which is conveyed through the control link can be exploited for beam tracking. Furthermore, ML-based methods are usually used for UAV movement prediction and channel angle estimation based on historical information \cite{chiang2021machin}.

A moving UAV on the edge of a cell requires a timely inter-cell beam handover. The most intuitive approach is to use sub-6\,GHz omnidirectional pilots to  detect the received signal power in order to make the handover decision. If this is not possible, beam scanning can also be used, but will introduce significant overhead and delay problems. Therefore, an effective approach is to reduce the scope of the scanning space by means of cooperation between BSs. In addition, intra-cell beam handovers potentially solve the issues of high overhead and delay in beam tracking. IBCS based on resource allocation has been shown to reduce the handover failure rate for high speed train applications \cite{oh2014aneffi,oh2015anenha}. However, beam handover strategies for aerial terminals need to be further studied.

	\section{MmWave-UAV Ad Hoc Networks}
	A multi-UAV system organized in a mesh manner is refered to as a flying ad hoc network (FANET), where multiple UAVs can collaboratively carry out complex missions. Due to the characteristics of high autonomy, flexibility, and self-healing, FANETs have broad applications in the military and civil domains \cite{zafar2016flying}. Compared to terrestrial ad hoc networks, the network organization and link maintenance for FANETs are more challenging because of fast changing links, especially for harsh environments with strong electromagnetic interference. Unique requirements, such as high throughput, low probability of intercept, and high anti-interference ability pose further challenges for FANETs. MmWave communication technologies with abundant spectrum resources are promising to support high-rate transmission in FANETs. Moreover, the directivity of mmWave channels and the narrow beamwidth provide significant potential for enhancing the security and anti-interference capabilities of FANETs. However, most of the existing works on FANETs place emphasis on employing sub-6 GHz frequency bands for networking \cite{gupta2016survey,cao2018airbor,zafar2016flying,zheng2018adapti,gankhuyag2017robust,yang2018rendez,wei2018neighb,lakew2020routin}.

	In this section, we comprehensively discuss and analyze crucial issues and corresponding potential solutions for mmWave-enabled FANETs.
		Firstly, the network architecture for mmWave-UAV FANETs is presented. In particular, SDN will play an important role in network management due to the resulting flexibility and programmability. Secondly, we analyze how to establish and maintain links in mmWave-UAV FANETs. Specifically, the challenges and solutions for directional neighbor discovery, the comparative analysis of existing UAV routing strategies, and the potential of and advanced technologies for resource allocation are comprehensively surveyed. Then, we reveal the significant benefits of joint mmWave and sub-6~GHz band networking. Finally, security threats and potential solutions to these threats are discussed.

	\subsection{Network Architecture}
	The network architecture for mmWave-enabled FANETs is a foundation for network management and application. In a mmWave-enabled FANET, the network topology and routing stability may change frequently due to the 3D motions of UAVs and the directional transmission of mmWave signals, and thus the efficiency of the network management may be affected. Architectures of mmWave-enabled FANETs are still under research. Some common architectures and corresponding solutions that may be adopted for FANETs are introduced as follows.
	\subsubsection{Network Topology}
\begin{figure}[t]
		\centering
		\includegraphics[width=8.5 cm]{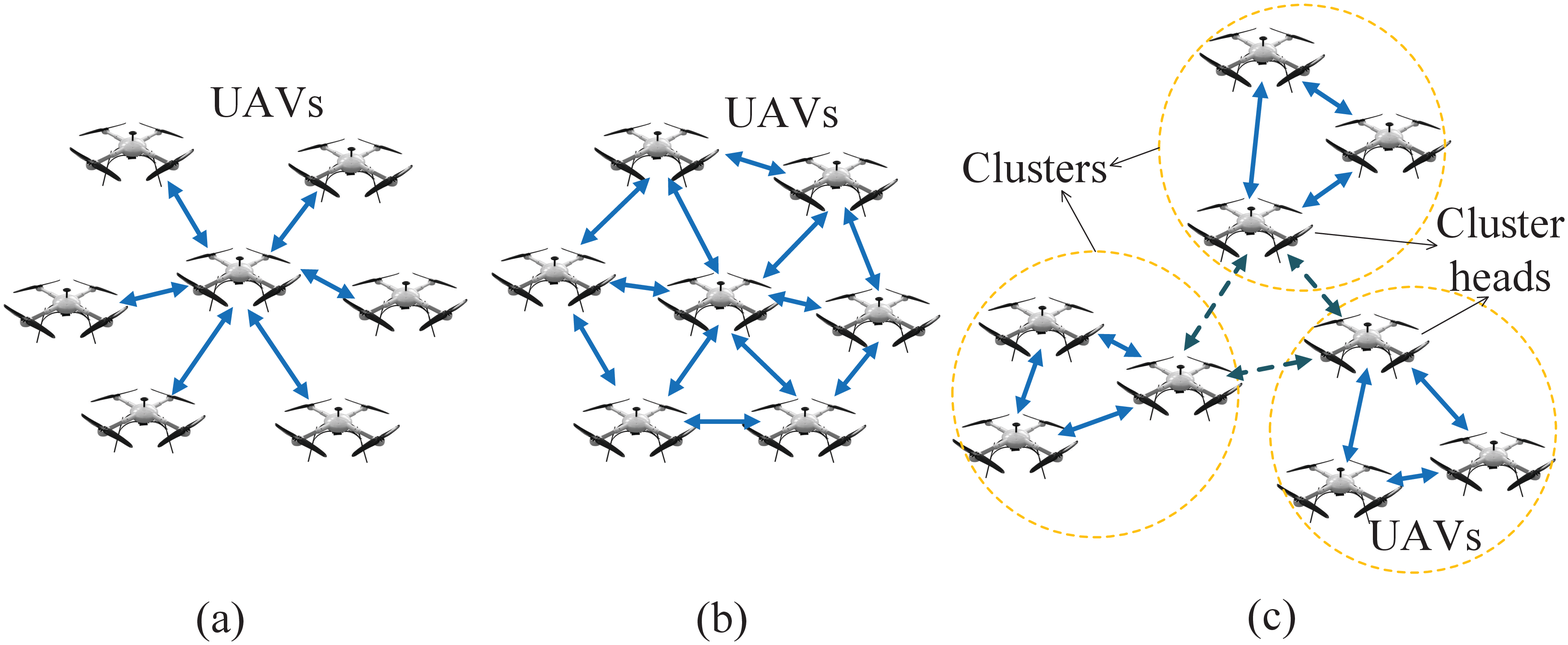}
		\caption{Three typical network topologies: (a) Star topology, (b) Mesh topology, (c) Hybrid hierarchical topology.}
		\label{fig:topo}
	\end{figure}

	Typical network topologies include star and mesh networks \cite{gupta2016survey}, as shown in Fig. \ref{fig:topo}.
	The star topology for mmWave-enabled FANETs is a relatively simple structure, where a UAV (or a ground BS) serves as the control and data-forwarding center, and other UAVs only connect with the control center. For a small-scale mmWave-UAV communication network, this structure is easy for control and management because of the short distance and less airframe shadowing. However, as the number and range of UAVs get larger, UAVs may suffer from severe path loss and much more airframe shadowing, and the center UAV may face nasty link congestion and interference. Once the center UAV breaks down, the entire network loses control and becomes paralyzed.
	Comparing to the star topology, the mesh topology has higher autonomy, flexibility, and invulnerability. A UAV node can associate with any other nodes via either single-hop or multi-hop routing to tackle the problems of high path loss and possible shadowing. Therefore, the mesh topology owns the features of highly-resilient reorganization and malfunction tolerance. However, the data packet transmission from a UAV node to the destination node may require high-complexity and high-overhead communication protocols, especially in large-scale mesh networks. In contrast, by employing a cluster-based scheme, the hybrid hierarchical topology is a good approach to reduce the complexity. Moreover, the inter-cluster and intra-cluster networks can select any type of topology according to application scenarios and management strategies. However, it also brings new problems. For example, the cluster size, cluster number, and cluster head should be carefully designed \cite{wang2017taking}. For all topology structures, the utilization of mmWave frequency bands can increase network throughput. Meanwhile, the directional transmission raises the difficulties for topology discovery and management, which will be discussed later.
	\subsubsection{SDN-Based Network}
\begin{figure}[t]
		\centering
		\includegraphics[width=8.5 cm]{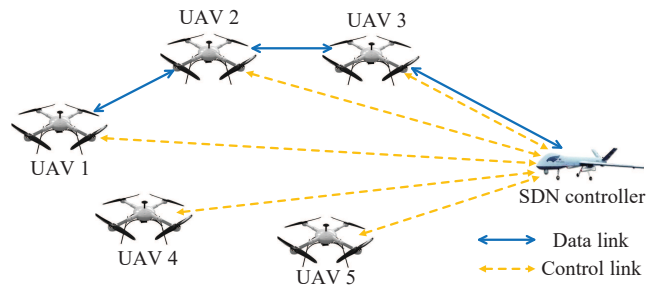}
		\caption{Architecture of a typical centralized SDN-based FANET.}
		\label{fig:SDN}
	\end{figure}

	Due to the heterogeneity of different aerial nodes, the reorganization of an FANET is limited by the hardware and protocol constraints. This problem can be resolved with the SDN architecture by programmatically controlling the network \cite{nunes2014asurve, kreutz2015softwa}. The introduction of SDN to FANETs helps different aerial nodes efficiently acquire the network state, and this scheme caters to the requirements of the routing and resource scheduling design in dynamic environments. Specifically, UAVs can equip programmable SDN switches (e.g., openflow \cite{hu2014asurve}), which contain flow tables and protocols for communicating with controllers. Some of the UAVs are installed with control facilities, which indicates that the control plane can be centralized, distributed, or hybrid. A typical centralized SDN-based FANET is shown in Fig. \ref{fig:SDN}.
	The control UAVs have a global view of the network, and thus global resources and traffic requests can be efficiently scheduled. Moreover, by decoupling control and data planes, the SDN architecture can increase the visibility and availability of the UAV network topology. It also enhances the abilities of the routing selection and network configuration. However, there are still some challenges to be addressed when using antenna array in mmWave frequency bands. First, in a large-scale mmWave-enabled FANET, the number of the control links is large, which makes the antenna configuration and beam management of the controller more difficult. The UAVs are usually equipped with limited number of RF chains, which means the number of the accessed nodes is limited for an SDN controller in a specific time slot. If a distributed architecture is used, the association assignment is an inevitable issue, but multi-beam may clash and lead to strong interference in the mmWave frequency bands. Besides, due to the separation of the control and data links, the spectrum efficiency may be affected. For different rate requirements of control and data information, careful channel allocation and bandwidth selection are required.

    \subsubsection{Summary and Discussion}
	In the above, we have focused on the issues and challenges arising for mmWave-enabled FANETs for various network architectures. The network architecture partly determines the characteristics and possible applications of the network. Although ad-hoc networks are not yet fully embedded in traditional cellular networks, there is no doubt that more and more new applications and intelligent technologies will facilitate the development of ad-hoc networks driven by 6G. The directionality of mmWave communication introduces both potentials and challenges for FANETs, while to the authors' best knowledge, research on the design of mmWave-enabled FANETs has not been conducted yet. As discussed before, compared to the centralized and distributed topologies, the hybrid topology is more suitable for large-scale mmWave-enabled FANETs, as it achieves a compromise between manageability and autonomy \cite{wang2017taking}. In addition, deploying SDN controllers on  UAVs has great potential. Specifically, the programmability and flexibility of SDN can support the UAVs to improve the communication protocols to meet different communication requirements \cite{nunes2014asurve, kreutz2015softwa, hu2014asurve}. Therefore, further research on SDN-assisted mmWave-enabled FANETs is important.

	\subsection{Link Establishment and Maintenance }
	Although the introduction of mmWave communications to FANETs may tremendously enhance the data rate and the anti-jamming capability of communication links, the directional characters of mmWave beams and the 3D motions of UAVs greatly challenge the link establishment and maintenance .
	Significant performance degradation and link outage can be observed from a large-scale on-the-moving UAV network in a mmWave frequency band \cite{guan2019onthee}. Due to the directional antennas and dynamic network topology, both beam steering and transmission path selection should be taken into consideration for designing the communication protocols. So far, the research works on mmWave-enabled FANETs are very limited. Hence, we analyze the specific issues in the link establishment and maintenance  of mmWave-enabled FANETs and discuss the possible solutions in the following, in order to inspire future research.
	\subsubsection{Neighbor Discovery}
	Before establishing a communication link between two nodes in a network, it is necessary to perceive and maintain the connection with each other, which is known as neighbor discovery (also called routing discovery). An effective neighbor discovery accelerates the implementation efficiency of upper-layer protocols and serves as a key foundation of the topology and networking. It usually requires the communicating parties to complete the Hello package transmission as agreed. Due to the 3D-space and mobility characteristics, UAVs may need frequent neighbor detections to maintain network connectivity. The easiest way is to always do neighbor discovery throughout the mission. However, the long-term neighbor discovery consumes much energy and resource, and unnecessary substantial overhead is generated. Therefore, the frequency of the neighbor-discovery operations should be carefully designed to balance the efficiency and overhead. Using directional transmit antennas and omnidirectional receive antennas, the authors in \cite{wei2018neighb} developed a two-way handshaking discovery scheme in 3D UAV networks, considering the deployment and mobility of UAVs. The Markov process was adopted to analyze the efficiency of the proposed scheme, and extensive simulation results showed that the overhead of neighbor discovery can be balanced. In fact, when the future motion information of a UAV formation is already known or can be correctly predicted, the frequency of neighbor discovery can be reduced. In addition, most of the existing schemes for neighbor discovery adopt synchronous clock, such as TDMA. For FANET, the distributed nodes may not have perfect synchronization, and thus, it is necessary to support the asynchronous scenarios \cite{yang2018rendez}.

	\begin{figure}[t]
		\centering
		\includegraphics[width=\figwidth cm]{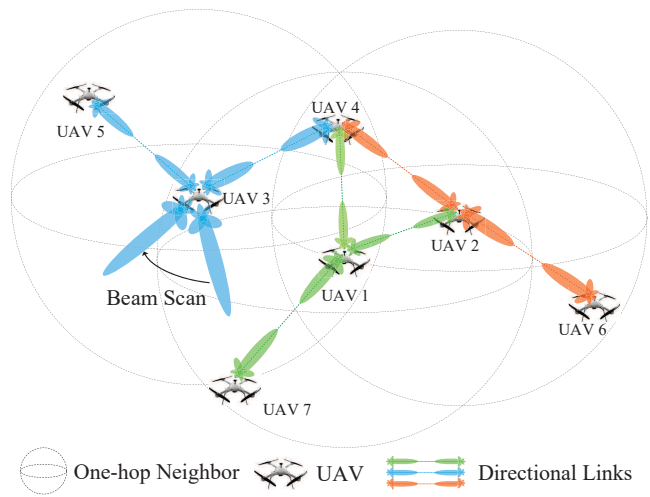}
		\caption{Neighbor discovery in mmWave-enabled FANET.}
		\label{fig:ND}
	\end{figure}
	The neighbor discovery in mmWave-enabled FANETs is shown in Fig. \ref{fig:ND}. Compared to the network employing omnidirectional antennas, the utilization of directional antennas can avoid signal conflicts and interference problems, and increases the distance of neighbor discovery. However, the characteristic of the directional transmission induces a serious misalignment problem (also called spatial rendezvous problem \cite{wang2017ondire}), which costs much more scanning time in a 3D space as compared to 2D. Specifically, all the control frames, e.g., Request-To-Send (RTS) or Clear-To-Send (CTS), have to transmit in a sector sweep manner. As a result, the real-time beam alignment causes serious delays and overheads. It is impractical to implement exhaustive beam scanning in a highly dynamic 3D UAV network. In addition to beam alignment in the spatial domain, perfect alignment and synchronization strategies in the time and frequency domains are also required, but further increase the overhead and delay. To reduce the adverse influence of the alignment problem, a possible solution is that the receiver nodes work in a quasi-omnidirectional manner in the mmWave transmission, similar to the directional neighbor discovery in the traditional sub-6\,GHz frequency bands \cite{gabriel2017neighb}. However, quasi-omnidirectional receivers may receive multiple beacons from different directions, resulting in collisions. For directional transmission in mmWave frequency bands, it is beneficial for nodes to have partial prior knowledge (e.g., the directions of the potential neighbor nodes) in order to rapidly discover neighbors and achieve fast convergence. However, acquiring real-time 3D positions and velocities of the moving UAVs is also difficult. Hence, the 3D motion prediction of UAVs is vital to simplify the neighbor discovery and maintain the network connection \cite{rodriguezfernandez2018positi}.
	
	\begin{table*}[t]\small
		\begin{center}
			\caption{Comparison of different schemes for neighbor discovery with directional antennas.}
			\label{Tab:discovery}
			\begin{spacing}{1.3} \begin{tabular}{|m{0.1\textwidth}<{\centering}|m{0.1\textwidth}<{\centering}|m{0.31\textwidth}<{\centering}|m{0.18\textwidth}<{\centering}|m{0.18\textwidth}<{\centering}|}
					\hline
					\textbf{Reference} &\textbf{Scheme} &\textbf{Main idea} &\textbf{Pros} &\textbf{Cons} \\
					\toprule[0.5pt]\hline
					\cite{wang2017ondire} &Deterministic &Steer beams in accordance with predefined sequences &Ensures successful discovery &High average delay \\
					\hline
					\cite{zhang2008neighb,an2011impact} &Probabilistic &Memoryless beam scanning &Low average delay compared to deterministic schemes &Possible discovery failure \\
					\hline
					\cite{chen2017onobli,riaz2018energy} &Pseudo-deterministic &Generate scanning sequences including both a deterministic
						and a random component &Tradeoff between low average delay and high successful discovery rate &May not fit in a 3D environment \\
					\hline
					\cite{huang2013aninte} &ML-based &Interact with the environment and learn from the experience direction based state space and successful discovery based reward &High successful discovery rate and low total delay for many neighbors &Possibly low scalability and high learning cost \\
					\hline
				\end{tabular}
			\end{spacing}
		\end{center}
	\end{table*}
	
	In addition, the problems of the deafness and hidden terminals challenge the design of neighbor discovery and upper-layer protocols for directional communications \cite{yan2019machin,dai2013anover}. When the nodes are working in a directional transmission mode, they can only receive the signals from the mainlobe direction. Due to the weak signal gain, the destination node may not receive RTS from the sidelobe direction and reply CTS in time. Thus, the sender nodes that locate in the sidelobe direction of the receiver become deaf nodes. On the other hand, for two sender nodes located in the same sector/beam of a destination node, one sender may not accomplish the directional RTS/CTS handshake with the destination node because the other sender is communicating with the destination node. This circumstance results in a deafness problem for the first sender node. The deafness problem leads to the short-term inequity and seriously influences the utilization of the space resource \cite{takata2007amacpr}. The hidden terminal problem in directional ad hoc networks is caused by two conflicting nodes that cannot listen to each other and send data to the same terminal at the same time \cite{alam2013neighb,choudhury2006ondesi}. Due to the directional transmission, some nodes may be unaware of the existing communication links in the network, and the conflict is induced when a node tries to send RTS to another node who is communicating. The hidden terminal problem may also be caused by the asymmetry in antenna gain because the communication ranges between the directional and omnidirectional modes are different \cite{alam2013neighb,choudhury2006ondesi}.	
	
	For a network using directional antennas in mmWave frequency bands, the existing neighbor discovery schemes can be categorized into two main classes, i.e., deterministic and probabilistic schemes. The probabilistic approach is memoryless, where the nodes randomly choose directions to steer their beams for neighbor discovery \cite{zhang2008neighb,an2011impact}. This approach usually performs better in terms of average discovery delay \cite{zhang2008neighb}. However, the main problem is that it does not ensure the successful discovery. In deterministic approach, the nodes steer their beams in accordance with predefined sequences \cite{wang2017ondire,chen2017onobli,riaz2018energy}. Although the average delay is usually larger than that in probabilistic approach, the neighbor discovery is guaranteed within a bounded time.
	
	To tackle the spatial rendezvous and deafness problems, the authors in \cite{wang2017ondire} proposed a deterministic approach, namely hunting-based directional neighbor discovery algorithm. The nodes rotate their beams to scan the potential neighbors in a 2D plane and the nodes with a faster angular velocity will catch up with the slower nodes. The condition of successful beacon-ACK handshake and an upper-bound on the worst discovery time were derived. To achieve a desirable tradeoff between the average and worst-case delays, the authors in \cite{chen2017onobli} and \cite{riaz2018energy} developed pseudo-deterministic schemes such that the process of generating scanning sequences includes both a deterministic and a random component in 60\,GHz networks. Under the case with heterogeneous beamwidths of the nodes and without any prior coordination or synchronisation, the authors in \cite{chen2017onobli} proposed an oblivious neighbor discovery algorithm. This strategy adopts a short extended address for the nodes, which means that the latency of scanning operations is small. Subsequently, the authors in \cite{riaz2018energy} used the above method for constructing extended identity sequences, and then utilized the Polya's enumeration theorem and Fredricksen, Kessler and Maiorana algorithm to find the shorter and efficient scanning sequences for the nodes. In \cite{hashima2020neighb}, the authors proposed a stochastic multi-armed bandit (MAB) online learning solution to resolve the neighbor discovery and selection in mmWave D2D networks. To reduce the discovery time and maximize the average throughput, a group of energy-constraint MAB based algorithms were developed. Since the networks are assumed quasi-static in 2D space, these solutions \cite{zhang2008neighb,an2011impact,wang2017ondire,chen2017onobli,riaz2018energy,hashima2020neighb} should be improved for being applied in a mmWave-enabled FANET. In Table \ref{Tab:discovery}, the different neighbor discovery schemes with directional antennas are compared. More research efforts and advanced technologies for neighbor discovery are required to meet the requirements of high throughput and low latency under complex 3D environments.

	ML emerges as a powerful tool and has been used for various purposes in UAV networks. Specifically, Q-Learning theory was applied in~\cite{huang2013aninte} for directional neighbor discovery in ad hoc networks. The Q-Learning based algorithm takes the antenna steering direction as the \emph{state}, the transmitting/receiving strategy as the \emph{action}, and the successful neighbor discovery as the \emph{reward}. By interacting with the environment and learning from the experience, the proposed algorithm shows better performance than conventional sequential scanning. However, the learning-based neighbor discovery needs a certain of successful samples, which tests the timeliness. Besides, it may be hard to update the changing topology for UAV nodes in FANETs. An alternative way is that each UAV only maintains the state of its neighbors and employs distributed intelligent decisions according to the available information. Hence, ML-based neighbor discovery methods in mmWave-enabled FANETs are worthy of more studies.

\begin{table*}[ht]\scriptsize
		\begin{center}
			\caption{Classification, main idea, and comparison of routing protocols in FANETs \cite{lakew2020routin}.}
			\label{Tab:route}
			\begin{spacing}{1.1}
\begin{threeparttable}
	\begin{tabular}{|m{1.6cm}<{\centering}|m{1.8cm}<{\centering}|m{3.55cm}<{\centering}|m{1.2cm}<{\centering}|m{1.6cm}<{\centering}|m{1.4cm}<{\centering}|m{1.35cm}<{\centering}|m{2.1cm}<{\centering}|}
		\hline
		\textbf{Classification}      &\textbf{Subclassification} &\textbf{Main idea}  &\textbf{Signaling overhead}  &\textbf{Communication latency}  &\textbf{Bandwidth and energy consumption}  &\textbf{Memory requirement}  &\textbf{Application scenarios}  \\ \toprule[0.5pt]\hline
		\multirow{4}{*}{\textbf{Topology-based}}
		&\textbf{Proactive}      &Each UAV locally stores and periodically refreshes a routing table  &Very high (periodical update)  &Very low  &Very high (frequent update)  &Very high (storage of network information)  &Small-scale and real-time applications  \\ \cline{2-8}
		&\textbf{Reactive (on-demand)}       &UAVs generally use the flooding technique to find routes only when there is a data transmission requirement &High (flooding) &High (route discovery)  &High (flooding)  &High (on-demand storage)  &Data collection or remote sensing for small-scale to medium-scale UAV networks \\ \cline{2-8}
		&\textbf{Hybrid}         &Depending on the type, characteristics and requirements of UAVs, different proactive or reactive protocols are adopted by different UAVs  &Medium  &Medium to high  &Medium  &High  &Reconnaissance search and rescue for small-scale to large-scale UAV networks   \\ \cline{2-8}
		&\textbf{Cluster-based}  &Aiming at the network scalability issue, this routing strategy integrates the cluster head selection, clustering and real-time management schemes into traditional protocols  &High (cluster maintenance)  &Medium  &Medium (different inter and intra protocols)  &Medium  &Military confrontation and network coverage for small-scale to large-scale UAV networks \\ \hline
		\multirow{2}{*}{\textbf{Geographic}}
		&\textbf{Non DTN}        &Using the local mobility prediction and the neighbor node closest to the destination as the routing metric, the UAV makes the decision on selecting the next hop    &Low  &Low  &Low  &Low (only neighbor information) &Cooperative monitoring, reconnaissance, and battlefield applications for small-scale to large-scale UAV networks  \\ \cline{2-8}
		&\textbf{DTN}            &Aiming at the problem of intermittent network connections, the UAV selects the appropriate next hop based on node movement and the store-carry-forward mechanism  &High  &High  &High  &Low (only neighbor information)  &Delay tolerant applications, such as video making and data collection, for small-scale to large-scale UAV networks  \\ \hline
		\multicolumn{2}{|m{3.41cm}<{\centering}|}{\textbf{Hybrid (topology-based and geographic})}
		&Based on the local node information and movement prediction obtained by the reactive routing of local flooding, the failure links use geographic routing to select alternative routing options  &Medium  &Medium  &Medium  &High  &Network coverage and multi-task cooperation for small-scale to large-scale UAV networks  \\ \hline
		\multicolumn{2}{|m{3.41cm}<{\centering}|}{\textbf{Bio-inspired}}
		&Through local communication with less complex interactions and the cooperative response ability to internal and external disturbances, evolutionary or swarm-based bionic algorithms are used to make routing decisions  &High  &Medium  &High  &Medium  &Intelligent searching and battlefield applications for for small-scale to large-scale UAV networks \\ \hline
	\end{tabular}
	\begin{tablenotes}
		\footnotesize
		\item DTN = delay tolerant networking.
	\end{tablenotes}
\end{threeparttable}
            \end{spacing}
		\end{center}
\end{table*}
	\subsubsection{Routing}
	In mmWave-enabled FANET, the routing design owns unique requirements. When a UAV needs to transmit its data to other UAVs or ground BSs, the data transfer paths should be selected under the consideration of the QoS requirements, associated data traffic, and network topology. Unlike traditional ad hoc networks, the 3D-mobility character of UAVs has to be considered in the routing design for FANETs. Moreover, the SWAP limitation, the unstable link management, and the frequent removal and addition of UAV nodes are relevant for routing design. Especially for mmWave-enabled FANET, it is essential to consider the information from different layers, such as the mmWave channel conditions and interference from physical layer, fault tolerance and hop count from the network layer, throughput and delay from the data link layer, and QoS requirements and reliability from the application layer \cite{fan2017asurve}. Since the high-speed UAV may cause high-dynamic topology, attention should also be paid to the link stability in routing. Under this circumstance, the authors in~\cite{Zhao2019RouteD} introduced Gaussian Markov moving model to describe the movement of UAVs, which achieved better packet delivery rate and lower end-to-end delay.

	The potential routing methods for FANETs can be classified into four types based on the strategies used, namely, topology-based, geographic, hybrid (topology-based and geographic), and bio-inspired routing, as shown in Table \ref{Tab:route}.
Specifically, in topology-based routing, the routing information from the source to the destination must be obtained from the network topology information before data transmission starts. Based on the assumption that each UAV knows its own location from an onboard positioning system, geographic routing utilizes the local geographic locations of the UAVs to make the data packet forwarding decisions. Hybrid routing generally combines reactive routing and geographic routing. Bio-inspired routing is inspired by biological systems. Based on \cite{lakew2020routin}, we illustrate the different routing categories in more detail, along with their main ideas, performance comparisons, and application scenarios in Table \ref{Tab:route}.
	For more information on FANET routing protocols, we refer to \cite{lakew2020routin}.
	However, the applicability of these routing methods in mmWave communications needs further study.
	In mmWave-enabled FANETs, beam scanning for directional neighbor discovery may increase the signaling overhead, delay, and energy consumption of proactive routing and flooding-based reactive routing. Geographic routing strategies require additional hardware at the UAVs \cite{lakew2020routin}. In addition, how to match the intermittent contact time of high-speed UAVs with the beam scanning time is also an important issue. To address these challenges, routing combined with reliable mobility prediction strategies may be a potential solution, but requires further investigation. In addition, for control signaling, a low-frequency omnidirectional strategy for routing is preferable.
	We note that, compared to traditional routing methods, bio-inspired routing algorithms can utilize their own self-organizing means to manage the dynamic features of FANETs. So far, there have been only a few works on bio-inspired routing protocols. As an example, artificial bee colony and ant colony algorithms have been used for routing in FANETs \cite{Leonov2016Modeli,Zhao2019RouteD,Leonov2016Applic}, where the routing discovery process in FANETs is modeled as the honey collection process in a bee colony or the food finding process in an ant colony. The obtained performance are very promising. However, these methods have a relatively high computational complexity. Thus, bio-inspired strategies need to be explored more in detail to verify their true potential.

	In mmWave-enabled FANET, the network topology and link quality are known to UAVs after accomplishing neighbor discover. That is to say, some of traditional routing protocols can be applied in mmWave-enabled FANET. The remaining issue is to select the data transfer paths, which can be modeled as a multi-commodity flow problem \cite{kolar2006amulti}. Since the FANET is a non-delay tolerant network, the commodity flows are time-dependent, which is an NP-hard problem and is different from that in static networks. A possible method is to use graph theory based strategies \cite{zhang2018adynam}. However, the modeling process of the routing path selection may be difficult because routing and resource allocation are coupled in general \cite{weeraddana2011resour, eisherif2014jointr}. Hence, how to decouple the corresponding problem and find a global solution is valuable to investigate.
	
	\subsubsection{Resource Allocation}	
	To enhance the performance of a network, the wireless resources need to be carefully allocated for mitigating the interference and improving the throughput. Compared to the sub-6\,GHz frequency bands, the design of the MAC protocols for mmWave-enabled FANETs is more challenging due to the directional transmission mode, dynamical link fluctuation, beam management, time-consuming beam alignment, etc. To maintain the high-efficiency link connection, it is vital to realize reasonable resource allocation and sharing of the space, time, frequency, and other limited resources for different nodes. IEEE 802.11aj \cite{ieee2018ieeest} and IEEE 802.11ay \cite{ieee2019ieeedr} have provided detailed MAC designs of wireless networks for supporting the mmWave frequency bands above 45\,GHz, but they are more suitable in low-mobility indoor environments. The existing research works in mmWave-enabled UAV networks mainly focus on the physical layer design, while the MAC layer research is still in its infancy.
	
	In the time domain, the mmWave transmission challenges the frame design. Most present mmWave-enabled UAV communication networks adopt the HD mode. Thus, in ad hoc networks, it is important to reasonably design the frame to guarantee efficient transmission and avoid collisions as multi-path and multi-hop routing generates \cite{niu2016boosti}. Applying the FD mode doubles the spectrum efficiency and decreases the network delay, but the scheduling algorithm should be carefully designed to handle the interference \cite{chang2018effici}. In the space domain, although beamforming technologies in mmWave frequency bands bring high spectrum efficiency and anti-jamming ability, the beam management is challenging due to the frequent change of the topology and connection. Specifically, as the relative direction and communication range change, the transmit beams must be realigned in a real time and the number of connected neighbor UAVs may change. Thus, fast beam tracking and resource reconfiguration methods should be used. Besides, the multiplexing of the beams in ad hoc networks will cause serious collision and interference, which should be properly designed. In the frequency domain, frequency-division assisted SDMA can avoid interference, but it reduces the spectrum efficiency and transmission bandwidth. Moreover, to maximize the network throughput, the global spectrum management and real-time allocation should be optimized according to the 3D network topology and time-varying interference. In addition, the transmit power control is also very crucial for mmWave-enabled FANETs in terms of the energy-efficiency. As the scale of a mmWave-enabled FANET increases, the computational complexity for resource optimization exponentially increases, which challenges the timeliness of the network management.

	In order to minimize the number of time slots for multi-path multi-hop transmissions, the authors in \cite{niu2016boosti} utilized two heuristic algorithms for traffic distribution and transmission scheduling to determine the frame structure. The scheme achieves a superior performance in terms of the delay and throughput comparing to the other directional MAC protocols, subject to the minimum traffic demand of all flows. Similarly, to improve the efficiency of concurrent transmission for mmWave networks, the authors in \cite{chang2018effici} redesigned the time slot to enlarge the scheduling space and proposed an efficient time-slot adjustment scheduling algorithm in the multi-hop packet forwarding process. However, the above works in mmWave wireless personal area networks (WPANs) may not be perfectly suited to mmWave-enabled FANET. In \cite{zhou2019beamma}, the authors proposed a fast beam tracking algorithm in mmWave-UAV mesh networks, where a self-healing request/response frame was designed to ensure the network robustness, and an efficient algorithm for the re-selection of the UAV group leader was developed to ensure high link quality between the group leader and ground BS. The proposed self-healing mechanism improves the robustness of the mmWave-UAV mesh networks and reduces the overhead in establishing the directional communication links comparing to existing MAC protocols. In \cite{zheng2018adapti}, using directional antennas in FANET, the authors proposed a position-prediction-based directional MAC protocol, including the position prediction, communication control, and data transmission phases. In the first phase, each UAV can be a position sender and directionally transmits its GPS-coordinate vector clockwise. The position packet brings only an extra 17 bytes of overhead. When a node acts as a sender, the other nodes are working as listeners until receiving the position packets. In the second phase, three control packet handshakings, i.e., RTS, CTS and Wait-To-Send (WTS), are executed. In the third phase, the transmitter UAVs steer their antennas and transmit data to the receivers. For channel scheduling, the authors in \cite{feng2019spectr} modeled the interaction of adjacent links as a 3D time-varying interference graph and utilized graph coloring method to allocate the mmWave channels in UAV swarm networks. This approach is a potential way to solve the channel allocation in mmWave-enabled FANET. However, each UAV needs to periodically carry out the interference measurement and channel estimation, which may require high system overhead.
	
	Traditional approaches for allocating the wireless resource are usually based on optimization techniques, e.g., greedy heuristic search \cite{wu2013flashl}, iterative methods for local optimum \cite{shen2017fplinq}, hyper-graph coloring \cite{feng2019spectr}, matching theory \cite{liu2021resour}, polyblock-based optimization \cite{qian2010smapel}. However, all these optimization methods require accurate CSI and may not perform well for a large-scale mmWave-enabled FANET. In contrast, ML is a promising approach for network optimization. In \cite{cui2019spatia}, the authors developed a deep learning approach, which bypassed the channel estimation and scheduled the links efficiently based on the geographic spatial information. The generalization ability of the neural network was demonstrated for different link density, which revealed the advantages compared to traditional optimization methods and heuristic algorithms. In a centralized wireless network with imperfect CSI, the authors in \cite{wadu2020federa} proposed a joint user scheduling and resource-block allocation scheme via federated learning, in which a Gaussian process regression based method and Lyapunov optimization framework were utilized to learn and track the wireless channel and to solve the stochastic optimization problem, respectively.

    \subsubsection{Summary and Discussion}
    Neighbor discovery is the basis for achieving self-organization, and is also the premise of routing and resource scheduling decisions. First, the frequency of neighbor discovery in FANETs should be optimized to balance efficiency and overhead. Although directional neighbor discovery can reduce interference and increase the detection range, it will introduce a serious spatial rendezvous problem \cite{wang2017ondire}, which can cause intolerable latency because of beam scanning in 3D space. A potential solution is to use prior knowledge of location or mobility prediction to assist directional neighbor discovery based on a pseudo-deterministic approach. In particular, neighbor discovery based on ML is promising for predicting the mobilities of the UAVs \cite{huang2013aninte}. However, this method needs a certain amount of data samples. Besides, the existing directional neighbor discovery strategies, as shown in Table \ref{Tab:discovery}, are based on 2D scenes, which cannot be well extended to 3D scenes. Moreover, the topology of the ad hoc network changes frequently, which implies that highly time efficient directional neighbor discovery schemes are needed.
    
    The design of routing and resource allocation schemes for mmWave-enabled FANETs is a coupled decision-making process. We have compared different routing strategies for FANETs in Table \ref{Tab:route}. Since directional beams pose new challenges, it remains to be seen whether these routing strategies are applicable to mmWave-UAV communications. When routing discovery is finished, the transmission path selection of the data links evolves into a network flow problem. However, different from traditional ad hoc networks, beamforming and the dynamic topology should be taken into account in mmWave-enabled FANETs. In fact, resource allocation greatly affects the performance of multi-flow and multi-hop data transmission. The resource allocation for mmWave-enabled FANETs needs to consider not only the original decision domains of time, frequency, and power, but also the beam domain. Although this will improve the network performance, it will make resource scheduling more complex. In addition, since routing and resource allocation significantly influence each other, it is usually difficult to obtain a globally optimal solution. Instead, suboptimal heuristic methods with low complexity are of interest for joint routing and resource allocation  in mmWave-enabled FANETs.

	\subsection{Integration of Sub-6\,GHz and MmWave Bands}
	Although the use of mmWave frequency bands brings new potentials for FANET, there are some challenging problems induced by the directional transmission as we discussed before. In practice, different frequency bands can be jointly utilized according to their characteristics and the requirements of different applications, such as the control link, data link, and target detection. The advantages of sub-6\,GHz and mmWave frequency bands can be combined for networking. In the following, the integration of sub-6\,GHz and mmWave frequency bands in FANETs will be discussed.
	
	First, the omnidirectional communications in sub-6\,GHz frequency bands can be utilized for network controlling. Due to the fluctuating communication links and complex beam-alignment operations in mmWave-enabled FANET, it is difficult to initialize and maintain stable connections. Meanwhile, since the directional beams of UAVs have to frequently scan, a high latency is induced and the delay-sensitive control messages cannot be delivered in time. Hence, utilizing the control channels under sub-6\,GHz is a good solution. For handling the mobility, UAVs can periodically exchange their location information via the control channel and conduct location prediction for neighbors \cite{peng2018aunifi,gankhuyag2017robust}. In addition, the resource allocation and routing commands can also be transmitted in the low-frequency control channel for improving the management efficiency of the network. Besides, the auxiliary information from the sub-6\,GHz channels can be used to assist to accomplish beam management and establish data links in the mmWave frequency bands. Second, the directional communications in mmWave frequency bands can be utilized to transmit high-rate and delay-tolerant data. Since the periodic prediction of the location information produces large amounts of control messages, the controlling on mobility management can be partly transferred to the data plane for releasing the overload on the low-capacity control channel \cite{zeng2019aduala}.

	In \cite{park2012cooper}, the authors proposed a cooperative neighbor discovery procedure, in which the 2.4\,GHz link was used to assist neighbor discovery and the 60\,GHz link was applied for high data transmission in ad hoc networks. Compared to the conventional directional neighbor discovery procedure, this scheme can reduce the average discovery time by 69\,\%-78\,\%. Furthermore, they expanded the scheme to integrate omnidirectional neighbor discovery and directional data transmission \cite{park2015multib}, namely the multi-band directional neighbor discovery. In particular, the proposed scheme provided compatible superframe structure to the IEEE 802.11a and IEEE 802.11ad specifications.
	
	However, resource allocation in such multi-band integration ad hoc networks still faces great challenges. First, resource allocation and interference control are imperative in the control plane. Channel allocation policies help avoid the interference by assigning different time slots or frequency channels to UAVs within the interference range. The channel allocation problem can be modeled as a graph coloring problem \cite{chlamtac1987distri,vishram2015achann}, where the colors (i.e., channels) are assigned to the UAVs to avoid conflict. In addition, game-theoretic approaches have attracted a lot of attention to resolve the resource allocation problem \cite{wang2008priceb,duarte2012onthep}. In \cite{wu2020colori}, the authors combined the graph coloring and game-theoretic approaches to reduce the co-channel interference and improve the channel reuse capability. Meanwhile, the authors proved the existence of a Nash equilibrium in the proposed graph coloring game and the convergence under the proposed distributed message-passing protocol. In addition to conventional optimization methods, online learning-based adaptive resource allocation approaches are also potential ways, but the dynamic modeling and low-complexity design are required for further investigation. Second, in the data plane, the beam management for achieving reliable and high-rate communications in multi-hop data transmission is challenging. In particular, due to the UAV jittering, there is a tradeoff between directional antenna gain and beamwidth \cite{dabiri2020analyt}, which poses a challenging requirement for beamforming.
	
\emph{Summary and Discussion:} Considering the advantages of  omnidirectional coverage and directional high-gain transmission, the integration of the sub-6\,GHz and mmWave bands holds great potential for FANETs. Omnidirectional transmissions of control information can enhance the efficiency of network management, and the high-rate, low-delay data can be transmitted through  mmWave links. However, it should be noted that the ground cellular network will introduce serious interference for aerial networks operating in sub-6\,GHz bands. Therefore, the proposed approach  needs to consider the potential unavailability of low-frequency links in real-world scenarios, and carry out resource scheduling and interference management in the control plane. In addition, the nature of mmWave-UAV links can also pose some challenges for directional data transmission, such as beam scheduling and beam design.

    \subsection{Security}
\begin{figure}[t]
		\centering  
		\includegraphics[width=\figwidth cm, trim=0 0 0 0 ,clip]{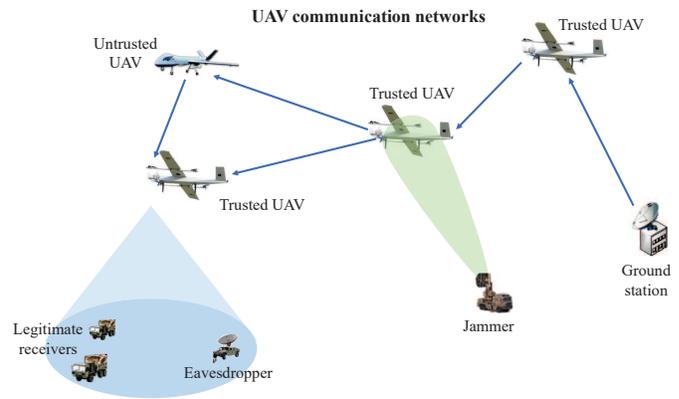}  
		\caption{Security attackers in UAV communication networks.}  
		\label{fig:security}   
	\end{figure}

	The utilization of mmWave frequency bands for FANETs improves the network security. In general, there are three primary types of attackers in wireless communication networks, i.e., eavesdroppers, jammers, and untrusted nodes \cite{sun2019physicMag, he2010cooper}. In the first scenario, given the open nature of wireless communications, eavesdroppers have the chance to intercept some confidential data, which leads to information leakage. In the second scenario, jammers intend to degrade the channel quality by transmitting jamming signals to the legitimate receivers, which leads to information loss. 
	In the third scenario, an untrusted node in the network may be unauthenticated or have a lower level of security than the other nodes, which provides an opportunity for eavesdropping if manipulated by criminals. 
	Fig.~\ref{fig:security} demonstrates an example of UAV communication network which includes three types of security attackers. In this scenario, UAVs form an aerial ad hoc network to transmit information to legitimate ground terminals, while the jammer transmits jamming signals to the relay UAV to degrade its channel, an eavesdropper is trying to intercept the classified information, and an untrusted UAV may try to decode the confidential information that they are relaying.
	
	\subsubsection{Security Metrics}
	\emph{Secrecy capacity} is the most widely used metric in physical layer security as shown in \eqref{secrecy_thr}, which is defined as the difference between the capacity of legitimate pair and that of the eavesdropping link. Another important metric is the \emph{secrecy outage probability}, which measures the probability that the rate redundancy of an information encoding scheme is smaller than the capacity of the eavesdropping link \cite{zheng2015optima}. Besides, there are three important metrics for measuring the performance of network-wide security, namely \emph{area secure link number (ASLN)}~\cite{wang2016physic}, \emph{network-wide secrecy throughput (NST)}~\cite{zhang2013enhanc}, and \emph{network-wide secrecy energy efficiency (NSEE)}~\cite{chen2013energy}. ASLN is defined as the average number of secure links per unit area, where a secure link refers to a communication link that neither connection outage nor secrecy outage occurs. Correspondingly, under the required connection outage and secrecy outage probabilities, the achievable rate of successful information transmission per unit area is called NST. Further, NSEE is defined as the ratio of NST to the power consumption per unit area, which is used to evaluate the energy efficiency.
	
	\begin{table*}[t]\small
		\begin{center}
			\caption{Description of and potential responses to main types of attackers in mmWave-enabled FANETs.}
			\label{Tab:security}
			\begin{spacing}{1.3} \begin{tabular}{|m{0.16\textwidth}<{\centering}|m{0.3\textwidth}<{\centering}|m{0.45\textwidth}<{\centering}|}
					\hline
					\textbf{Type of attackers} &\textbf{Description} &\textbf{Solutions for improving security} \\
					\toprule[0.5pt]\hline
					Eavesdropper &Intercepts data around legitimate receivers &Resource allocation and trajectory optimization, artificial noise, cooperative jamming, beamforming, CoMP \\
					\hline
					Jammer &Degrades channel
						quality by transmitting jamming signals to legitimate UAVs &CoMP, routing design\\
					\hline
					Untrusted UAV &May be unauthenticated or have a lower level of security than other nodes &Cooperative jamming, CoMP, routing design, IRS assisted methods \\
					\hline
				\end{tabular}
			\end{spacing}
		\end{center}
	\end{table*}
	
	\subsubsection{Anti-eavesdropping Techniques}
	Security issues for mmWave communication networks~\cite{wang2018hybrid,zhu2017secure,darwesh2020secrec,zhang2013enhanc,zhu2016physic}, UAV communication networks~\cite{wang2020energy,wu2019safegu,sun2019physicMag,sheng2020secure,ye2019secure}, and mmWave-UAV communication networks \cite{pang2021secrec,Yap1c12021physic} have attracted widespread attention. In addition to conventional anti-eavesdropping techniques, such as artificial noise and cooperative jamming \cite{wang2020energy,wu2019safegu,mukherjee2014princi}, the flexibility of UAV positioning and the directionality of mmWave transmission can be exploited for improving secure communications.
	
	For the links between two UAVs, the trajectory of UAVs can be cooperatively designed with the resource allocation to enhance the security~\cite{zhang2019securi}. It was demonstrated that the passive eavesdroppers may be detected by legitimate transceivers from the local oscillator powers that are inadvertently leaked from eavesdroppers' RF front end \cite{mukherjee2012detect}. With this information, a UAV could adjust its transmission power in different waypoints to increase secrecy rate and change its position and velocity to bypass or quickly pass through the eavesdroppers. Inevitably, this trajectory design leads to more energy consumption for propulsion. Considering mmWave communications, the A2A or A2G channels are very sparse in the angular domain, which entails highly directional transmission.
	Considering the distribution of the eavesdroppers, the secrecy rate can be effectively improved by setting a protected zone around users \cite{Yap1c12021physic}.
Besides, beamforming also improves transmission security by suppressing the signal power in the directions of the eavesdroppers~\cite{Wu2020energy,wang2021robust}. UAV mounted mmWave beamforming can make this process more efficient because the high altitude of the UAV can provide more elevation separation, while its horizontal position can be adjusted to achieve a better azimuth separation~\cite{wu2019safegu}. 
It was demonstrated that by adjusting the altitude of a UAV relay, the secrecy rate of the legitimate links can be improved \cite{pang2021secrec}. Furthermore, the deployment of IRSs also has a great potential for enhancing communication security \cite{direnzo2020smartr}. Specifically, an IRS can be deployed on the buildings to reflect the signals towards locations of eavesdroppers to create destructive interference. What's more, untrusted UAV can configure IRSs to relaying data of the high security requirement, because there are no information transmissions between untrusted UAVs and trusted UAVs.
	
	The network-layer technologies can also be used for the security design in a mmWave-enabled FANET. First, the multi-hop relaying transmission is an effective way to avoid information leakage by designing a route which bypasses eavesdroppers~\cite{rodriguez2015physic}. This strategy is also effective to antagonize jamming attacks, in which the most secure route needs to be selected. Second, UAV-assisted artificial jamming is a promising approach, where a number of UAVs in the network can be utilized to send artificial jamming signals to the eavesdroppers for deteriorating their wiretap channels~\cite{li2019uavena}. However, jamming signals not only interfere eavesdropping links but also impact legitimate transmissions. Thus, the transmit power of the cooperative jamming signal needs to be carefully designed to achieve high secrecy performance \cite{pang2021secrec}. Third, CoMP transmission can be utilized to enhance the network security, where multiple UAVs form virtual antenna arrays to improve the beam energy in the direction of legitimate receiver while degrading that in the other directions~\cite{sun2019physicMag}.

    \subsubsection{Summary and Discussion}
    In Table \ref{Tab:security}, we summarize the main types of attackers and potential responses for mmWave-enabled FANETs. In particular, the directionality of mmWave beams reduces the vulnerability of the network. Physical layer security techniques, such as artificial noise injection and cooperative jamming, have been proposed for UAV-assisted networks \cite{wang2020energy,wu2019safegu,mukherjee2014princi}. MmWave-UAV communications offers additional security. A single UAV can adjust its 3D position to avoid the jamming area. Meanwhile, beamforming can suppress the signal power in the direction of the eavesdroppers. Besides, cooperative jamming  with directional jamming signals can be used by multiple UAVs to improve the secrecy throughput. Furthermore, in mmWave-enabled FANETs, security aspects can be taken into account in the routing decisions in order to select a safe data transmission path. In the future, by combining UAVs with mmWave technology, novel anti-eavesdropping techniques can be developed to improve security.
	\section{Open Issues and Research Directions}
	Among the above descriptions and reviews, we have already mentioned some future researches. However, many of these problems for mmWave-UAV communications and networking have not been well addressed. In this section, we summarize the open issues and suggest some additional research directions.
	\subsection{Open Issues}
	\subsubsection{Channel Measurement and Modeling}
	Although mature modeling works and standards for mmWave channels have been established in the terrestrial networks, the A2A and A2G channel measurement and modeling in mmWave frequency bands are still in the initial stage. Particularly, the transmission characteristics of mmWave signals and 3D motions of UAVs doubly challenge the corresponding research. Most of the existing works about mmWave-enabled UAV communications focus on the performance evaluation and analysis, particularly with simplified static mmWave channels. Unfortunately, these works cannot fully capture the propagation characteristics and lack actual verifications in the real scenarios. The channel model is the foundation for the communication establishing and performance evaluating. However, overly complex channel models are tricky for theoretical analysis and optimization. Hence, it is vital to broadly measure the transmission parameters of mmWave-UAV communications and build universal channel models under different environments. On one hand, the precise modeling of the scatterers in different scenarios is important. Due to the high altitude of the UAV, the scatterers are usually distributed around the ground nodes, but it is worth noting that the airframes or aerofoils of big UAVs may also be scatterers. On the other hand, although there have been some works investigating the hovering and jittering of UAVs, the channel situation under the posture change of the moving UAVs is still valuable to study, especially in the presence of the turbulent wind field.

	\subsubsection{Robust Communication under Fast Moving and Jittering}
	Different from the terrestrial infrastructures, UAVs have a mobility character. Especially for a fixed-wing UAV, high-velocity navigation is a necessary condition for itself to obtain lift and maintain flying. Besides, the airflow disturbances and engine vibrations result in inevitable jittering of the fuselage. Due to the fast moving and jittering, the A2A channels between two UAVs and the A2G/G2A channels between a UAV and a ground equipment change rapidly over time. Furthermore, the moving and jittering of UAVs induce severe Doppler effect in the high-frequency bands. Large frequency shift and spread disrupt the orthogonality between the sub-carriers and thus result in inter-carrier interference. These problems lead to deterioration of the communication quality in mmWave-UAV communications. Hence, the robustness of mmWave-UAV communications should be guaranteed. Although there are some preliminary researches on this direction, such as performance analysis and beam tracking, there are still lots of real-world factors and open problems that have not been well addressed. The user discovery, initial access, handover, and Doppler compensation dedicated to mmWave-UAV communication systems require further study.
	
	\subsubsection{3D Placement and 3D Beamforming}
	Different from the traditional terrestrial networks, UAVs extend the communication scenario from 2D plane to 3D space. The horizontal position and altitude of a UAV can be flexibly adjusted to improve the channel quality, which provides new DoFs for the optimization of wireless communication systems \cite{you20193dtraj,you2020hybrid}. In addition, 3D beamforming is perfectly appropriate for mmWave-UAV communication systems. By employing large antenna arrays in the high frequency bands, the transceivers can perform flexible beamforming for compensation of the high propagation loss of the target signals and for mitigation the dominant interference at UAV platforms. Especially for mmWave-UAV communication systems, the joint 3D placement and 3D beamforming have promising potentials to enhance the performance metrics, such as the throughput, coverage, latency, and security.
	
	\subsubsection{Communication Under Imperfect CSI}
	For a UAV-BS equipped with a large antenna array, the beamforming has to be designed according to the knowledge of the channels between the UAV-BS and ground users. Given perfect CSI, the UAV positioning and beamforming can be jointly optimized for improvement of the communication service. However, perfect CSI is difficult to be acquired because of the high-dynamic channels, the high-overhead channel estimation, and the existence of noise. Furthermore, the channels between the UAV and users are influenced by the 3D position of the UAV, and the channels for different positions of the UAV are determined by the physical environment, which are more challenging to acquire. Therefore, the acquisition of the CSI for different environments and the communication design under inaccurate CSI are two important problems for mmWave-UAV communication systems. In general, to acquire more environment information, a UAV has to move to different positions and measure the channel qualities, which lead to higher system overhead. The tradeoff between the communication performance and the system overhead may be carefully investigated.

    \subsubsection{Fast Tracking and Handover for Next-Generation Mobile Networks}
	B5G/6G wireless systems will take advantage of mmWave frequencies and beamforming to enable data transmission at up to Tbps rates. However, communication based on beamforming is highly directional and is sensitive to movement. In order to ensure seamless connection, fast beam tracking and inter-cell and intra-cell beam handovers are required for ground mobile UEs and UAVs. Beam tracking and handover decisions are mainly determined based on the received signal power, which can be estimated by periodically sending probe frames omnidirectionally or quasi-omnidirectionally to determine the tracking direction and  to select a beam for data transmission. In addition, the ground stations can cooperate to locate the aerial/terrestrial UEs and schedule beams to realize fast tracking and handover. However, for A2A high-speed communications and next-generation mobile networks, new methods for reducing the tracking and detection signaling overhead and the time needed for beam training have to be developed.

	\subsubsection{Joint Resource Allocation and Routing}
	For mmWave-UAV ad hoc network, the resource allocation in physical and MAC layers and the routing in network layer are highly coupled. The physical resources, such as time slot, carrier frequency, beam, and power involve multiple dimensions and should be carefully scheduled according to the communication task. Besides, due to the high mobility of UAVs, the topology of UAV network is rapidly changing, which results in a time-sensitive requirement of routing programming. For different communication tasks, the UAV network should allocate the resources in an active manner to establish physical links, as well as update the data routing according to the network state. The cross-layer optimization for mmWave-UAV ad hoc network is rarely studied in the existing works and more research efforts may be devoted to this direction.
	
	\subsubsection{AI for MmWave-UAV Communications}
	Taking a wide variety of application tasks, mmWave-UAV communications and networking are becoming more complicated, decentralized, and autonomous. Some problems may not be tractable by employing traditional model-driven methods. In contrast, AI technologies can be utilized for creating intelligent mmWave-UAV communication systems and networks. The data-driven methodology exhibits unparalleled properties, such as model-free, adaptive, scalable, and distributed characteristics. For example, the accurate beam alignment is crucial for mmWave-UAV communications, while conventional beam sweeping methods require high system-level and network-level overheads. AI is ideal to develop quick response mechanisms and to select the optimal beam based on prior decisions and environment information. Besides, for UAV swarm, due to the high mobility and rapid-changing topology, routing and resource allocation are quite challenging. AI is promising to achieve cross-layer optimization in a distributed manner and reduce the computation delay. In summary, AI is a powerful approach to realize a mmWave-UAV communication system having rapid response, adaptive learning, and intelligent decision.
	
	\subsubsection{IRS/RIS Aided MmWave-UAV Communications}
	Conventionally, wireless channel is regarded as a stochastic and uncontrollable entity. The emergence of IRS/RIS has changed this status, allowing us to control the propagation environment via programmable elements. For mmWave communications, the signals are more vulnerable to blockages, while the employment of IRS and UAV can efficiently alleviate this issue. On one hand, a UAV may adjust its position to establish communication links with better channel conditions, such as LoS environment and low path loss. On the other hand, an IRS may assist to improve the channel condition via a virtual LoS path and passive beamforming. Furthermore, employing on-ground or on-board IRSs, the placement and trajectory of UAVs and the passive beamforming of IRSs may be jointly optimized to leverage more DoFs in mmWave-UAV communication systems. Considering the network realization, more efficient strategies are needed to achieve a good balance between computational complexity and system performance.
	
    \subsubsection{Hardware Design and Implementation}
    Although the research on mmWave beamforming enabled UAV communications has increased rapidly in recent years, most works focus on theoretical aspects. Nevertheless, hardware implementation is one of the key factors that affect the development of mmWave-UAV communications.
    	Onboard hardware devices must be lightweight, low-power, low-cost, high-integration, high-precision, and controllable. To this end, firstly, it is necessary to investigate advanced technologies and materials to design the required low-cost and  power-efficient multi-polarized antenna array modules. For instance, a cross-polarization leakage canceller may be needed to improve the isolation of the transmitter array and receiver array on the UAV. Besides, it is also important to develop high-integration and high-precision RF front ends to meet the space limitation of UAVs. Integrated passive device technology can provide ultra-miniatured and multifunctional RF chips \cite{yang2020anultr}. However, compared to other integrated processes, integrated passive device technology may result in a poor stopband rejection and frequency selectivity \cite{yang2020anultr}. Therefore, further research on this promising technology is needed to make it a viable hardware option for mmWave-enabled UAVs.

	\subsection{Additional Research Directions}
	\subsubsection{Space-Air-Ground Integrated Networks}
	As the traffic requirements of various services increase, space-air-ground integrated network is becoming a promising architecture to enhance the traditional terrestrial networks. Particularly, geostationary orbit, medium earth orbit, and low earth orbit satellites can provide long-term and seamless service for global areas, and mmWave technologies have been widely utilized in satellite communications. In addition, air networks consisting of UAVs, airships, and balloons can provide on-demand service and far-ranging coverage, where mmWave communication is a potential technology to support the large-capacity requirements. As a 3D and heterogeneous network, SAGIN involves different communication protocols and segments for high-efficiency and secure data transmission \cite{liu2018spacea}. High transmission delay is the bottleneck restricting satellite communications, especially for time sensitive applications. A potential solution is caching data streams through UAV or terrestrial nodes \cite{xu2018overco}. Besides, SAGIN faces other challenges, such as protocol design, mobility management, routing scheduling, load balancing, resource management and scheduling, QoS requirements, traffic control, and security, which require more research efforts.
	
	\subsubsection{Air-and-Sea Integrated Communications}
	With the rapid development of marine economy, reliable and high-throughput maritime wireless communication has become a more and more important requirement \cite{zhang2020placem}. Conventionally, satellite maritime communication faces the limitations of large propagation delay and high implementation cost, and the marine MF/HF/VHF-based radio communication has insufficient bandwidth. With the aid of UAV relays, the offshore ships and devices can be connected with the ground stations via low-latency and high-capacity communication links. Moreover, for far-ocean area, aircrafts and vessels may form air-and-sea integrated networks for efficient information exchanging and situation sharing. The mmWave frequency bands have potentials to be exploited for supporting high data rate in maritime communication. Due to the special physical environment over the sea surface, the electromagnetic characteristics of the air-to-sea channels are different from the conventional A2A and A2G channels. Since the study of air-and-sea integrated communication is still in an initial stage, more research efforts are needed to explore the communication and network design.

	\subsubsection{Joint Radar and Communication}
	Recently, mmWave ground station radars have been studied for locating and tracking UAVs and are expected to be used in future UAV traffic management systems \cite{rai2021locali}. In addition, the integration of onboard radar and communication has become a strong demand in some scenarios where space and power are extremely limited \cite{feng2020jointr}. For medium-scale and large-scale UAVs, an airborne radar is used to detect and identify target. The user discovery, channel estimation, and beam tracking problems for a communication subsystem may be effectively solved with the assistance of the detection information from a radar subsystem, especially for mmWave radar with a high resolution. On the other hand, the performance of a single on-board radar is limited, while the communication subsystem can help establish multi-radar networks for realizing a rapid fusion of numerous detection data. The joint radar and communication system is possible to reuse the spectrum and improve the automation level of a single UAV or a UAV network. However, there are still many problems to be addressed for joint radar and communication, such as architecture unification, waveform design, joint resource sharing, and multidimensional signal processing.
	
	\subsubsection{Energy Harvesting for MmWave-UAV Communications}
	As we have mentioned before, the high energy consumption is one of the challenging problems for mmWave-UAV communications. Although the battery capacity is gradually increasing with the emergence of new materials, the  immanent contradiction between the size and capacity has always been a bottleneck. In this scenario, energy harvesting enables UAVs with the ability of sustainable communication without power infrastructures. By harvesting energy from ambient environment, such as solar and wind energy, a UAV is able to realize perpetual flight and provide continuous communication service to ground users \cite{sun2019opttraj}. Under the constraint of harvesting energy, there is a basic tradeoff between propulsion energy and communication performance. Hence, the resource allocation for energy harvesting enabled mmWave-UAV communication systems requires future study.

	\section{Conclusions}
	In this paper, we provided a comprehensive survey on mmWave beamforming enabled UAV communications and networking. First, we analyzed the technical potentials of the combination of UAV and mmWave communications, and summarized the key challenges on mmWave-UAV communication systems. Then, we provided an overview on the antenna structures that may be employed for UAV platforms in the mmWave frequency bands. Whereafter, we reviewed the characteristics and modeling of mmWave-UAV communication channels. Subsequently, the key enabling technologies for UAV-connected mmWave cellular networks were discussed, including methods for performance analysis, beam coverage, access and backhaul, aerial relays, and techniques for supporting aerial UEs in mmWave cellular networks. Furthermore, important aspects for facilitating mmWave-UAV ad hoc networks were discussed, including the network architecture, neighbor discovery, resource allocation, routing, integrated sub-6 GHz and mmWave frequency bands, and communication security. Finally, we summarized open issues and potential research directions based on our survey for mmWave beamforming enabled UAV communications and networking, so as to motivate further research on these topics.

	\bibliographystyle{IEEEtran} 
	\bibliography{IEEEabrv,bib_zhu,bib_liu,bib_yi}
	
\end{document}